\documentclass[twocolumn,amsmath,amssymb,prd]{revtex4}

\def\al{\alpha}
\def\be{\beta}
\def\ga{\gamma}
\def\de{\delta}
\def\ep{\epsilon}
\def\ve{\varepsilon}
\def\ze{\zeta}
\def\et{\eta}
\def\th{\theta}

\def\ka{\kappa}
\def\la{\lambda}

\def\si{\sigma}

\def\ph{\phi}

\def\ch{\chi}
\def\ps{\psi}
\def\om{\omega}

\def\De{\Delta}

\def\La{\Lambda}

\def\Ps{\Psi}
\def\Om{\Omega}

\def\cl{{\cal L}}

\def\cP{{\cal P}}

\def\cT{{\cal T}}

\def\fr#1#2{{{#1}\over{#2}}}
\def\frac#1#2{{\textstyle{{#1}\over{#2}}}}
\def\half{{\textstyle{1\over 2}}}
\def\ol{\overline}
\def\prt{\partial}

\def\Re{\hbox{Re}\,}
\def\Im{\hbox{Im}\,}

\def\lsim{\mathrel{\rlap{\lower4pt\hbox{\hskip1pt$\sim$}}
    \raise1pt\hbox{$<$}}}
\def\gsim{\mathrel{\rlap{\lower4pt\hbox{\hskip1pt$\sim$}}
    \raise1pt\hbox{$>$}}}

\def\etal{{\it et al.}}

\def\vev#1{\langle {#1}\rangle}

\def\bra#1{\langle{#1}|}
\def\ket#1{|{#1}\rangle}

\def\sqr#1#2{{\vcenter{\vbox{\hrule height.#2pt
         \hbox{\vrule width.#2pt height#1pt \kern#1pt
         \vrule width.#2pt}
         \hrule height.#2pt}}}}

\newcommand{\beq}{\begin{equation}}
\newcommand{\eeq}{\end{equation}}
\newcommand{\bea}{\begin{eqnarray}}
\newcommand{\eea}{\end{eqnarray}}
\newcommand{\rf}[1]{(\ref{#1})}

\newcommand{\bM}{\begin{pmatrix}}
\newcommand{\eM}{\end{pmatrix}}

\def\w{w}
\def\f{w}

\def\psfb{\ol{\ps_\f}{}}

\def\mbf#1{\boldsymbol #1}

\def\syjm#1#2{{}_{#1}Y_{#2}}

\def\Q{\mathcal Q}

\def\V{\mathcal V}

\def\T{\mathcal T}

\def\K{\mathcal K}

\def\pvec{\mbf p}

\def\sivec{\mbf\si}

\def\bevec{\mbf\be}

\def\pmag{|\pvec|}

\def\punit{\hat p}

\def\epunit{\hat\ep}
\def\thunit{\hat\th}
\def\phunit{\hat\ph}

\def\phat{\mbf\punit}

\def\ephat{\mbf\epunit}
\def\thhat{\mbf\thunit}
\def\phhat{\mbf\phunit}

\def\Qhat{\widehat\Q}

\def\gt{\widetilde g}
\def\Ht{\widetilde H}

\def\Heff#1{{\Ht}_{\f,{\rm eff}}^{#1}}
\def\geff#1{{\gt}_{\f,{\rm eff}}^{#1}}

\def\X{X}
\def\Y{Y}
\def\Z{Z}
\def\Xhat{\widehat\X}
\def\Yhat{\widehat\Y}
\def\Zhat{\widehat\Z}

\def\nr{{\rm NR}}

\def\nrtemplate#1#2#3{#1^{\nr#3}_{#2}}

\def\anr#1{\nrtemplate{a}{#1}{}}

\def\cnr#1{\nrtemplate{c}{#1}{}}

\def\gzBnr#1{\nrtemplate{g}{#1}{(0B)}}
\def\goBnr#1{\nrtemplate{g}{#1}{(1B)}}

\def\HzBnr#1{\nrtemplate{H}{#1}{(0B)}}
\def\HoBnr#1{\nrtemplate{H}{#1}{(1B)}}

\def\Vnr#1{\nrtemplate{\V}{#1}{}}

\def\ToEnr#1{\nrtemplate{\T}{#1}{(1E)}}
\def\Vnrf#1#2{\nrtemplate{{\V_{#1}}}{#2}{}}
\def\TzBnrf#1#2{\nrtemplate{{\T_{#1}}}{#2}{(0B)}}
\def\ToEnrf#1#2{\nrtemplate{{\T_{#1}}}{#2}{(1E)}}
\def\ToBnrf#1#2{\nrtemplate{{\T_{#1}}}{#2}{(1B)}}

\def\anrf#1#2{\nrtemplate{{a_{#1}}}{#2}{}}
\def\cnrf#1#2{\nrtemplate{{c_{#1}}}{#2}{}}
\def\gzBnrf#1#2{\nrtemplate{{g_{#1}}}{#2}{(0B)}}
\def\goBnrf#1#2{\nrtemplate{{g_{#1}}}{#2}{(1B)}}

\def\HzBnrf#1#2{\nrtemplate{{H_{#1}}}{#2}{(0B)}}
\def\HoBnrf#1#2{\nrtemplate{{H_{#1}}}{#2}{(1B)}}

\def\widecheck#1{\hskip#1pt\huge$\check{}$}
\def\bighacek#1#2{\vbox{\ialign{##\crcr\widecheck#2\crcr
  \noalign{\kern-9.5pt\nointerlineskip}
   $\hfil\displaystyle{#1}\hfil$\crcr}}}

\def\mr{m_{\rm r}}

\def\k{k}

\def\atm{{\rm H}}
\def\atmb{{\overline{\rm H}}}
\def\AzB{\La^{(0B)}}
\def\AoB{\La^{(1B)}}
\def\AzE{\La^{(0E)}}
\def\Agen{\La^{(qP)}_j}

\def\chM{\vartheta}
\def\phM{\varphi}
\def\AM{A}

\def\eb{{\overline{e}}}
\def\pb{{\overline{p}}}
\def\epb{{\overline{\ep}}}

\def\Tb{{\overline{\mathcal T}}}
\def\ab{{\overline{a}}}
\def\cb{{\overline{c}}}
\def\fb{{\overline{\f}}}

\def\nub{{\overline{\nu}}}
\def\AMb{{\overline{\AM}}}
\def\Rb{{\overline{R}}}
\def\Hb{{\overline{H}}}
\def\Tb{{\overline{T}}}

\def\ring#1{{\mathaccent'27 #1}}

\def\nrfctemplate#1#2{\nrtemplate{\ring{#1}}{#2}{}}
\def\anrfc#1{\nrfctemplate{a}{#1}}
\def\cnrfc#1{\nrfctemplate{c}{#1}}

\def\fctemplate#1#2#3{\ring{#1}^{(#2)}_{#3}}
\def\afc#1#2{\fctemplate{a}{#1}{#2}}
\def\cfc#1#2{\fctemplate{c}{#1}{#2}}

\def\afcb#1#2{\fctemplate{\ab}{#1}{#2}}
\def\cfcb#1#2{\fctemplate{\cb}{#1}{#2}}

\def\pd{{pd}}

\begin{document}
\title{
Lorentz and CPT tests with hydrogen, antihydrogen, and related systems
}

\author{V.\ Alan Kosteleck\'y and Arnaldo J.\ Vargas}

\affiliation{Physics Department, Indiana University, 
Bloomington, Indiana 47405, USA}

\date{IUHET 592, June 2015}

\begin{abstract}

The potential of precision spectroscopy as a tool
in systematic searches for effects of Lorentz and CPT violation
is investigated.
Systems considered include 
hydrogen, antihydrogen, deuterium, positronium, 
and hydrogen molecules and molecular ions.
Perturbative shifts in energy levels
and key transition frequencies are derived,
allowing for Lorentz-violating operators of arbitrary mass dimensions.
Observable effects are deduced 
from various direct measurements,
sidereal and annual variations,
comparisons among species,
and gravitational responses.
We use existing data to place new and improved constraints
on nonrelativistic coefficients for Lorentz and CPT violation,
and we provide estimates for the future attainable reach
in direct spectroscopy of the various systems
or tests with hydrogen and deuterium masers. 
The results reveal prospective sensitivities
to many coefficients unmeasured to date,
along with potential improvements of a billionfold or more
over certain existing results.

\end{abstract}

\maketitle

\section{Introduction}

Hydrogen spectroscopy has been intimately linked
with precision tests of the foundations of relativity 
since the exact solution of the Dirac equation for hydrogen
\cite{dirac,darwingordon}
matched relativistic quantum mechanics with experiments.
Indeed,
a famous classic test of special relativity,
the Ives-Stilwell experiment confirming time dilation
\cite{ivesstilwell},
was first performed using a hydrogen clock. 
Another classic experiment,
Gravity Probe A
\cite{gpa},
verified the relativistic frequency shift in a gravitational field 
using a hydrogen maser launched on a suborbital rocket.

The underlying symmetry of relativity,
Lorentz invariance,
can naturally be broken in some approaches
to the unification of gravity with quantum physics
such as string theory
\cite{ksp}.
This possibility opens the door 
to the experimental detection of new physics 
emerging from the Planck scale 
$M_P \simeq 10^{19}$ GeV,
and it has led to
numerous sensitive tests of relativity
using techniques from various subdisciplines of physics
\cite{tables}.
In the present work we further this program,
studying the prospects for signals of Lorentz violation
using spectroscopy of hydrogen, antihydrogen,
and related systems,
including deuterium, positronium, 
and hydrogen molecules and molecular ions.

The methods of effective field theory offer a powerful and general approach 
to describing physical phenomena at accessible scales 
when the fundamental theory at a larger scale is unknown
\cite{sw}. 
The general realistic effective field theory for Lorentz violation,
the Standard-Model Extension (SME)
\cite{ck,akgrav},
is built from General Relativity and the Standard Model of particle physics
by adding to the action all coordinate-independent contractions 
of Lorentz-violating operators with controlling coefficients.
Operators of larger mass dimension $d$
can be viewed as higher-order effects
in a large-distance expansion of the underlying physics.
Since CPT violation in effective field theory breaks Lorentz symmetry 
\cite{ck,owg},
the SME also provides a general description of CPT violation.
The limit of the SME restricted to operators with $d\leq 4$ 
is called the minimal SME,
and it is power-counting renormalizable in Minkowski spacetime
\cite{reviews}.

The minimal-SME terms generate striking effects 
in the spectra of hydrogen and antihydrogen,
including CPT-violating signals 
and shifts in the hyperfine and $1S$-$2S$ transitions
that depend on sidereal time
\cite{bkr}.
Published searches for these effects
have measured the hyperfine splitting using a hydrogen maser 
\cite{hu00,maser,hu03}
and compared the $1S$-$2S$ transition in atomic hydrogen
to a cesium fountain clock
\cite{ba10,ma13}.
Related experiments with antihydrogen are being developed 
\cite{ga02,alpha11,asacusa15,ya13},
and experiments with hydrogen molecules and molecular ions
have been proposed as well
\cite{mu04}.
In the context of the minimal SME,
theoretical modifications to the spectra of hydrogen and antihydrogen
have been widely studied
\cite{bkr,ch00,sh05,fe07,kh07,fr08,yo12,go14,ma14},
while spectral shifts are also known to arise from 
specialized nonminimal SME interactions with $d=5$
\cite{d5H}
and from $d=6$ terms originating in noncommutative quantum field theory
\cite{chklo,ncqft}.
The minimal SME also introduces
CPT-violating effects in positronium decay
\cite{ve04,ad10}.

Here,
we investigate the prospects for spectroscopic searches
for Lorentz and CPT violation 
using hydrogen, antihydrogen, deuterium, positronium, 
and hydrogen molecules and molecular ions.
We focus on effects that arise from general Lorentz and CPT violation
in the propagation of electrons, protons, neutrons, and their antiparticles.
An analysis of this type has recently become feasible 
following the detailed classification and enumeration
of Lorentz-violating modifications to the Dirac equation at arbitrary $d$
\cite{km13},
which includes operators 
of both renormalizable and nonrenormalizable dimensions.
Operators of higher $d$ are of crucial interest 
in several contexts including,
for example,
foundational perspectives
such as causality and stability 
\cite{akrl,causality}
or the underlying pseudo-Riemann-Finsler geometry 
\cite{finsler1,finsler2},
practical issues such as 
the mixing of operators of different $d$ through radiative corrections
\cite{clp14},
and phenomenological effects arising in certain theories
such as supersymmetric Lorentz-violating models 
\cite{susy}
or noncommutative quantum electrodynamics
\cite{chklo,ncqft,ncqft2}.
The spectroscopic experiments proposed here
therefore have potential to bear on many aspects
of Lorentz and CPT violation.

Dimensional analysis reveals that operators with larger $d$ 
can be expected to produce signals growing with energy,
whereas the spectroscopic experiments of interest here 
typically involve nonrelativistic species.
Remarkably,
however,
the nonrelativistic observables for Lorentz violation 
turn out to be combinations of operators of arbitrary $d$
\cite{km13},
while spectroscopic methods can achieve high sensitivity,
so the experiments proposed here are competitive
with other techniques.
Our treatment disregards possible Lorentz-violating interactions,
as these produce suppressed effects.
For instance,
the dominant contributions to the various spectra obtained below
are independent of the internal electromagnetic four-vector potential
in Coulomb gauge,
while any applied external electromagnetic fields 
are minuscule compared to the electron and proton masses
and so their Lorentz-violating effects are heavily suppressed.
This approach is consistent with other studies 
of both minimal and nonminimal effects
in conventional and muonic atoms 
\cite{kla,bkl00,gkv14}.
We also disregard possible flavor-changing effects,
which involve simultaneous lepton- or baryon-number violation
with Lorentz violation
and so can reasonably be taken as smaller
than the effects considered here. 

Including the present introduction,
the main text of the paper is divided into eight sections.
In Sec.\ \ref{Theory},
we initiate the explicit discussion of Lorentz and CPT violation
in hydrogen by presenting the underlying theoretical calculations
required to analyze spectroscopic experiments.
Some basic background information
about perturbation theory involving operators
of arbitrary $d$ is provided in Sec.\ \ref{Basics},
followed by a discussion in Sec.\ \ref{Coefficient selection rules}
establishing the coefficients for Lorentz violation
relevant for hydrogen spectroscopy.
Sec.\ \ref{Matrix elements} contains the derivation 
of the matrix elements for the calculation of the perturbative energy shift
due to Lorentz and CPT violation,
including both general results
and analytical expressions for special cases.
In Sec.\ \ref{Weak magnetic field},
we address the modifications arising
from the presence of a external magnetic field,
including the key equations underlying
the resulting sidereal and annual variations 
of the Lorentz- and CPT-violating energy-level shifts. 

With the theory in hand,
the analysis of various experimental scenarios 
for hydrogen spectroscopy becomes feasible.
This is addressed in Sec.\ \ref{Applications}.
We first consider the case of free hydrogen
in the absence of applied fields.
The effects of Lorentz and CPT violation
on the transition probabilities and line shapes
are discussed in Sec.\ \ref{no field},
along with the prospects for measuring signals.
We then turn in Sec.\ \ref{$1S$ hyperfine-Zeeman}
to the hyperfine Zeeman spectroscopy of hydrogen,
presenting the perturbative frequency shift
and studying signals from sidereal variations and
from changes in the orientation of the applied magnetic field,
corrections due to boosts,
and the prospects for a space-based mission. 
This is followed 
in Secs.\ \ref{The nSnP transitions} and \ref{The nSnD transitions}
by an investigation of potentially observable effects 
in various $nL$-$n'L'$ transitions
for which precision measurement in hydrogen is experimentally feasible.
Where possible in all these applications,
we use existing data to extract first or improved constraints
on nonminimal coefficients 
and estimate sensitivities attainable in future experiments. 

Following the discussion of hydrogen,
we turn attention in Sec.\ \ref{Antihydrogen}
to searches for Lorentz and CPT violation using antihydrogen.
We begin in Sec.\ \ref{hbar basics}
with an overview of the perturbation theory
and effects on the spectrum.
Signals in hyperfine transitions are the subject of
Sec.\ \ref{hbar Hyperfine Zeeman transitions},
while effects on the $1S$-$2S$ and similar transitions are considered 
in Sec.\ \ref{hbar 1S2S}.
We conclude the treatment of antihydrogen 
in Sec.\ \ref{hbar gravity}
with a discussion of the prospects 
for an anomalous gravitational response of antihydrogen.

Three sections are devoted to signals of Lorentz and CPT violation
in other related systems.
Deuterium spectroscopy is considered in Sec.\ \ref{Deuterium}.
The perturbative approach adopted is presented 
in Sec.\ \ref{Isotropic Lorentz-violating perturbations},
followed in Sec.\ \ref{Deuterium transitions}
by a discussion of frequency shifts
relevant to high-sensitivity spectroscopy.
Observable effects from isotropic coefficients
are considered in Sec.\ \ref{Deuterium self consistent},
while Sec.\ \ref{Deuterium maser} contains
a discussion of the prospects for hyperfine measurements 
using a deuterium maser.
Positronium spectroscopy is the subject of
Sec.\ \ref{Positronium},
while spectroscopy of hydrogen molecules
and related species is considered in Sec.\ \ref{Hydrogen molecules}.
We conclude with a summary in Sec.\ \ref{Summary}.
Throughout this paper
we follow the notation of Ref.\ \cite{km13},
with a few exceptions noted in the text.

\section{Theory}
\label{Theory}

In this section,
we present the general theoretical framework and calculations
for determining the perturbative shifts in the hydrogen spectrum
arising from Lorentz and CPT violation.
The basic framework for the calculation is discussed,
and then the symmetries of the system are used
to identify the subset of coefficients for Lorentz and CPT violation
that can contribute to modifications of the hydrogen spectrum. 
The general matrix elements of the perturbative hamiltonian are calculated,
and analytical expressions for the resulting energy shifts
are presented in simple cases.
We finally address generic effects arising 
in the presence of an applied magnetic field,
including in particular the time dependence 
of the energy-level shift due to sidereal and annual variations.

\subsection{Basics}
\label{Basics}

The dominant Lorentz-violating perturbations to the spectrum of hydrogen 
arise from corrections to the propagation 
of the electron $e$ and the proton $p$.
Introducing a flavor index $\f$ taking values $e$ and $p$,
the Lagrange density for the quantum fermion field $\ps_\f$ of mass $m_\f$
including all kinetic effects from Lorentz and CPT violation
can be written as
\cite{km13}
\beq
\cl \supset 
\half \psfb (\ga^\mu i\prt_\mu - m_\f + \Qhat_\f) \ps_\f 
+ {\rm h.c.}, 
\label{lag}
\eeq
where $\Qhat_\f$ is the sum of all possible terms
formed by contracting SME coefficients for Lorentz and CPT violation
with derivatives $i\prt_\mu$.
The operator $\Qhat_\f$ is a spinor matrix.
It can be expanded in Dirac matrices,
converted to momentum space,
and decomposed in spherical coordinates,
which permits the classification and enumeration
of the corresponding effects.
At each mass dimension $d$ and for each flavor $\f$,
only certain combinations of coefficients for Lorentz violation 
are observable,
due to the freedom to redefine the spinor basis 
without affecting the physics.
Each of these combinations,
called effective coefficients,
controls a physically distinct Lorentz-violating effect.
Spectroscopy of the hydrogenic systems considered in this work
offers access in principle to about half of these effective coefficients,
including some with sensitivities corresponding
to Planck-suppressed signals.

The leading-order perturbation $\de h_\atm$ 
to the free Dirac hamiltonian $h_\atm$
for the hydrogen atom can be obtained from the Lagrange density \rf{lag}
by adding the Lorentz-violating contributions from the electron and the proton.
In the center-of-mass frame,
the kinetic energies of the electron and proton are small
compared to their masses,
so it suffices to consider the hamiltonian perturbation 
in the nonrelativistic limit,
\beq 
\de h_\atm^\nr = \de h^\nr_e + \de h^\nr_p .
\label{pert}
\eeq 
For calculational purposes,
the operator $\de h^\nr_e$ is understood to represent
the tensor product of an operator acting on $e$ states
with the identity operator acting on $p$ states,
and similarly for $\de h^\nr_p$.
Note that the operators $\de h^\nr_\f$ 
depend on the fermion momentum $\pvec$.

For much of the spectroscopic analysis that follows,
it is useful to perform a spherical decomposition
of the hamiltonian perturbation $\de h_\atm^\nr$
because tests of rotation symmetry
are the predominant focus of many searches for Lorentz violation. 
The perturbation $\de h^\nr_\f$ can be decomposed as
\cite{km13} 
\beq
\de h_\f^\nr
= h_{\f 0}+ h_{\f r} \sivec\cdot\ephat_r
+ h_{\f +} \sivec\cdot\ephat_-
+ h_{\f -} \sivec\cdot\ephat_+,
\label{nr}
\eeq
where $\sivec = (\si^1, \si^2, \si^3)$ 
is the vector of Pauli matrices.
The unit basis vectors 
$\ephat_r = \phat \equiv \pvec/\pmag$,
$\ephat_\pm = (\thhat \pm i\phhat)/\sqrt{2}$
are defined by introducing the usual unit vectors
$\thhat$ and $\phhat$ for the polar angle $\th$ and azimuthal angle $\ph$
in momentum space,
so that 
$\phat = (\sin\th\cos\ph,\sin\th\sin\ph,\cos\th)$.
The component hamiltonians 
$h_{\f 0}$, $h_{\f r}$, $h_{\f \pm}$ 
can be expanded in a series of terms involving products of 
powers of $\pmag$,
spin-weighted spherical harmonics $\syjm{s}{jm}(\phat)$
of spin weight $s$,
and nonrelativistic spherical coefficients for Lorentz violation.
For the spin-independent term
this gives
\beq
h_{\f 0} =
-\sum_{kjm} \pmag^k 
~\syjm{0}{jm}(\phat) 
\Vnrf{\f}{\k jm},
\label{FEHSI}
\eeq
while for the spin-dependent terms the result is
\bea
h_{\f r} &=&
-\sum_{kjm} \pmag^k 
~\syjm{0}{jm}(\phat) 
\TzBnrf{\f}{kjm},
\nonumber \\
h_{\f \pm} &=&
\sum_{kjm} \pmag^\k 
~\syjm{\pm 1}{jm}(\phat) 
\left(i\ToEnrf{\f}{kjm} \pm \ToBnrf{\f}{kjm}\right).
\quad
\label{FEHSD}
\eea
The quantities $\Vnrf{\f}{kjm}$ and ${\T_\f}^{\nr(qP)}_{kjm}$,
where the superscripts $qP$ take the values $0B$, $1B$, $1E$,   
are the nonrelativistic spherical coefficients for Lorentz violation,
which we denote generically by ${\K_\f}_{kjm}^\nr$. 
Each of these can be separated into two pieces,
controlling either CPT-even or CPT-odd effects,
\cite{km13}
\bea
\Vnrf{\f}{kjm} &=&
\cnrf{\f}{kjm} - \anrf{\f}{kjm},
\nonumber \\
{\T_\f}^{\nr(qP)}_{kjm} &=&
{g_\f}^{\nr(qP)}_{kjm} - {H_\f}^{\nr(qP)}_{kjm},
\label{cpt}
\eea 
following the standard convention 
\cite{ck}
in which $a$- and $g$-type coefficients 
are associated with CPT-odd operators
and $c$- and $H$-type coefficients 
with CPT-even ones. 
Expressions involving antiparticles
can therefore be obtained by reversing the sign
of the $a$- and $g$-type coefficients.
The reader is cautioned that the $a$- and $H$-type coefficients
contain contributions only from operators of odd mass dimensions $d$,
while the $c$- and $g$-type coefficients
contain ones only from operators of even $d$.
Note that the mass dimension of each nonrelativistic
coefficient is $1-k$.

A primary target of spectroscopic experiments
is measurements of the nonrelativistic spherical coefficients \rf{cpt}. 
These coefficients are linear combinations
of the complete set of spherical coefficients for Lorentz violation,
given in Eqs.\ (111) and (112) of Ref.\ \cite{km13}.
The allowed range of the indices $k$, $j$, $m$ 
and the counting of independent coefficient components
are provided in Table IV of Ref.\ \cite{km13}.
The subscript index $k$ is used in the present work instead of $n$
to avoid confusion with the principal quantum number of the atom.
Note that the indices $j$, $m$ 
determine the rotational behavior 
of the spin-weighted spherical harmonics and
hence of the corresponding operators for Lorentz violation,
so these indices are distinct from 
the angular-momentum quantum numbers $J$, $M$ 
associated with the atomic states.
The basic properties of the spin-weighted spherical harmonics
are presented in Appendix A of Ref.\ \cite{km09}.
The usual spherical harmonics are recovered when $s=0$, 
so $Y_{jm}(\th,\ph) \equiv \syjm{0}{jm}(\phat)$.

For the special index choices $jm=00$ 
the corresponding physical effects are isotropic,
and following Eq.\ (114) of Ref.\ \cite{km13}
it is convenient to adopt a ring-diacritic notation for the
associated coefficients.
For the applications in this work,
it suffices to define
\beq
\anrf{\f}{k00} \equiv \sqrt{4\pi} ~\anrfc{\f,k} ,
\qquad 
\cnrf{\f}{k00} \equiv \sqrt{4\pi} ~\cnrfc{\f,k} .
\label{ring}
\eeq
We emphasize that the isotropic nonrelativistic coefficients 
$\anrfc{\f,k}$ and $\cnrfc{\f,k}$ 
contain isotropic spherical coefficients 
$\afc {d}{k}$ and $\cfc {d}{k}$ of arbitrarily large $d$.
For example,
using Eqs.\ (93) and (111) of Ref.\ \cite{km13} gives
\bea
\anrfc{\f,0} &=& 
\afc{3}{0} + m_\f^2 \afc{5}{0} +  m_\f^4 \afc{7}{0} + \ldots,
\nonumber\\
\anrfc{\f,2} &=& 
\afc{5}{0} + 2 m_\f^2 \afc{7}{0} + \ldots
+ \afc{5}{2} + m_\f^2 \afc{7}{2} + \ldots,
\quad
\eea
and 
\bea
\cnrfc{\f,0} &=& 
m_\f \cfc{4}{0} + m_\f^3 \cfc{6}{0} +  m_\f^5 \cfc{8}{0} + \ldots,
\nonumber\\
\cnrfc{\f,2} &=& 
\fr 1 {2m_\f} \cfc{4}{0} + \frac 32 m_\f \cfc{6}{0} 
+ \frac 52 m_\f^3 \cfc{8}{0} + \ldots
\nonumber\\
&&
+ \fr 1 {m_\f} \cfc{4}{2} + m_\f \cfc{6}{2} 
+ m_\f^3 \cfc{8}{2} + \ldots.
\eea

The Lorentz-violating perturbative shifts in the spectrum of atomic hydrogen
are obtained by calculating the matrix elements 
of the perturbation hamiltonian \rf{pert}
with respect to the unperturbed states of the system.  
We take the latter to be the Schr\"odinger-Coulomb eigenstates 
for a reduced mass $\mr \equiv m_e m_p/(m_e+m_p)$,
coupled to Pauli spinors for each particle.  
When the perturbative shifts are smaller than the hyperfine structure,
the total angular momentum $J$ of the electron and 
the total angular momentum $F$ of the atom are good quantum numbers.  
Other relevant quantum numbers for the system 
include the principal quantum number $n$ 
and the orbital angular momentum $L$.  
  
The scales of the perturbative frequency shifts
are controlled by the nonrelativistic coefficients ${\K_\f}_{kjm}^\nr$.
The latter can be viewed as background fields
in the chosen inertial frame,
which in the above equations for $\de h_\f^\nr$ 
is the zero-momentum frame for the hydrogen atom. 
However,
an Earth-based laboratory for spectroscopic experiments
is poorly suited to report coefficient measurements
because it represents a noninertial frame
due to the Earth's rotation about its axis
and its revolution around the Sun.
Instead,
a specified inertial frame can be used,
widely chosen to be the canonical Sun-centered frame
\cite{sunframe,tables}.
This frame adopts coordinates $T,X,Y,Z$
with the origin of the time $T$ chosen as the vernal equinox 2000, 
the $X$ axis pointing towards the vernal equinox,
and the $Z$ axis aligned along the Earth's axis of rotation.
The Sun-centered frame is inertial to an excellent approximation
on the timescale of laboratory experiments,
so it provides a standard and conveniently accessible frame
for reporting and comparing experimental results.

The nonrelativistic spherical coefficients ${\K_\f}_{kjm}^\nr$
can reasonably be taken as uniform and constant 
on the scale of the solar system
\cite{ck,akgrav}
and hence are constants when expressed in the Sun-centered frame.
The Earth's rotation and revolution therefore introduces 
variations with sidereal time 
in many coefficients expressed in the laboratory frame,
which implies time variations in physical signals
\cite{ak98}.
Since the Earth's orbital speed $\be_\oplus\simeq 10^{-4}$ is small,
the orbital motion can be disregarded 
for experimental analyses focusing on searches for rotation violations.
The transformation between the Sun-centered and laboratory frames
then reduces to a simple rotation,
so the spherical decomposition summarized above
offers definite calculational simplifications.
Suppose for convenience the laboratory frame coordinates $x,y,z$
are specified with the $z$ axis pointing towards the zenith
and the $x$ axis lying at an angle $\ph$ measured east of south.
Then,
the coefficients 
${\K_\f}_{kjm}^{\rm NR,lab}$ 
in the laboratory frame 
are related to the coefficients 
${\K_\f}^{\rm NR,Sun}_{kjm}$
in the Sun-centered frame by 
\beq
{\K_\f}_{kjm}^{\rm NR,lab} = 
\sum_{m'} e^{i m'\om_\oplus T_\oplus + im\ph}
d^{j}_{mm'}(-\ch)
{\K_\f}^{\rm NR,Sun}_{kjm'},
\label{tos}
\eeq
where $\om_\oplus\simeq 2\pi/(24\text{ h }56\text{ m})$ 
is the Earth's sidereal frequency,
$T_\oplus$ is the local Earth sidereal time,
and $\ch$ is the colatitude of the experiment.  
The little Wigner matrices $d^{j}_{mm'}$ 
are defined in Eq.\ (136) of Ref. \cite{km09}.
The result \rf{tos} reveals
that the sidereal dependence of the transition frequencies
is controlled by the azimuthal indices
on the coefficients contributing to the perturbation.

\subsection{Coefficient selection rules}
\label{Coefficient selection rules}

\renewcommand\arraystretch{1.5}
\begin{table*}
\caption{
Contributing nonrelativistic spherical coefficients.} 
\setlength{\tabcolsep}{6pt}
\begin{tabular}{ccccccc}
\hline
\hline
${\K_\f}_{kjm}^\nr$	&	$kjm$	&	number	&	condition	&	C	&	PT	&	CPT	\\
\hline													
$\anrf{\f}{kjm}$	&	$200$, $22m$, $400$, $42m$, $44m$	&	21	&	$j\leq 2J-1$	&	$-$	&	$+$	&	$-$	\\
$\cnrf{\f}{kjm}$	&	$200$, $22m$, $400$, $42m$, $44m$	&	21	&	$j\leq 2J-1$	&	$+$	&	$+$	&	$+$	\\
$\gzBnrf{\f}{kjm}$	&	$01m$, $21m$, $23m$, $41m$, $43m$, $45m$	&	34	&	$j\leq 2F-1$	&	$+$	&	$-$	&	$-$	\\
$\goBnrf{\f}{kjm}$	&	$01m$, $21m$, $23m$, $41m$, $43m$, $45m$	&	34	&	$j\leq 2F-1$	&	$+$	&	$-$	&	$-$	\\
$\HzBnrf{\f}{kjm}$	&	$01m$, $21m$, $23m$, $41m$, $43m$, $45m$	&	34	&	$j\leq 2F-1$	&	$-$	&	$-$	&	$+$	\\
$\HoBnrf{\f}{kjm}$	&	$01m$, $21m$, $23m$, $41m$, $43m$, $45m$	&	34	&	$j\leq 2F-1$	&	$-$	&	$-$	&	$+$	\\
\hline\hline
\end{tabular}
\label{rest}
\end{table*}

Before calculating explicitly the matrix elements of $\de h_\atm^\nr$
in the unperturbed states,
it is useful to study the symmetries of the system.
We show here that various symmetries imply vanishing values 
for many matrix elements of operators in the decomposition
\rf{nr}--\rf{FEHSD}.
This identifies a subset of effective spherical coefficients
that are inaccessible at leading order via spectroscopy.

A first observation is that the unperturbed states are parity eigenstates. 
It follows that only even-parity perturbations can contribute 
to the energy shift at first order. 
Since all operators in the decomposition 
\rf{nr}--\rf{FEHSD} have definite parity
\cite{km13},
it is straightforward to identify the inaccessible coefficients.
For the coefficients $\anrf{\f}{kjm}$, $\cnrf{\f}{kjm}$,
which are associated with the usual spherical harmonics,
$j$ must be even to contribute,
which turns out to imply that $k$ must be even as well.
For the coefficients 
${g_\f}^{\nr(qP)}_{kjm}$, ${H_\f}^{\nr(qP)}_{kjm}$,
the parity is even if $k$ is even
and either $P=E$ with even $j$ 
or $P=B$ with odd $j$.
 
A second observation is that 
the sums over $j$ in Eqs.\ \rf{FEHSI} and \rf{FEHSD}
can be truncated according to the angular momenta
of the unperturbed state of interest.
The key to implementing this truncation is the following proposition:
if $T_{jm}$ transforms as a spherical-tensor operator 
under the transformation generated by an angular-momentum operator $\mbf K$
with associated quantum numbers $K$ and $m_K$,
then the matrix element $\vev{Km_K'|T_{jm}|Km_K}$ 
vanishes unless $j\leq 2 K$.
This result is a direct consequence of the triangular condition 
$|j_1-j_2|\leq j_3\leq j_1+j_2$ 
of the Clebsch-Gordan coefficients $\vev{j_1m_1j_2m_2|j_3m_3}$
and the Wigner-Eckart theorem
\cite{we}.

To illustrate the use of this proposition to truncate the sums over $j$,
consider first the spin-independent terms in Eq. $\eqref{FEHSI}$.  
These operators transform as spherical operators 
with $\mbf K$ identified as $\mbf L$, $\mbf J$, or $ \mbf F$.  
Now, if $K$ is good quantum number,
then a matrix element in the unperturbed state 
can be expressed as a linear combination 
of matrix elements in the states $\ket{K m_K}$ with $K$ fixed.  
As a result, 
if the matrix elements in the states $\ket{K m_K}$ vanish,
then so does the matrix element in the unperturbed state. 
For $K=J$,
the proposition then implies 
that only operators satisfying the inequality $j\leq 2J$ 
can contribute to the energy shift.   
Since $2J$ is an odd number
and since only even values of $j$ contribute 
to the energy shift as noted above, 
we can express the condition on $j$ for the case $K=J$ as $j < 2J$. 
For $K=L$ or $K=F$,
no stronger constraints on the allowed values of $j$ are obtained.
For example,
if $F=J-1/2$ then $j\leq 2F=2J-1$,
which is equivalent to $j<2J$ because $j$ is an integer.  
To summarize,
among the spin-independent operators only 
those satisfying the condition $j<2J$ 
can contribute to the energy shift of a state with angular momentum $J$. 
A similar argument applies to the spin-dependent terms with $B$-type parity,
leading to the conclusion that
among this set of terms only those satisfying $j\leq 2F-1$ 
can contribute to the energy shift of a state 
with total angular momentum $F$.  

Invariance under time reversal is another symmetry of the system.
This symmetry can be used along with the Wigner-Eckart theorem 
to show that the spin-dependent terms 
with $E$-type parity in Eq.\ \rf{FEHSD}
cannot contribute to the energy shift at first order in perturbation theory.  
To see this,
we begin by considering a spin-dependent operator of $E$-type parity 
having fixed $j$ and $m=0$,
which takes the schematic form 
\beq
T_{j0}^{E} = 
\sqrt{2}\sum_{k}
\pmag^k\ToEnr{kj0}
\syjm{+1}{j0}(\phat)
(\si^2 \cos{\ph}-\si^1 \sin{\ph}).
\label{etype}
\eeq
Under time reversal
$\pvec\rightarrow -\pvec$ and $\mbf \si \rightarrow- \mbf \si$,
so the operator transforms as $T_{j0}^E\rightarrow(-1)^{j+1}T_{j0}^E$.  
Also,
states transform as
$\ket{Fm_F}\rightarrow \ket{F(-m_F)}$ 
up to a phase factor,
which implies 
\beq
\vev{Fm_F|T_{j0}^E|Fm_F} = (-1)^{j+1} \vev{F(-m_F)|T_{j0}^E|F(-m_F)}.
\label{trev}
\eeq
The Wigner-Eckart theorem 
and the properties of the Clebsch-Gordan coefficients 
permit the replacement $-m_F\rightarrow m_F$ 
on the right-hand side of this equation,
accompanied by a phase factor $(-1)^j$.
We thus obtain the equality
\beq
\vev{Fm_F|T_{j0}^E|Fm_F}=-\vev{Fm_F|T_{j0}^E|Fm_F},
\eeq
revealing that all the matrix elements of $T_{j0}$ vanish.  
Using the Wigner-Eckart theorem again, 
this result can be extended to the general matrix elements
$\vev{Fm_F'| T_{jm}^E|Fm_F}$
via the identity 
\beq
\vev{Fm_F'|T_{jm}^E|Fm_F} =
\dfrac{\vev{FF|T_{j0}^E|FF }}
{\vev{FFj0|FF}} 
\vev{Fm_F jm|Fm_F'},
\label{wiap}
\eeq
because the Clebsch-Gordan coefficient $\vev{FFj0|FF}$ 
always differs from zero when $2F\geq j$.  
We can thus confirm that $\vev{Fm_F'|T_{jm}^E|Fm_F }$ vanishes 
and hence that the spin-dependent terms with $E$-type parity 
in Eq.\ \rf{FEHSD} 
cannot contribute to perturbative energy shifts.

While the various constraints above restrict the sums over $j$
in the decomposition \rf{nr}--\rf{FEHSD},
the sums over $k$ remain unconstrained.
Evaluation of the matrix elements of operators $\pmag^k$ with $k>4$
reveals that they diverge when $n$ and $L$ are small.
This technical problem might be resolved by a suitable regularization.
However,
on dimensional grounds the size of the matrix elements
is governed by a factor $(\al \mr)^k$,
where $\al$ is the fine-structure constant.
This factor heavily suppresses the resulting energy shifts
even for sizeable coefficients for Lorentz violation.
For instance,
a large $k=6$ coefficient of order 1 GeV$^{-5}$
produces a spectroscopic frequency shift of only about a nanohertz.
We therefore limit attention to $k\leq 4$ in this work.
This choice further restricts the allowed values of $j$
\cite{km13},
with the maximum allowed value $j_{\rm max}$ given as 
$j _{\rm max} = k$ for the spin-independent terms and 
$j _{\rm max} = k+1$ for the spin-dependent terms of $B$-type parity.

Combining all the results in this subsection,
we can identify the subset of nonrelativistic spherical coefficients
of interest for spectroscopic experiments.
Table \ref{rest} summarizes the situation 
for each type of coefficient.
The first column of the table lists coefficients
that in principle can contribute at first order in perturbation theory
to a spectral shift of a state with quantum numbers $F$ and $J$.
The second column shows the allowed values of the triplet $kjm$ of indices,
where $-j \leq m \leq j$ as usual.
The third column gives the total number of independent components
for each type of coefficient.
Since there are two flavors for each coefficient for hydrogen, 
$\f=e$ and $\f=p$,
a total of 356 independent nonrelativistic spherical components 
are measurable in principle via spectroscopic experiments on hydrogen, 
with each corresponding to a distinct physical effect. 
The fourth column lists the constraint on $j$
for a measurement involving a state of angular momentum $J$ or $F$.  
The final three columns display the C, PT, and CPT handedness
of the corresponding operators.

Note that among the listed coefficients 
only the eight with $jkm=200$ or $jkm=400$ 
govern isotropic effects that are spectroscopically observable.
They correspond to the coefficients 
$\anrfc{\f,2}$, $\anrfc{\f,4}$, $\cnrfc{\f,2}$, and $\cnrfc{\f,4}$
given by the definitions \rf{ring}.
The coefficients with $jkm=000$,
which satisfy all the above criteria,
have been omitted from Table \ref{rest} 
because they produce only constant energy shifts in a given frame
and hence are undetectable via spectroscopy.
Detecting them might be feasible,
for example,
by studying anisotropies of the dispersion relation for hydrogen
in a boosted frame,
but investigating this lies outside our present scope. 
They also become detectable in principle
in the presence of interactions such as gravity
\cite{akgrav}.
This issue is revisited in the context
of studies of the gravitational response of antihydrogen
in Sec.\ \ref{hbar gravity} below.

\subsection{Matrix elements}
\label{Matrix elements}

In this subsection,
we present some explicit results for 
matrix elements of the perturbation hamiltonian $\de h_\atm^\nr$
introduced in Eq.\ \rf{pert}.
For each nonrelativistic spherical operator in $\de h_\atm^\nr$,
a given matrix element can be decomposed as the product of two pieces,
one depending on the principal quantum number $n$
and the other independent of it.
In what follows,
we provide general expressions for each of these two pieces.
The first piece can be given in analytical form.
For free-atom states with arbitrary total quantum number $F$,
the second piece cannot be expressed in closed form
due to the algebraic complexities of degenerate perturbation theory.
However,
we can obtain an analytical result for the cases $F=0$ and $F=1$.
The situation in the presence of an applied magnetic field
is discussed in Sec.\ \ref{Weak magnetic field}.

\subsubsection{General case}
\label{Angular expectation values}

The piece of the matrix element
that depends on the principal quantum number $n$
is given by the expectation value of the operator $\pmag^k$ 
in the unperturbed states.
For the cases $k=0$, 2, and 4 of interest here,
we find
\bea
\vev{\pmag^0}_{nL}&=& 1,
\nonumber\\
\vev{\pmag^2}_{nL}&=& \left(\dfrac{\al \mr}{n}\right)^2,
\nonumber\\
\vev{\pmag^4}_{nL}&=&
\left(\dfrac{\al \mr}{n}\right)^4
\left(\dfrac{8n}{2L+1}-3\right).
\label{radialExp}
\eea
The first of these equations reflects the normalization 
of the unperturbed states.
Note that only the case $k=4$ depends
on the angular-momentum quantum number $L$.

The second piece of the matrix element
is independent of $n$.
Using the spherical decomposition,
it can be expressed
in terms of the Clebsch-Gordan coefficients.  

To illustrate this fact,
we begin by consider the spin-independent term \rf{FEHSI}.  
Denoting the unperturbed states by the ket $\ket{n F J L m_F}$, 
a calculation of the expectation value 
of the usual spherical harmonics gives 
\begin{widetext}
\beq
\vev{n F J L m_F|\,\syjm{0}{jm}(\phat)|n F J L m_F}
= \sqrt{\fr{2j+1}{4\pi}}
\sum_{m_L m_J}\sum_{m_1 m_2}
\vev{L 0 j m|L 0}\vev{L m_L j m|L m_L}
\vev{\half m_1 L m_L|J m_J}^2\vev{J m_J\half m_2|F m_F}^2 .
\quad
\label{shex}
\eeq
Note that the result is independent of the flavor $\f$.
In this equation,
the Clebsch-Gordan coefficients involving $J$ and $F$ 
arise from the addition of orbital and spin angular momenta.
The other Clebsch-Gordan coefficients originate 
in the triple integral of the spherical harmonics.
This result 
and the Wigner-Eckart theorem in the generic form \rf{wiap}
can be combined to derive the spin-independent matrix elements 
in the fixed-$F$ subspace.  
We find
\beq
\vev{n F J L m'_F|h_{e0}+h_{p0}|n F J L m_F} =
-\sum_{\f kjm} 
\Vnrf{\f}{kjm} 
\vev{\pmag^k}_{nL} 
\vev{F m_F j m|F m'_F} 
\fr{\vev{F J L F|~\syjm{0}{j 0}(\phat)|F J L F}}
{\vev{F F j 0|F F}},
\label{siex}
\eeq
which demonstrates that the calculation of the matrix elements
for the spin-independent term
can be reduced to a determination of Clebsch-Gordan coefficients.
A similar result holds for the spin-dependent term.

More generally,
the above methods can be used to determine 
the matrix elements of the full perturbation $\de h_\atm^\nr$.  
Explicitly, 
after some calculation we find
\beq
\vev{n F J L m'_F|\de h_\atm^\nr|n F J L m_F} = 
\sum_{jm}\AM_{jm}
\vev{F m_F jm|F m'_F},
\label{mel}
\eeq 
where the weights $\AM_{jm}(nFJL)$ are given by
\beq
\AM_{jm}^{\rm }=
-\sum_{\f k}
\vev{\pmag^k}_{nL} 
\left[ 
\AzE_j 
\Vnrf{\f}{kjm}
+ \fr{\AzB_j}{2J+1} 
\TzBnrf{\f}{kjm}
-\AoB_j \left(\fr{\de_{\f e}}{2(L-J)} + \fr{\de_{\f P}}{2(J-F)} \right)
\ToBnrf{\f}{kjm}
\right],
\label{Ajm}
\eeq
\end{widetext}
with $\de_{ab}=1$ if $a=b$ and $\de_{ab} = 0$ otherwise. 
In this expression,
the factors $\Agen$ are related 
to the ratio of the expectation value of the operator 
and the corresponding Clebsch-Gordan coefficient.
This can be verified for the spin-independent term by comparing 
Eqs.\ \rf{siex} and \rf{Ajm}.
Note that the weights obey the identity 
\beq
A_{jm}^* = (-1)^m A_{j(-m)},
\eeq
by virtue of the properties of the coefficients for Lorentz violation
\cite{km13}.

The factors $\Agen$ can be expressed explicitly
in terms of the quantum numbers of the state involved.
For the factors $\AzE_0(F,J)$ associated with the spin-independent terms,
we find
\beq
\AzE_0(F,J) =
\fr 1{\sqrt{4\pi}}
\label{AzE0}
\eeq 
when $j=0$,
and 
\bea
\AzE_j(F,J) &=& 
i^{j}\fr{j-1}{2^{2j}}
\left(1-2j\fr{F-J}{2J+1}\right)
\fr{(J-j/2)!}{(J+j/2)!}
\nonumber \\
&& 
\times \sqrt{\fr{(2j+1)(2F+j+1)!}{\pi(2F+1)(2F-j)!}} 
\qquad
\label{AzE}
\eea
when $j=2$ or $j=4$.
For the factors $\AoB_j(F)$ associated with the spin-dependent terms,
we obtain
\bea
\AoB_j(F) &=& 
\left(\frac{1}{2}(j-1)\right)!  
\sqrt{\dfrac{2^{(1-j)/2}}{j!!}} \AzB_j(F) 
\nonumber \\
&&
\hskip-50pt
=i^{j-1}
\sqrt{\fr{j!!(2j+1)\left(F+\half(j+1)\right)!(2F-j)!!}
{2^{(j-1)/2}\pi(2F+1)\left(F-\half(j+1)\right)!(2F+j)!!}} 
\nonumber \\
\label{AzB}
\eea
for $j = 1,3,5$.
In these expressions,
the double-factorial symbol $!!$ is defined as usual 
by $N!!=N(N-2)\cdots 1$ for odd $N$ 
and $N!!=N(N-2)\cdots 2$ for even $N$.  

For convenience,
Table \ref{LaCoeff} presents
the numerical values for some instances of the factors $\Agen$.
The table lists the numerical values of the factors 
for energy levels with orbital angular momentum $L\leq 2$.
The left-hand side of the table concerns 
the factors $\AzE_j(F,J)$ associated with the spin-independent perturbation, 
displaying their values for
$j = 2,4$, $J=\frac 32, \frac 52$, and $F=1,2,3$.
The right-hand side
gives values of the factors $\AzB_j(F)$ and $\AoB_j(F)$
for spin-dependent effects,
for the ranges $j=1,3,5$ and $F=1,2,3$. 

\renewcommand\arraystretch{2.5}
\begin{table}
\caption{
Some numerical values of the factors $\Agen$.}
\setlength{\tabcolsep}{7pt}
\begin{tabular}{cccc|cccc}
\hline
\hline
$j$ & $J$  & $F$  & $\AzE$& $j$ & $F$  & $\AzB$  & $\AoB$\\\hline
2 & $\dfrac 32$  & $1$ & $-\dfrac{1}{2\sqrt{2\pi}}$  
& 1 &  1 & $\sqrt{\dfrac{2}{3\pi}}$  & $\sqrt{\dfrac{2}{3\pi}}$ \\
& & $2$ & $-\dfrac{1}{2}\sqrt{\dfrac{7}{10\pi}}$ &
& 2  & $\dfrac{3}{5}\sqrt{\dfrac{2}{\pi}}$ &$\dfrac{3}{5}\sqrt{\dfrac{2}{\pi}}$\\
& $\dfrac 52$  &$2$ & $-2\sqrt{\dfrac{2}{35\pi}}$ &
& 3  & $\dfrac{6}{7\sqrt{\pi}}$ &  $\dfrac{6}{7\sqrt{\pi}}$\\ 
& & $3$ & $-\dfrac{2}{7}\sqrt{\dfrac{3}{\pi}}$  &   
3 &  2  &$-\dfrac{6}{5}\sqrt{\dfrac{2}{\pi}}$ & $-\dfrac{2}{5}\sqrt{\dfrac{3}{\pi}}$\\
4 & $\dfrac 52$  & $2$ & $\dfrac{1}{\sqrt{14\pi}}$ &
& 3 &$-2\sqrt{\dfrac{6}{7\pi}}$ & $-\dfrac{2}{\sqrt{7\pi}}$\\
& & $3$ & $\dfrac{1}{7}\sqrt{\dfrac{11}{2\pi}}$ & 
5 &  3 &$\dfrac{10}{7}\sqrt{\dfrac{3}{\pi}}$ & $\dfrac{2}{7}\sqrt{\dfrac{5}{\pi}}$\\
\hline\hline
\end{tabular}
\label{LaCoeff}
\end{table}

To illustrate the methods described in this subsection,
we construct the matrix element 
for the perturbative energy shift in the $F=1$ subspace
of the ground state $J=1/2$.
Inspecting Table \ref{rest} reveals that
only the spin-independent terms with $j=0$ 
and the spin-dependent terms with $j=1$ can contribute,
while $k$ can take the values 0, 2, and 4.
From the general expression \rf{mel},
the matrix element $\de h(m'_F,m_F)$ in the $F=1$ subspace
takes the form
\beq
\de h(m'_F,m_F) = 
\AM_{00}\de_{m'_Fm_F} 
+ \sum_{m}\AM_{1m} \vev{1 m_F 1 m|1 m'_F}.
\eeq
To determine the weight $\AM_{00}(1 1 \half 0)$ using Eq.\ \rf{Ajm},
we need the expectation values \rf{radialExp} 
evaluated at $n=1$ and $L=0$
and the factor $\AzE_{0}(1,\half)$ 
obtained from Eq.\ \rf{AzE0}. 
To calculate the weight $\AM_{1m}(1 1 \half 0)$ 
requires the values of $\AzB_{1}(1)$ and $\AoB_{1}(1)$ 
obtained from Table \ref{LaCoeff}.
Collecting all the pieces,
we find 
\beq
A_{00}(1 1 \half 0) = 
\sum_{q=0}^2
-(\al \mr)^{2q}(1+4\de_{q2})
\sum_\f 
\fr{\Vnrf{\f}{(2q)00}} {\sqrt{4\pi}}
\label{A001S}
\eeq
for the spin-independent weight,
and 
\bea
A_{1m}(1 1 \half 0) &=& 
\fr{1}{\sqrt{6\pi}}
\sum_{q=0}^2
-(\al \mr)^{2q}(1+4\de_{q2})
\nonumber\\
&&
\hskip 20pt
\times
\sum_\f 
\left( 
\TzBnrf{\f}{(2q)1m}+2\ToBnrf{\f}{(2q)1m}
\right)
\qquad
\label{A1m1S}
\eea
for the spin-dependent one.

\subsubsection{Analytical energy shifts for $F=0$ and $F=1$}
\label{Energy shifts for $F=0$ and $F=1$}

Since the unperturbed hydrogen energy levels are $(2F+1)$-fold degenerate,
the perturbative corrections to the energy levels for fixed $F$ 
are obtained by the eigenvalues of a $(2F+1)\times(2F+1)$ matrix,
which is specified by Eq.\ \rf{mel}.  
In general,
these eigenvalues are determined
by the roots of a polynomial of degree $2F+1$ 
corresponding to the secular equation of the matrix \rf{mel}.  
This implies a closed-form expression
for the energy shifts at arbitrary $F$
is unattainable.
However,
for the special cases $F=0$ and $F=1$
the secular polynomial can be solved in closed form.
An analytical expression for the energy shifts can therefore be found,
as we demonstrate next.
For simplicity,
we suppress the arguments $nFJL$ of the weights $\AM_{jm}(nFJL)$ 
in what follows.

Consider first the energy shift $\de \ep(n,L)$ for the case $F=0$.
Since this energy state is nondegenerate,
the shift can be obtained directly from Eq.\ \rf{mel}.
The result is
\beq
\de \ep(n,L) = 
\AM_{00}
= -\sum_{\f k} \vev{\pmag^k}_{nL}
\fr{\Vnrf{\f}{k00}} {\sqrt{4\pi}}. 
\label{iso}
\eeq
From Eqs.\ \rf{Ajm} and \rf{AzE0},
we can infer that the weight $A_{00}$ depends only 
on the quantum numbers $n$ and $L$.
In fact,
this feature holds for any $F$ and $J$
because the identity $\vev{F m_F 0 0|F m'_F}=\de_{m_Fm'_F}$ 
implies that the contribution 
involving the isotropic coefficients with $jm=00$
to the matrix element of the perturbation in the fixed-$F$ subspace 
is given by $\AM_{00}$  times the identity matrix.  
The energy shift can therefore always be expressed 
as the sum of contributions from the isotropic coefficients 
with ones from the anisotropic coefficients,
with the former being given by $\AM_{00}(nL)$ 
independent of the values of $F$ and $J$.  
One consequence of this observation 
is that isotropic coefficients 
can only contribute to frequency shifts 
for transitions with $\De n\ne 0$ or $\De L\ne 0$.

The expression for the $F=1$ case is more involved 
because it is obtained from the solution of a cubic equation.  
The energy shift $\de \ep(n,L,J,\xi)$ for this case takes the form
\bea
\de \ep(n,L,J,\xi) = 
\AM_{00}
+\fr{1-i \xi \sqrt{3}}{9\xi^2-3} 
\sqrt[3]{\dfrac{\De_1-\sqrt{\De_1^2-4\De_0^3}}{2}}
\nonumber\\
+\fr{1+i \xi \sqrt{3}}{9\xi^2-3} \fr{\De_0}
{\sqrt[3]{\dfrac{\De_1-\sqrt{\De_1^2-4\De_0^3}}{2}}},
\quad
\label{f1c}
\eea
where $\xi=-1,0,1$.  
The quantities $\De_0$ and $\De_1$ 
can be written in terms of the weights $\AM_{jm}(n1JL)$ 
and Clebsch-Gordan coefficients.  
The expression for $\De_0$, 
\bea
\De_0 = \frac{9}{2} \sum_{jm}
\fr{\AM_{jm}\AM^*_{jm}}{2j+1},
\label{twoa}
\eea
explicitly shows that it is a rotational scalar 
because $\AM_{jm}$ transforms dually to $\AM^*_{jm}$ 
under observer rotations.  
Similarly,
we can conclude from the structure of $\De_1$,
\bea
\De_1 &=& 
-\sum_{j m_3} \sum_{m_1 m_2}
\fr{27\sqrt{(2j+1)(2j+3)}}{2\times 5^{j-1/2}} 
\vev{jm_1jm_2|2m_3}
\nonumber \\
&&
\hskip 20pt
\times
\vev{2m_32(-m_3)|00}\AM_{jm_1}\AM_{jm_2}\AM_{2(-m_3)},
\label{threea}
\eea
that it too is a rotational scalar.  
One way to understand this is to notice 
that the weights $\AM_{jm}$ transform under observer rotations
like $\bra{jm}$,
and the equation for $\De_1$ can be viewed 
as the sum of singlets $\bra{00}$ 
obtained by the angular-momentum coupling 
of $\bra{jm_1}$ with $\bra{jm_2}$ and then to $\bra{2m_3}$.  

The result \rf{f1c} holds for both allowed values of $J$.
However, 
its complexity reduces significantly for $J=1/2$.  
As can be seen from Table \ref{rest},
the $j=2$ coefficients provide no contribution for $J=1/2$ 
and so the weight $\AM_{2m}$ vanishes.
This implies that $\De_1=0$,
thereby reducing Eq.\ \rf{f1c} to
\bea
\de \ep(n,L,\half,\xi) = 
A_{00} +
\fr 1{\sqrt{2}} \xi A,
\label{f1cr}
\eea
where $A\equiv \sqrt{\sum_{m} \AM^*_{1m} \AM_{1m}}$.
The contribution from the anisotropic coefficients to the energy shift 
thus takes the form of a linear Zeeman shift,
where $\xi$ can be interpreted as the eigenvalues 
of the component of the total angular momentum $\mbf F$ 
in the direction of the pseudovector $\AM^*_{1m}$.  
In terms of the unperturbed state $\ket{n1\half Lm_F}$,
the corresponding eigenvectors $\ket{nL\half \xi}$ take the form
\bea
\ket{nL\half 0} &=& 
\fr 1 A
\sum_m A_{1m} \ket{n1\half Lm},
\nonumber\\
\ket{nL\half (\pm 1)} &=&
\fr 1 {N_\pm}
\Big[
A_{11}^* (A_{10} \mp A)^2 
\ket{n1\half L(-1)}
\nonumber\\
&&
\qquad
+ 2(A_{10} \mp A) |A_{11}|^2 
\ket{n1\half L0}
\nonumber\\
&&
\qquad
+ 2A_{11} |A_{11}|^2 
\ket{n1\half L(+1)}
\Big],
\label{eigenvectors}
\eea
where the factors $N_\pm$ are normalizations.

In the expressions \rf{iso}, \rf{f1c}, and \rf{f1cr}
for the energy shifts,
the weights $\AM_{jm}$ containing
the nonrelativistic coefficients for Lorentz violation
appear only in combinations that are observer rotation scalars.
This is a general feature of energy shifts for any $F$,
which can be understood as follows.
Recall that an observer transformation amounts merely to changing a basis,
without changing the physics
\cite{ck,akgrav}.
However,
to specify completely a quantum observer transformation 
requires also defining its effect on the basis of states in the Hilbert space.
The definition can be chosen freely,
and it is convenient for the argument here
to require quantum observer rotations 
to leave the basis states $\ket{nFJLm_F}$ invariant.
This choice has similarities to the adoption of the Heisenberg picture 
in quantum mechanics.
By construction,
the operator $\de h_\atm^\nr$ 
is a scalar under observer rotations. 
The matrix elements 
$\vev{n F J L m'_F|\de h_\atm^\nr|n F J L m_F}$ 
then explicitly form a rotation scalar,
consistent with the notion that the perturbed energy of the atom
should be invariant under observer rotations.
However,
the weights $\AM_{jm}$ with $jm\neq 00$
transform nontrivially under observer rotations,
so in the final expression for the energy shift
they can appear only in combinations that are rotational scalars.
In fact,
the combinations are also scalars under particle rotations,
which transform the system while leaving unchanged
the coefficients for Lorentz violation.
As a result,
neither observer nor particle rotations 
affect the expressions for the energy shifts.
The physical manifestation of Lorentz violation appears 
as the lifting of the degeneracy of the unperturbed energy levels
of the free atom,
reflected in the appearance of the parameter $\xi$,
with the size of the splitting determined
by the magnitude of the coefficients for Lorentz violation.

\subsection{Applied magnetic field} 
\label{Weak magnetic field} 

The complications in calculating the spectral shifts for free hydrogen
arise in part from the rotational symmetry of the unperturbed states. 
Applying an additional known perturbation to the system
can break this symmetry
and can thereby considerably simplify the analysis.  
As an example with crucial relevance to many experimental situations,
we study here some consequences 
of applying a constant uniform magnetic field.
We assume the associated energy shift 
is small compared to the scale of the hyperfine structure 
but large compared to any Lorentz-violating shifts.
In this scenario,
the applied magnetic field lifts the $(2F+1)$-fold degeneracy,
so nondegenerate perturbation theory can be used
to determine the overall energy shifts. 

Choosing for convenience the laboratory frame 
so that the applied magnetic field is aligned with the $z$ axis, 
the energy shifts $\de \ep(n F J L m_F)$ of the Zeeman levels 
are determined by the diagonal components
of the matrix elements \rf{mel},
which have $m'_F = m_F$.
This gives
\beq
\de \ep(n F J L m_F) = 
\sum_{j} \AM_{j0}(n F J L) \vev{F m_F j 0|F m_F}, 
\label{defe}
\eeq
where the weights $\AM_{j0}$ are defined in Eq.\ \rf{Ajm}.  
Only weights with $m=0$ contribute,
as the Clebsch-Gordan coefficients $\vev{F m_F j m|F m_F}$ 
vanish unless $m=0$.  

The Clebsch-Gordan coefficients $\vev{F m_F j 0|F m_F}$ 
are even functions of $m_F$ for even $j$ 
and are odd functions of $m_F$ for odd $j$.  
This implies that $\vev{F 0 j 0|F 0}=0$ 
for odd values of $j$.
However,
a glance at Table \ref{rest} shows that
the only Lorentz-violating operators with even $j$
producing spectroscopic contributions are spin independent.
As a result,
spin-dependent Lorentz-violating terms cannot contribute at leading order
to the shift of any energy level with $m_F=0$.
This means,
for example,
that the transition frequency for any two levels with $m_F=0$
can at most depend on spin-independent terms. 

A key feature of an applied magnetic field
is that it sets a definite orientation for the experimental system.
Since nonzero coefficients for Lorentz violation
imply a fixed orientation in the background,
generic changes of direction of the magnetic field
alter its alignment with the coefficients and so
can produce corresponding changes in the perturbative energy shifts.
Possible origins of a changing magnetic-field orientation 
relative to the coefficients
include the rotation of the Earth,
the revolution of the Earth around the Sun,
and any effects in the laboratory due,
for example,
to placing the apparatus on a turntable.
In the laboratory frame,
these appear as a consequence of time-dependent
coefficients for Lorentz violation,
as outlined in Sec.\ \ref{Basics}.
The motion of the Earth thus naturally produces
sidereal and annual variations in some energy levels
and hence in certain spectroscopic frequencies.
Next,
we present some general considerations for these variations.
More explicit experimental applications
are presented in Sec.\ \ref{Applications}.

\subsubsection{Sidereal variations}
\label{Sidereal variations}

First,
consider effects arising from the Earth's rotation about its axis.
In the laboratory frame 
with the magnetic field along the $z$ direction as above,
the relevant nonrelativistic spherical coefficients 
${\K_\f}_{kjm}^{\rm NR,lab}$ 
have $m=0$
and vary with sidereal time
due to the rotation.
The relationship between the coefficients 
${\K_\f}^{\rm NR,Sun}_{kjm}$
in the Sun-centered frame and the coefficients 
${\K_\f}_{kj0}^{\rm NR,lab}$ 
in the laboratory frame is given by 
\beq
{\K_\f}_{kj0}^{\rm NR,lab} = 
\sum_{m} e^{i m\om_\oplus T_\oplus}
d^{j}_{0m}(-\chM)
{\K_\f}^{\rm NR,Sun}_{kjm}.
\label{ltos}
\eeq
This differs from Eq.\ \rf{tos}
due to the choice of the laboratory frame coordinates.
In particular,
the angle $\chM$ is now the relative angle 
between the applied magnetic field and the Earth's axis of rotation. 
A possible constant phase factor shifting $\om_\oplus T_\oplus$ 
has been chosen by setting the orientation of the magnetic field 
in the $XZ$ plane of the Sun-centered frame at $T_\oplus=0$.

Combining Eqs.\ \rf{defe} and \rf{ltos}
yields the energy shift
$\de \ep(n F J L m_F)$
in the presence of an applied magnetic field
expressed in the Sun-centered frame.
We find 
\bea
\de \ep(n F J L m_F) &=&
\sum_{jm} 
d^{(j)}_{0|m|}(-\chM)
\vev{Fm_Fj0|Fm_F}
\nonumber\\
&&
\hskip 10pt
\times 
\big[ \Re{\AM_{j|m|}^{\rm Sun}} \cos{(|m| \om_\oplus T_\oplus )} 
\nonumber\\
&&
\hskip 30pt
- \Im{\AM_{j|m|}^{\rm Sun}} \sin{(|m| \om_\oplus T_\oplus )} \big],
\qquad
\label{rotgen}
\eea
where the weights $\AM_{jm}^{\rm Sun}(nFJL)$ 
are defined in the Sun-centered frame 
by an expression of the same form as the weights \rf{Ajm}.

The sidereal variations \rf{rotgen}
induce oscillations of the spectroscopic lines
as a function of sidereal time.
The frequencies of these oscillations are harmonics $m\om_\oplus$ 
of the Earth's sidereal frequency $\om_\oplus$,
where $-j_{\rm max}\leq m \leq j_{\rm max}$ 
and $j_{\rm max}$ is the maximum $j$ value 
for the two energy levels involved in the transition.  
As shown in Sec.\ \ref{Coefficient selection rules},
$j_{\rm max}$ is determined by the quantum numbers $J$ and $F$.  
Denoting by $K_{\rm max}$ 
the maximum among $F$ and $J$ for both energy levels,
the harmonics that appear are given by 
$-2K_{\rm max}+1\leq m \leq 2K_{\rm max}-1$.  

The above result for the harmonic frequencies of spectroscopic lines
holds for general Lorentz violation.
However, 
as discussed in Sec.\ \ref{Coefficient selection rules},
we are limiting attention in the present work to terms with $k\leq 4$.
This corresponds to the restriction $j\leq 5$,
as can be confirmed from Table \ref{rest},
and hence involves only harmonic frequencies $\om \leq 5\om_\oplus$.
An example generating fifth-harmonic oscillations
is the transition $2S_{1/2}$-$3D^{F=3}_{J=5/2}$,
which is one of a family of transitions considered
in Sec.\ \ref{The nSnD transitions} below.
Higher harmonics are also signals of Lorentz violation
and could be sought experimentally,
but they involve more suppressed effects.

\subsubsection{Annual variations}
\label{Annual variations and parity-odd operators}

The transformation $\eqref{ltos}$
between the Sun-centered frame and the laboratory frame 
holds at zeroth order in the laboratory speed.
However,
the orbital motion of the Earth around the Sun
offers another source of variations
for tests of Lorentz and CPT symmetry,
which to date has been 
used to extract constraints on SME coefficients 
in comparatively few analyses
\cite{sunframe, boost,ma13}.
Next,
we consider some leading-order boost effects.

The instantaneous Lorentz transformation 
from the Sun-centered frame to the laboratory frame 
can be viewed as the combination 
of a boost from the Sun-centered frame 
to a frame comoving with the instantaneous laboratory frame, 
followed by a rotation to align the latter two frames 
\cite{sunframe}.
The required boost velocity $\bevec$ is the vector sum
$\bevec=\bevec_\oplus +\bevec_L$
of the instantaneous Earth orbital velocity $\bevec_\oplus$
in the Sun-centered frame 
and the instantaneous velocity $\bevec_L$
of the laboratory frame relative to the Earth's rotation axis.  
The Earth's orbital speed $\be_\oplus\simeq 10^{-4}$
is much greater than the typical rotation speed 
$\be_L \approx r_\oplus \om_\oplus \sin{\ch} \simeq 10^{-6}$
for a laboratory at colatitude $\ch \simeq 45^\circ$,
but both motions are considered here
as they yield distinct phenomenological effects.
To a sufficient approximation
the Earth's orbit can be taken as circular,
so the velocity $\bevec_\oplus$ can be written as 
\beq
\bevec_\oplus=
\be_\oplus \sin{\Om_\oplus T} ~\Xhat
-\be_\oplus \cos{\Om_\oplus T}
( \cos\et~\Yhat + \sin\et~\Zhat ) ,
\label{vorb}
\eeq
where $\Om_\oplus\simeq 2\pi/(365.26 \text{ d})$ 
is the Earth's orbital frequency,
$T$ is the time in the Sun-centered frame,
and $\et\simeq23.4^\circ$ is the angle between the
$XY$ plane and the Earth's orbital plane.
Similarly,
the velocity $\bevec_L$ takes the form
\beq
\bevec_L=
- \be_L \sin{\om_\oplus T_\oplus}~\Xhat
+ \be_L \cos{\om_\oplus T_\oplus}~\Yhat ,
\label{vrot}
\eeq
where $T_\oplus$ is the local Earth sidereal time. 
Note that the difference $T-T_\oplus$ 
is merely a phase 
that physically represents a convenient choice of local time zero 
for a specified tangential velocity. 

One advantage to considering boost effects arises 
because the boost and parity operators fail to commute,
implying that parity-even operators in the laboratory frame
incorporate parity-odd ones in the Sun-centered frame.
The connection between the two sets of operators
is provided by the boost velocity,
which changes sign under parity.
The experimental sensitivity to parity-odd Lorentz violation
is therefore suppressed by at least a factor of $10^{-4}$,
but as shown below the observable signals are distinct.
Another advantage arises because
in addition to mixing operators of different parity,
the transformation between the two frames
also mixes the irreducible rotation representations.
This can enrich the expected signals for Lorentz violation.
For example,
laboratory measurements of an isotropic Lorentz-violating effect
can also test anisotropic effects in the Sun-centered frame,
which then appear combined with the boost velocity.
The mixing of irreducible rotation representations does,
however,
imply a significant calculational issue for boost effects
because performing the spherical decomposition is no longer natural,
yielding cumbersome transformation rules for the spherical operators.
To avoid this issue,
we work here with the cartesian basis,
for which calculations are more direct.

In the context of the applications 
discussed in Sec.\ \ref{Applications} below,
two types of parity-even laboratory observables are of particular interest,
scalars and axial 3-vectors.
For example,
the former is relevant for the 1$S$-2$S$ transition,
while the latter is relevant to hyperfine Zeeman transitions.
Consider first the simplest case
involving the laboratory-frame measurement
of a parity-even observer rotational scalar $S^{\rm lab}$ 
such as the weight $\AM_{00}(n L)$ in Eq.\ \rf{defe}.
The scalar $S^{\rm lab}$ 
can be expressed in the Sun-centered frame as 
\beq
S^{\rm lab} = S^{\rm Sun} + V^J \bevec^J,
\label{Suniso}
\eeq
where $V^J$ is defined in the Sun-centered frame 
and transforms as a vector under observer rotations.
Note that $V^J$ can receive only contributions 
from anisotropic parity-odd Lorentz-violating operators 
in the Sun-centered frame.  
Substituting for $\bevec^J$ using Eq.\ \rf{vorb}
then reveals that in the laboratory frame 
the measurement of $S$ exhibits annual variations.
Similarly,
Eq.\ \rf{vrot} predicts sidereal variations of $S$.
The two effects have distinct experimental signatures
and are sensitive to different combinations
of the components of $V^J$ in the Sun-centered frame.
As a result,
experiments performing a boost analysis on a scalar observable
can achieve interesting and distinctive sensitivities
to coefficients for Lorentz violation.

Next,
we consider a measurement of the $z$ component $A^z$
of an observer axial 3-vector $A^J$ in the laboratory frame,
such as the weight $\AM_{10}(nFJL)$ in Eq.\ \rf{defe}.
At first order in $\bevec^J$
and in terms of quantities in the Sun-centered frame,
$A^z$ can be written as 
\beq
A^z = R^{zJ} A^{{\rm Sun},J} 
+ R^{zJ} \bevec^{K} T^{JK},
\label{Sunvec}
\eeq
where $T^{JK}$ is defined in the Sun-centered frame 
and transforms as a rank-2 pseudotensor under spatial rotations.
The quantity
$R^{zJ}$ is the $z$th row of the rotation matrix $R^{jJ}$
between the boosted frame and the laboratory frame,
with entries given by
\bea
R^{zX} &=&
\sin{\chM}\cos{(\om_\oplus T_\oplus + \phM)},
\nonumber\\
R^{zY} &=&
\sin{\chM}\sin{(\om_\oplus T_\oplus + \phM)}, 
\nonumber\\
R^{zZ} &=&
\cos{\chM},
\label{Rz}
\eea
where as before $\chM$ is the angle between the magnetic field 
and the Earth's rotation axis. 
The phase $\phM$ is the angle between the $X$ axis 
and the projection of the magnetic field on the $XY$ plane at $T_\oplus =0$. 
A useful perspective is to view $R^{zJ}$ as a unitary vector 
pointing in the direction of the applied magnetic field.  

The first term in Eq.\ \rf{Sunvec}
is just the $j=1$ component of the right-hand side of Eq.\ \rf{rotgen},
expressed in the cartesian basis.
This produces sidereal signals as discussed 
in Sec.\ \ref{Sidereal variations},
so it suffices here to consider the second term 
$R^{zJ} \bevec^{K} T^{JK}$
in Eq.\ \rf{Sunvec}.
To compare the pseudotensor $T^{JK}$
to the spherical decomposition,
it is convenient to decompose $T^{JK}$
into irreducible rotation representations.
This decomposition gives 
\bea
R^{zJ} \bevec^{K} T^{JK}
&=&
\frac 13 R^{zJ}\bevec^{J} T^{KK}
+ R^{zJ}\bevec^{K} T^{[JK]} 
\nonumber \\
&&
+ R^{zJ}\bevec^{K}
\big( T^{(JK)} - \frac 13\de^{JK}T^{LL} \big),
\quad
\label{ired}
\eea
where indices in brackets and parentheses
indicate antisymmetrization and symmetrization,
respectively,
both with a factor of 1/2.

The first term in Eq.\ \rf{ired}
contains the trace $T^{KK}$,
which in the spherical basis corresponds 
to combinations of nonrelativistic coefficients 
of $B$-type parity with $jm=00$.
Its contribution is proportional to $R^{zJ}\bevec^{J}$,
so the corresponding signals can be altered significantly 
by manipulating the direction of the magnetic field.  
For example,
if the magnetic field is chosen orthogonal to $\bevec_L$,
then $R^{zJ}\bevec_L^J=0$
and only annual variations arise from this term.
If instead the magnetic field is parallel to $\bevec_L$,
then $R^{zJ}\bevec^J = \be_L + R^{zJ}\bevec_\oplus^J$.
For more generic orientations the signal can be complicated,
with coupled sidereal and annual variations.

The contribution in Eq.\ \rf{ired} 
involving the antisymmetric representation $T^{[JK]}$ 
contains nonrelativistic spherical coefficients
of $E$-type parity with $j=1$.
This term can be viewed as being contracted with 
a factor $R^{z[J}\bevec^{K]}$,
which represents the components of the cross product 
of the vector $R^{zJ}$ with the velocity $\bevec^K$.
If the magnetic field is parallel to $\bevec_L$, 
then the contributions from 
$SR^{zJ}\bevec^{K} T^{[JK]}$ 
vary only at the annual frequency $\Om\oplus$.
If the magnetic field is parallel to the Earth's rotation axis,
then the configuration is insensitive
to the combination of coefficients
contained in $\ep^{ZJK}T^{JK}$.
Generic orientations of the magnetic field
again lead to coupled sidereal and annual variations. 

The final term in Eq.\ \rf{ired} 
involves the traceless symmetric part of the pseudotensor,
which corresponds to nonrelativistic spherical coefficients
of $B$-type parity with $j=2$.
If the magnetic field is chosen along the Earth's rotation axis,
the term of order $\be_\oplus$ exhibits only annual variations
and the experiment is sensitive 
to the combinations of coefficients contained in
$T^{(ZJ)}$ and $T^{JJ}$.
In this configuration,
the term of order $\be_L$ involves 
only the fundamental frequency $\om_\oplus$. 
For other orientations of the magnetic field,
the signal also incorporates variations
at the second harmonic $2\om_\oplus$.

\section{Applications}
\label{Applications}

This section discusses observable experimental signals
for Lorentz and CPT violation
that could appear in spectroscopic studies of hydrogen. 
We begin by addressing the case of free hydrogen
in the absence of applied fields,
characterizing the resulting level splitting
and possible experimental signals.
Experiments with hyperfine Zeeman transitions
are treated next.
We obtain the energy-level and frequency shifts
due to Lorentz and CPT violation
and study several types of time variations in experimental signals,
including sidereal and annual modulations 
together with turntable and orbital effects.
Existing experimental data are used to place
constraints on nonrelativistic coefficients.
Two subsections treat spectroscopy involving
the transitions $nL$-$n'L'$,
including in particular the $1S$-$2S$ transition.
We obtain the associated frequency shifts
and discuss new constraints and the future reach 
available via a self-consistent analysis
or via sidereal and annual variations.

\subsection{Signals without background fields}
\label{no field}

In the absence of applied fields,
the physical manifestation of Lorentz violation in free atomic hydrogen
is a splitting of otherwise degenerate energy levels.
In the Sun-centered frame,
the perturbed states can be constructed by diagonalizing 
the Lorentz violation in the degenerate subspace.
For example,
as discussed in Sec.\ \ref{Energy shifts for $F=0$ and $F=1$},
Lorentz violation causes the ground state of free hydrogen 
to split into four sublevels 
in a pattern analogous to hyperfine Zeeman splitting,
despite the absence of a magnetic field.
At leading order,
the ground states are eigenstates of the operator $\mbf A \cdot \mbf F$ 
restricted to the corresponding subspace, 
where $\mbf A$ is the vector formed from the $A_{1m}$ coefficients 
in Eq.\ \rf{f1cr}
and $\mbf F$ is the total angular momentum.
In this subsection,
we consider prospects for experimental investigations 
of this degeneracy lifting.

\subsubsection{Transition probabilities}
\label{Level splitting}

In the Sun-centered frame,
the splitting of the energy levels in free hydrogen is time independent
because the coefficients for Lorentz and CPT violation 
can be taken as constant in this frame. 
Suppose an experiment is designed to excite transitions between these states 
using a laser with a fixed polarization in the laboratory frame.  
In the Sun-centered frame,
the laboratory is rotating due to the Earth's spin 
and the laser polarization rotates with it.
The relative orientation between the polarization of the laser 
and the split hydrogenic states 
therefore changes as a function of sidereal time,
affecting the transition probabilities.  
This represents a unique signal of Lorentz violation.

To see more explicitly the effect,
we restrict attention to the comparatively simple case with $J=1/2$.
Suppose the laser has linear polarization aligned 
along the laboratory $z$ axis,
and suppose we want to excite the transition 
between the eigenstates $F=0$ and $F=1$, $\xi=0$
with energies fixed by Eq.\ \rf{f1cr}.
Using the dipole approximation,
the transition probability is proportional 
to the squared magnitude of the dipole matrix element $\cT_{fi}$ 
between the initial state $\ket{i}$
and the final state $\ket{f}$,
$|\cT_{fi}|^2 \propto |\bra{i}z\ket{f}|^2$.
In terms of the basis states $\ket{nFJLm_F}$
and the weights \rf{A1m1S} 
introduced in Sec.\ \ref{Matrix elements},
we have 
\beq
\ket{i} = \ket{n0\half L0},
\qquad
\ket{f} = \fr 1A \sum_{m} \AM_{1m} \ket{n'1\half L'm}.
\label{ada}
\eeq
These expressions are valid in any frame.
The basis $\ket{nL\half Fm_F}$ can be taken as quantum observer invariant
under frame transformations,
as discussed in Sec.\ \ref{Energy shifts for $F=0$ and $F=1$},
but the weights $A_{1m}$ transform under rotations.
In the laboratory frame,
the squared magnitude of the dipole matrix element becomes
\beq
|\cT_{fi}|^2 \propto |\bra{i}z\ket{f}|^2 =
\fr{|\AM_{10}^{\rm lab}|^2}{A^2}
|\vev{nL\half 00|z|n'L'\half 10}|^2.
\label{trans}
\eeq
We assume here an adiabatic rotation
so that the perturbation method is valid.
This is reasonable as the Earth's sidereal period
is much greater than the timescale for photon absorption.
Finally,
converting this result to the Sun-centered frame 
using Eq.\ \rf{tos}
reveals the time variation of the transition probabilities
at harmonics of the sidereal frequency $\om_\oplus$.

An interesting insight obtained from Eq.\ \rf{trans}
is that the sidereal variation of the transition probability
can be an unsuppressed effect,
as it depends only on the ratio
of coefficients for Lorentz and CPT violation
rather than their absolute values.
This distinctive feature has no parallel
in typical experiments performed in applied fields,
such as the observations of Zeeman hyperfine transitions
discussed in Sec.\ \ref{$1S$ hyperfine-Zeeman} below.
The catch here is that an experiment measuring
the unsuppressed transition probabilities
must be able to resolve the energy splitting
due to the Lorentz and CPT violation,
which itself is a suppressed effect.

\subsubsection{Line shapes}
\label{Applied oscillatory magnetic field}

Since the transition probabilities vary with sidereal time,
so do the observed line shapes.
To illustrate this,
we assume 
that the Lorentz-violating splitting is detectable
and that an ensemble of particles 
in the state $F=1$, $\xi=0$ can be produced.  
If this system is exposed to an oscillating magnetic field 
$\mbf B = \mbf B_0 \cos \om t$,
the time-dependent perturbation can be taken as 
\beq
\De h(t)= \mu_B 
(g_e \mbf{S_e} \cdot \mbf B_0 + g_p\mbf{S_p} \cdot \mbf B_0) 
\cos{\om t},
\eeq
where $\mu_B$ is the Bohr magneton 
and $g_\f$ is to the gyromagnetic ratio of the particle of flavor $\f$.  
The time $t$ coincides with the local Earth sidereal time $T_\oplus$
up to a phase.
Note that the frequency for the transition $\De F=-1$, $\De \xi=0$ 
is unaffected by Lorentz violation,
so it coincides with the ground-state hyperfine-splitting frequency $\om_{0}$.  

Suppose now the oscillation frequency $\om$ of the magnetic field
is tuned near resonance, 
$\om= \om_0 + \De\om$,
where $\De\om \ll \om_0$.
For generic orientations of the field,
the setup can then be approximated as a two-state system.
Using the rotating-field approximation 
\cite{ra56},
the transition probability is given by
\beq
\cP(t) = 
\fr{\ga^2}{\ga^2+\De\om^2}
\sin^2\big(\half {\sqrt{\ga^2+\De\om^2}~t}\big),
\label{ls}
\eeq
where $\ga = (g_e-g_p) (\mbf B \cdot \widehat{\mbf A})$.  
The magnetic field rotates with the laboratory
and the Earth at angular frequency $\om_\oplus$,
so in the Sun-centered frame
the product $\mbf B \cdot \widehat{\mbf A}$
and hence $\ga$ depends on sidereal time.
The explicit dependence is
\bea
\widehat{\mbf B} \cdot \widehat{\mbf A} &=& 
A_{10} \cos{\chM}
- \sqrt{2} \sin{\chM} \Re{A_{11}} \cos(\om_\oplus T_\oplus + \ph_0)
\nonumber\\
&&
+ \sqrt{2} \sin{\chM} \Im{A_{11}} \sin(\om_\oplus T_\oplus + \ph_0),
\label{gasi}
\eea
where $\chM$ is the angle between the magnetic field
and the rotation axis of the Earth.
The phase $\ph_0$ is the azimuthal angle of the magnetic field 
in the Sun-centered frame at time $T_\oplus =0$.  
Substituting this expression into Eq.\ \rf{ls}
determines the line shape for the transition,
including its variation with sidereal time.

The two-state approximation used above fails for configurations
with orthogonal or near-orthogonal $\mbf B$ and $\mbf A$
because the probability of the stimulated transition with $\De \xi =0$ 
then becomes smaller and comparable to other allowed transitions.  
None of the transitions are on resonance in this scenario,
so the probability for a stimulated transition can be disregarded.
For example,
if $A_{10} \cos{\chM}$ is negligible compared to the other terms
in Eq.\ \rf{gasi},
then  $\widehat{\mbf B} \cdot \widehat{\mbf A}$
fluctuates to zero and back,
implying that the signal for Lorentz violation includes peaks and valleys 
in the transition probability as a function of the sidereal time. 

\subsubsection{Prospects}
\label{Prospects}

The vector $\mbf A$ is determined 
by the coefficients for Lorentz and CPT violation.
Assuming this vector is nonzero and known,
the amusing possibility arises that a hydrogen maser
could be created based on the Lorentz-violating level splitting.
Constructing the oscillating magnetic field $\mbf B$ 
to be aligned with $\mbf A$
would generate an approximate two-state system
without the need for the usual applied external field
to break the system degeneracy. 
The necessary population inversion could be produced,
for example,
by overlapping the Lorentz-violating background field
with an inhomogeneous magnetic field 
to select the states seeking low field.
A possible advantage of a Lorentz-violation maser 
is that the vector $\mbf A$ is expected to be highly homogeneous
because the coefficients for Lorentz violation
can be assumed uniform and constant in the Sun-centered frame
\cite{ck},
so the issues for conventional masers
arising from the inhomogeneity of the applied magnetic field
would be irrelevant.  
However,
realizing a Lorentz-violation maser 
in an Earth-based laboratory would face the challenge
of overcoming the effective sidereal oscillation of $\mbf A$
in the laboratory due to the Earth's rotation.
This tends to skew its alignment with $\mbf B$
and hence would permit the excitation
of transitions between the ground state
and the levels $F=1$, $\xi = \pm 1$,
destroying the two-state approximation
and reducing the emission of coherent microwaves. 

More generally,
the similarities between the Lorentz-violating splittings
and conventional linear Zeeman shifts 
provide an intuitive guide to prospective experimental options.
For example, 
transitions between the different $F=1$ sublevels 
could be investigated using tools like those 
adopted for studies of $F=1$ Zeeman transitions 
in the presence of a uniform magnetic field.  
Current sensitivities to the $F=1$ Lorentz-violating splittings
have attained about 1 mHz
using measurements of hyperfine Zeeman transitions
\cite{hu00,maser,hu03},
so we can assume the resonance frequency between the $F=1$ levels
lies below this value.
One option to improve the sensitivity might be
to prepare an ensemble of atoms in the $F=1$ state 
and probe them with a magnetic field oscillating at a frequency below 1 mHz.  
Assuming a mechanism to monitor induced transitions can be implemented,
then sweeping over decreasing frequencies could lead to better sensitivities
to the coefficients for Lorentz violation.
Note, however, 
that for $J>1/2$ the Lorentz-violating splitting
lacks a Zeeman-type structure,
so studies of the various associated transitions
would require developing the corresponding phenomenology.

Another possibility is to search for line separation or broadening
arising from the Lorentz-violating level splittings.
Suppose a transition between two states with $J=1/2$ is studied.  
Ideally,
the Lorentz-violating splitting would be detected 
in the form of multiple resonance peaks.  
Even if individual peaks cannot be resolved,
the modified line shapes could be calculated
and the minimum value of the effect leading to resolvable peaks 
within the particular experimental scenario could be determined.
This would correspond to a constraint 
on the coefficients for Lorentz violation.
Note that the parallel to the Zeeman hyperfine splitting
implies that the putative signal for Lorentz violation 
can be approximated experimentally
by applying to the ensemble of hydrogen atoms
a uniform external magnetic field 
rotating with sidereal frequency.
Moreover,
the Lorentz violation also produces line broadening,
which could be studied directly.
Consider,
for example,
the transitions $F=0$ to $F=1$ under the assumption
that levels with all values of $\xi$ are excited with equal probability.  
An estimate of the Lorentz-violating line broadening $\De E$
can be found by calculating the statistical deviation
arising from the availability of levels of different $\xi$.
Ignoring the natural linewidth,
this gives
\beq
(\De E)^2 = \frac 13 |\mbf{A}|^2.
\eeq
The result indicates that the Lorentz-violating broadening 
is related to the magnitude of the vector $\mbf{A}$.

\subsection{Hyperfine Zeeman transitions}
\label{$1S$ hyperfine-Zeeman}

In this subsection,
we consider effects of Lorentz and CPT violation 
on the hyperfine levels of hydrogen in the presence of a weak magnetic field.
The $1S_{1/2}$ level in hydrogen is split into two sublevels,
a ground state with total atomic angular momentum $F=0$ 
and an excited state with $F=1$. 
Applying a weak magnetic field further splits the $F=1$ hyperfine level 
into three Zeeman sublevels
with energies determined by the eigenvalue $m_F$ 
of the component of $\mbf F$ along the magnetic field.  
We determine the frequency shifts from Lorentz and CPT violation,
and we discuss some signals involving sidereal variations,
changes of the orientation of the magnetic field, 
and boosts.
Signals in space-based missions
and in other hydrogenic systems are also described.

\subsubsection{Frequency shift}
\label{Frequency shift}

The Lorentz-violating energy shifts of the hyperfine Zeeman sublevels
for $J=1/2$, $L=0$ or 1, and any $n$
can be obtained from Eq.\ \rf{defe}
along with the expression \rf{Ajm}
for the weights $A_{00}(n 1 \half L)$.
The result is
\bea
\de\ep(m_F) &=& 
- \sum_{q=0}^2
\left(\fr {\al \mr}{n}\right)^{2q} 
\left(1 + \left( \fr{8n}{2L+1} - 4 \right) \de_{q2}\right) 
\nonumber\\
&&
\hskip -40pt
\times
\sum_\f 
\left[
\fr{\Vnrf{\f}{(2q)00}} {\sqrt{4\pi}}
+\fr{m_F}{2\sqrt{3\pi}}
\left( \TzBnrf{\f}{(2q)10} + 2\ToBnrf{\f}{(2q)10}\right) 
\right],
\nonumber\\
\label{1S}
\eea
where the quantities
$\Vnrf{\f}{(2q)00}$, $\TzBnrf{\f}{(2q)10}$, $\ToBnrf{\f}{(2q)10}$ 
are expressed in terms of nonrelativistic spherical coefficients
by Eq.\ \rf{cpt}.
Note that this extends the known result for the minimal SME
\cite{bkr}
to include contributions from the $d=4$ coefficients $g_{\la\mu\nu}$ 
along with ones involving operators of arbitrary $d$.

The frequency shifts for the hyperfine Zeeman transitions
of the ground state follow from this result.
Denoting by $\De m_F$ the difference between the values of $m_F$ 
for the initial sublevel and the final one,
we obtain
\bea
2\pi\de\nu &=&
-\fr{\De m_F}{2\sqrt{3\pi}}
\sum_{q=0}^2
(\al \mr)^{2q}(1+4\de_{q2})
\nonumber\\ 
&&
\hskip 40pt
\times
\sum_\f 
\big[ \gzBnrf{\f}{(2q)10}-\HzBnrf{\f}{(2q)10}
\nonumber\\ 
&&
\hskip 60pt
+2\goBnrf{\f}{(2q)10}-2\HoBnrf{\f}{(2q)10}
\big].
\qquad
\label{1Sf}
\eea
Note that the isotropic coefficients,
which are contained in $\Vnr{k00}$ and modify the energies
according to Eq.\ \rf{1S},
are absent from this frequency shift.
This agrees with the result 
obtained in Sec.\ \ref{Energy shifts for $F=0$ and $F=1$}
that isotropic coefficients only contribute to transitions
with $\De n\ne 0$ or $\De L \ne 0$. 
Note also that the result \rf{1Sf}
contains the minimal-SME limit via the restriction
\bea
&&
\hskip -10pt
\gzBnrf{\f}{010} + 2\goBnrf{\f}{010} - \HzBnrf{\f}{010} - 2\HoBnrf{\f}{010} 
\nonumber\\
&& 
\to
2\sqrt{3\pi} [b^\f_3 - m_\f d^\f_{30} - H^\f_{12}
- m_\f {g^{\f(A)}_3} + m_\f {g^{\f{(M)}}_{120}}],
\nonumber\\
\eea
where the superscripts (A) and (M) indicate 
the irreducible axial and irreducible mixed-symmetry
combinations of the minimal-SME coefficients $g^{\f}_{\ka\la\nu}$,
respectively
\cite{krt08,fittante}. 
The frequency shift \rf{1Sf} thereby
matches the result reported in Ref.\ \cite{bkr}
with only $b^\f_\mu$, $d^\f_{\mu\nu}$, and $H^\f_{\mu\nu}$
contributing at leading order,
which neglects the $g$-type minimal-SME coefficients 
as suppressed by the necessary accompanying breaking 
of the electroweak SU(2)$\times$U(1) symmetry
\cite{ck}.

The expression \rf{1Sf} reveals that only transitions with $\De m_F \ne 0$ 
are sensitive to Lorentz violation at leading order,
independent of the operator mass dimension $d$. 
One implication of this observation 
is that the standard transition used in hydrogen masers,
$F=0\rightarrow F=1$ with $\De m_F=0$,
is insensitive to Lorentz violation.  
The Lorentz violation considered in this work 
involves only propagation effects,
which cannot shift the standard transition frequency
because the reduced density matrices for the spin singlet 
and entangled triplet are identical
and so yield identical expectation values for any operator 
that acts on only one subsystem.

\subsubsection{Sidereal variations}
\label{Sidereal variations at zero order in the boost parameter}

At zeroth order in the boost,
the nonrelativistic spherical coefficients in the laboratory frame 
can be expressed in terms of coefficients 
in the canonical Sun-centered frame as
\bea
{\K_\f}_{\k 10}^{\rm NR,lab} &=&
{\K_\f}_{\k 10}^{\rm NR,Sun} 
\cos\chM
\nonumber\\
&& 
-{\sqrt{2}} ~\Re {\K_\f}_{\k 11}^{\rm NR,Sun} 
{\sin\chM}\cos\om_\oplus T_\oplus
\nonumber\\
&& 
+{\sqrt{2}} ~\Im {\K_\f}_{\k 11}^{\rm NR,Sun} 
{\sin\chM}\sin\om_\oplus T_\oplus, 
\label{rot}
\eea
which is a special case of Eq.\ \rf{ltos}. 
As before, $\om_\oplus$ is the Earth's sidereal rotation frequency, 
$T_\oplus$ is the sidereal time,
and $\chM$ is the angle between the applied magnetic field 
and the Earth's rotation axis.  
Together with the expression \rf{1Sf} for the frequency shift,
the above relation predicts that 
the hyperfine Zeeman transition frequencies oscillate 
with frequency $\om_\oplus$ in the presence of Lorentz violation.
This result is in agreement with the discussion  
in Sec.\ \ref{Sidereal variations},
with the identification $K_{\rm max}=F=1$ 
and hence obtaining $|m_{\rm max}|=2K_{\rm max}-1=1$.  

An experiment performed with a maser located 
at the Harvard-Smithsonian Center for Astrophysics 
searched for sidereal variations 
of the hyperfine Zeeman transitions with $F=1$ and $\De m_F=\pm 1$,
finding no signal to within $\pm 0.37$ mHz at one standard deviation 
\cite{hu00,maser,hu03}.  
Using the frequency shift \rf{1Sf} for colatitude $\ch\simeq 48^\circ$,
this implies the bound
\bea
\bigg|
\sum_{q=0}^2
(\al \mr)^{2q} (1+4\de_{q2})
\sum_\f 
\big[ \gzBnrf{\f}{(2q)10}-\HzBnrf{\f}{(2q)10}
&&
\nonumber\\
&&
\hskip -120pt
+2\goBnrf{\f}{(2q)10}-2\HoBnrf{\f}{(2q)10}
\big]
\bigg|
\nonumber \\
< 9 \times 10^{-27} {\rm ~GeV},
\quad
\label{1Sb}
\eea
which constrains a subset of the nonrelativistic spherical coefficients 
in the Sun-centered frame.

\renewcommand\arraystretch{1.5}
\begin{table}
\caption{Constraints on the moduli of the real and imaginary parts 
of electron and proton nonrelativistic coefficients 
determined from hyperfine Zeeman transitions in hydrogen.}
\setlength{\tabcolsep}{7pt}
\begin{tabular}{cl}
\hline
\hline
Coefficient & \qquad Constraint on 
\\
[-4pt]
$\K$ & \qquad $|\Re\K|$, $|\Im\K|$ 
\\
\hline
$\HzBnr{011}$,	$\gzBnr{011}$	&\quad	$<9 \times10^{-27}\,$GeV\\
$\HoBnr{011}$,	$\goBnr{011}$	&\quad	$<5\times10^{-27}\,$GeV\\
$\HzBnr{211}$,	$\gzBnr{211}$	&\quad	$<7 \times10^{-16}\,\text{GeV}^{-1}$\\
$\HoBnr{211}$,	$\goBnr{211}$	&\quad	$<4 \times10^{-16}\,\text{GeV}^{-1}$\\
$\HzBnr{411}$,	$\gzBnr{411}$	&\quad	$<9 \times10^{-6}\,\text{GeV}^{-3}$\\
$\HoBnr{411}$,	$\goBnr{411}$	&\quad	$<5 \times10^{-6}\,\text{GeV}^{-3}$\\
\hline\hline
\end{tabular}
\label{tanr1S}
\end{table}

Intuition about the implications of this constraint can be gained 
by adopting the assumption that only one coefficient is nonzero at a time
and extracting the resulting limits.
Table \ref{tanr1S} presents the constraints 
on individual nonrelativistic spherical coefficients 
obtained in this way.
The results hold equally for electron and proton coefficients.
Note that several of these lie well below the level 
at which Planck-scale signals 
might be expected to arise in some models.

We emphasize that this type of measurement 
bounds the effects of certain Lorentz-violating operators at arbitrary $d$,
even though only the subset of nonrelativistic spherical coefficients 
with $k\leq 4$ and $j=1$ is accessible.
This is because a nonrelativistic spherical coefficient with $j=1$ 
is a linear combination of an infinite subset 
of the basic spherical coefficients with $d\geq 3$
\cite{km13}.
Note also that 
in terms of the basic spherical coefficients
the experiment has greater reach for protons than for electrons
due to the mass factors that enter the relevant linear combinations.
For example, 
the sensitivity to the basic spherical coefficient 
at mass dimension $d$ and with $k=0$ 
is numerically that of the corresponding nonrelativistic spherical coefficient 
suppressed by a factor of $(0.94)^{3-d}/3$ for the proton 
and a factor of $(5.1\times 10^{-4})^{3-d}/3$ for the electron.

\subsubsection{Changes of magnetic-field orientation}
\label{Variations of the orientation of the magnetic field}

The bound \rf{1Sb} is insensitive 
to nonrelativistic spherical coefficients with $m=0$ 
in the Sun-centered frame
because these coefficients enter the frequency \rf{rot}
without the dependence on $T_\oplus$
necessary for the experimental signal.
However,
Eq.\ \rf{rot}
predicts that these coefficients do change with the angle $\chM$
between the magnetic field and the Earth's rotation axis.
An experiment involving a changing magnetic-field orientation
is therefore of interest.
One possibility along these lines 
would be to place the apparatus on a rotating turntable.

For simplicity,
suppose the rotation axis of the turntable points towards the zenith
and the magnetic field is perpendicular to it. 
Defining the laboratory-frame $z$ axis to lie along the magnetic field,
the laboratory-frame coefficients are given
in terms of Sun-frame coefficients by
\bea
{\K_\f}_{\k 10}^{\rm NR,lab} 
&=&
- {\K_\f}_{\k 10}^{\rm NR,Sun} 
\sin\ch 
\cos{T_r\om_r} 
\nonumber\\
&&
- \sqrt{2}~
\Re {\K_\f}_{\k 11}^{\rm NR,Sun} 
\cos{\ch} \cos{\om_r T_r} 
\cos{\om_\oplus T_\oplus} 
\nonumber \\
&&
+ \sqrt{2}~
\Im {\K_\f}_{\k 11}^{\rm NR,Sun} 
\cos{\ch} \cos{\om_r T_r} 
\sin{\om_\oplus T_\oplus} 
\nonumber \\
&&
+ \sqrt{2}~ 
\Re {\K_\f}_{\k 11}^{\rm NR,Sun} 
\sin{\om_r T_r} 
\sin{\om_\oplus T_\oplus}
\nonumber\\
&&
+ \sqrt{2}~ 
\Im {\K_\f}_{\k 11}^{\rm NR,Sun} 
\sin{\om_r T_r} 
\cos{\om_\oplus T_\oplus},
\quad
\eea
where $\ch$ is the colatitude of the experiment 
and $\om_r$ is the angular rotation frequency of the turntable.  
For convenience,
we have introduced a time $T_r$ shifted relative to $T_\oplus$,
with the origin $T_r=0$ chosen  
to be the time when the magnetic field points south. 
In this scenario,
the coefficients with $m=0$ in the Sun-centered frame 
are independent of the sidereal frequency,
producing variations at the turntable angular frequency $\om_r$ 
of the coefficients in the laboratory frame
and hence of the measured transition frequencies.
The attainable sensitivity to $m=0$ coefficients 
in an experiment of this type 
is expected to be similar to the sensitivities 
presented in Table \ref{tanr1S} 
for the corresponding $m=1$ coefficients.

\subsubsection{Annual variations}
\label{Boost corrections and parity-odd operators}

\renewcommand\arraystretch{1.7}
\begin{table}
\caption{
Values of the pseudotensors $T_{\f}^{(d)JK}$ for $3\leq d \leq 8$.} 
\setlength{\tabcolsep}{7pt}
\begin{tabular}{cc}
\hline
\hline
$d$ & $V^{(d)J}_\f$ 
\\
\hline
3   &   $ \Heff{JK} $\\
4   &   $-2m_\f \geff{K(TJ)}$\\
5   &   $3 m_\f^2\Heff{J(TKT)}+(\al \mr)^2\Heff{J(KLL)}$\\
6   &   $-4 m_\f^3\,\geff{J(TTTK)}-4(\al \mr)^2m_\f\,\geff{J(TLLK)}$ \\
 7  &    $5m_\f^4 \,\Heff{J(TTTTK)}+10 (\al \mr)^2m_\f^2\,\Heff{J(TTLLK)}$\\
      &   $+5 (\al \mr)^4\Heff{J(KLLMM)}$\\
8   &     $-6 m_\f^5 \,\geff{J(TTTTTK)}-20 (\al \mr)^2m_\f^3 \,\geff{J(TTTLLK)}$\\
      &   $-30 (\al \mr)^4m_\f\geff{J(KTLLMM)}$\\
\hline\hline
\end{tabular}
\label{Amm}
\end{table}

As described in
Sec.\ \ref{Annual variations and parity-odd operators},
the inclusion of boosts in the analysis
implies the appearance of contributions from parity-odd operators. 
The frequency shift in the laboratory frame 
transforms as the $z$ component of a vector,
so for present purposes we denote it as $\de\nu^z$.
Its expression in the Sun-centered frame
takes the generic form \rf{Sunvec}
at first order in the boost velocity $\bevec$,
\beq
2\pi \de\nu^z =
2 \pi R^{zJ}\de\nu^{{\rm Sun,}J}
+ \De m_F
\sum_{d\,f} R^{zJ} T^{(d)JK}_\f (\bevec^K_\oplus + \bevec^K_L),
\label{hfboost}
\eeq
where 
$\bevec_\oplus$, $\bevec^J_L$, and $R^{zJ}$ 
are defined in 
Eqs.\ 
$\eqref{vorb}$, $\eqref{vrot}$, and $\eqref{Rz}$, 
respectively.
The form of the pseudotensor $T^{(d)JK}_\f$
depends on the operator mass dimension $d$ and the particle flavor $\f$.

Table \ref{Amm} provides explicit expressions for $T^{(d)JK}_\f$
with $3\leq d\leq 8$
in terms of 
the particle rest masses $m_\f$,
the fine-structure constant $\al$,
the reduced mass $\mr$ of the system,
and the effective cartesian coefficients for Lorentz violation
defined in Eqs.\ (27) and (28) of Ref.\ \cite{km13}.
Only leading-order nonrelativistic
contributions from each coefficient are included.
In the table,
parentheses around sets of $n$ indices indicate total symmetrization
with respect to all indices enclosed,
including a factor of $1/n!$.

Examining Eq.\ \rf{hfboost} and Table \ref{Amm} reveals that
only the first index of each effective cartesian coefficient
is contracted with the rotation matrix.
This feature arises because only the first two indices 
of the $g$- and $H$-type effective cartesian coefficients 
are coupled to the particle spin via contraction
in the original Lagrange density
\cite{km13}.
The applied magnetic field interacts with the magnetic dipole moment
of the particle 
and so fixes the particle's spin orientation in the laboratory frame.
The spin follows any adiabatic rotation of the magnetic field,
which thereby changes the value 
of the contraction between the coefficients and the spin.
However,
the $g$- and $H$-type effective cartesian coefficients 
are antisymmetric on the first two indices,
so the result can always be interpreted as a rotation
associated with the first index on any coefficient. 

The cartesian basis is convenient for boost corrections.
However,
as outlined in Sec.\ \ref{Annual variations and parity-odd operators},
we can decompose $T_{\f}^{(d)JK}$ 
in terms of irreducible representations of the rotation group.  
These irreducible representations are associated with spherical coefficients.
For example,
for mass dimensions $d=3$ and $d=4$,
we obtain 
\bea
\ep_{zJK} T^{(3)JK}_\f &=&
\sqrt{\fr 3 \pi} {H_\f}^{(3)(1E)}_{110},
\nonumber\\
\ep_{zJK}T^{(4)JK}_\f &=&
- m_\f \sqrt{\fr 3 \pi} {g_\f}^{(4)(1E)}_{110},
\nonumber\\
T^{(4)JJ}_{\f} &=&
- m_\f \sqrt{\fr 9 \pi} {g_\f}^{(4)(0B)}_{100},
\nonumber\\
T^{(4)zz}_\f - \fr 13 T^{(4)JJ}_\f &=&
- m_\f \sqrt{\fr 5 \pi} {g_\f}^{(4)(0B)}_{120}.
\eea
These and similar expressions for $d>4$ can be used
to relate results in the cartesian and spherical bases.

As discussed in Sec.\ \ref{Annual variations and parity-odd operators},
orienting the magnetic field parallel to the Earth's rotation axis 
decouples the sidereal and annual variations of the measured frequency,
with sidereal variations associated to terms of order $\be_L$ 
and annual variations to terms of order $\be_\oplus$.  
At first order in the boost parameter,
the frequency shift for this orientation of the magnetic field 
is given in the Sun-centered frame by
\bea
\de \nu^z &=& 
\de \nu^{{\rm Sun},Z}
- \fr{\De m_F}{2\pi} 
\sum_{\f d} 
\Big[T^{(d)ZZ}_\f \be_\oplus \sin{\et} \cos{\Om_\oplus T}
\nonumber \\
&&
+ T^{(d)ZY}_\f 
(\be_\oplus \cos{\et} \cos{\Om_\oplus T}
- \be_L \cos{\om_\oplus T_\oplus})
\nonumber \\
&&
+ T^{(d)ZX}_\f
(\be_L \sin{\om_\oplus T_\oplus}
- \be_\oplus\sin{\Om_\oplus T})
\Big],
\nonumber\\
\eea
where $\Om_\oplus\simeq 2\pi/(365.26 {\rm ~d})$ 
is the Earth orbital frequency, 
$\et\simeq23.5^\circ$ is the Earth's orbital tilt,
and $\ch$ is the colatitude of the experiment. 

As an illustration of the expected sensitivity 
of an experiment in this configuration,
suppose a search finds no signal for the annual variations $\de\nu^z$ 
at the level of $\pm1$ mHz.
Then,
constraints of order $10^{-23}$ GeV 
would be implied on $T^{(d)JK}_{\f}$.
For minimal coefficients,
this corresponds to limits of order $10^{-23}$ GeV on $H^{TX}$ and $H^{TY}$ 
in the electron and proton sectors,
limits of order $10^{-19}$ on 
$g^{XYZ}$, $g^{YXX}$, $g^{XYY}$, $g^{XTT}$ and $g^{YTT}$ 
in the electron sector,
and limits of order $10^{-23}$ 
on the corresponding coefficients in the proton sector.
For the nonminimal sector,
the disparity between the electron and proton masses 
implies that the experiment is more sensitive to proton coefficients.
For example,
limits on the $d=8$ coefficients proportional to $m_\f^5$
would be about $10^{-23} {\rm ~GeV}^{-4}$ in the proton sector 
and about $10^{-6} {\rm ~GeV}^{-4}$ in the electron sector.

For other orientations of the magnetic field,
one readily isolated signal of Lorentz violation 
associated with the boost correction 
is the twice-sidereal variation of the frequency,
which decouples from other variations.  
We can express this term as
\bea
\de \nu_{2\om_\oplus}&=&
\fr 1{4\pi}\be_L \De m_F\sin{\chM}
\nonumber \\
&&
\times\sum_{\f d} \Big[
\cos 2\om_\oplus T_\oplus
(T^{(d)XY}_\f+ T^{(d)YX}_\f)
\nonumber\\
&&
\hskip 25pt
+\sin 2\om_\oplus T_\oplus
(T^{(d)YY}_\f- T^{(d)XX}_\f)
\Big],
\qquad
\label{sur}
\eea
where $\chM$ is the angle between the magnetic field
and the Earth's rotation axis.
For simplicity,
the direction of the magnetic field at $T_\oplus=0$
is taken to lie in the $XZ$ plane in the Sun-centered frame,
which eliminates a phase shift.  
Assuming an experimental search establishes no signal 
for the second-harmonic sidereal variations $\de \nu_{2\om_\oplus}$
at the level of $\pm1$ mHz,
then constraints of order $10^{-21}$ GeV 
would be implied for $T^{(d)JK}_{\f}$.  
For minimal coefficients,
this corresponds to sensitivities of order $10^{-18}$ 
to $g^{ZXY}$, $g^{YZX}$, $g^{ZXX}$ and $g^{ZYY}$ 
in the electron sector 
and of order $10^{-21}$ for the same coefficients in the proton sector.

\subsubsection{Space-based experiments}
\label{Space-based experiments}

Laboratory measurements of boost effects provide no control over 
the orbital and rotational motion of the Earth. 
As a result,
space-based experiments offer broader options 
for studies of the full range of possible boost effects
due to the choice and variability of orbital and rotational motions 
for various space platforms.
Here,
we consider some prospects for measurements 
with a space-based hydrogen maser.
For example,
the Atomic Clock Ensemble in Space (ACES) mission
\cite{aces}
incorporates a hydrogen maser in the payload to be delivered
and operated on the International Space Station (ISS).
As discussed in Sec.\ \ref{Frequency shift},
conventional maser transitions with $\De m_F=0$
provide no leading-order sensitivity to effects from Lorentz violation.
We therefore assume the maser is configured instead to achieve
sensitivity to transitions with $\De m_F\ne 0$,
perhaps using a double-resonance technique
\cite{an68}
similar to that already successfully implemented in the laboratory 
for tests of Lorentz and CPT invariance 
\cite{hu00,maser,hu03}.

In the context of the minimal SME,
specifics for analyzing data from space missions
studying Lorentz violation 
are discussed in Ref.\ \cite{space}.
In the presence of nonminimal operators,
the expected experimental signals have the same generic behavior 
because they too are governed by the form \rf{Sunvec}.
As a result,
using the information in Table \ref{Amm} 
permits the measurement of nonminimal coefficients as well.
Satellite experiments offer a particular advantage for this purpose
because the boost of the space platform
differs from the boost of laboratory experiments on the Earth. 
More explicitly,
consider the term $R^{zJ}\bevec^K T^{JK}$ in Eq.\ \rf{Sunvec}.  
The rotation matrix can be altered 
by changing the orientation of the magnetic field
in both Earth- and space-based experiments,
producing sensitivity to the combinations $\bevec^K T^{JK}$.  
However,
space-based experiments can vary the boost $\bevec^K$ 
more broadly as well, 
which offers the potential to disentangle
more components of $T^{JK}$.  

For space-based experiments,
the frequency shift takes the same form as Eq.\ \rf{hfboost}
but with modified expressions 
for the rotation matrix and boost parameter.
For simplicity,
suppose the applied magnetic field is oriented 
parallel to the direction of propagation of the satellite 
relative to the center of mass of the Earth,
and approximate the orbit as circular.
This configuration could in principle be realized
in an experiment on the ISS, 
for example. 
We can then view the instantaneous components $R_{zK}$ of the rotation matrix
as forming a unitary vector
parallel to the satellite velocity $\bevec_s$. 
It follows that $R^{zJ}\bevec^J_s=\be_s$,
where $\be_s$ is the average satellite speed,
and also that $\ep_{JKL}R^{zJ}\bevec^K_s=0$.
Comparing these results with Eq.\ \rf{ired}
reveals that only the symmetric piece of $T^{JK}$ varies
with the direction of the boost relative to the Earth.
Note that other orientations of the magnetic field
relative to the direction of motion would introduce variations 
involving the trace and antisymmetric pieces of $T^{JK}$.

For the parallel configuration,
an explicit calculation reveals 
that the term in the frequency that depends on $\bevec_s$
varies only at the second harmonic $2\om_s$
of the mean satellite orbital angular frequency $\om_s$.
The form of this term is
\beq
\de \nu_{2\om_s}=
\fr{\De m_F}{16\pi}
\sum_{\f d} \be_s 
({A_s}_\f^{(d)} \sin{2\om_s T_s}+{A_c}_\f^{(d)} \cos{2\om_s T_s}),
\label{twooms}
\eeq
where $T_s$ is a reference time in the Sun-centered frame 
chosen such that $T_s=0$ 
when the satellite crosses the equatorial plane on an ascending orbit. 
The amplitudes $A_s$ and $A_c$ are given by
\bea
{A_s}_\f^{(d)}&=&
4\cos\ze \sin 2\al (T^{(d)XX}_\f-T^{(d)YY}_\f)
\nonumber \\
&&
-8 \cos\ze \cos 2\al ~T^{(d)(XY)}_\f 
\nonumber \\
&&
-8\sin\ze \cos \al ~T^{(d)(ZX)}_\f
\nonumber \\
&&
+8\sin\ze \sin \al ~T^{(d)(YZ)}_\f,
\nonumber \\
{A_c}^{(d)}_\f&=&
-2 \sin^2\ze
(T^{(d)XX}_\f+T^{(d)YY}_\f-2 T^{(d)ZZ}_\f)
\nonumber \\
&&
-2 (3+\cos 2\ze) \sin 2\al ~T^{(d)(XY)}_\f
\nonumber \\
&&
-(3+\cos 2\ze) \cos 2\al (T^{(d)XX}_\f-T^{(d)YY}_\f) 
\nonumber \\
&&
+4 \sin 2\ze \cos\al ~T^{(d)(YZ)}_\f
\nonumber \\
&&
-4 \sin 2\ze \sin\al ~T^{(d)(XZ)}_\f,
\eea
where $\ze$ is the angle between 
the satellite orbital axis and the Earth's rotation axis,
and $\al$ is the azimuthal angle 
between the satellite's orbital plane and the $X$ axis
in the Sun-centered frame.

Direct inspection of the result \rf{twooms} for the satellite experiment
demonstrates that distinct combinations of coefficients appear
relative to the frequency shift \rf{sur} involving the second harmonic
of the sidereal frequency for an Earth-based experiment.
Note that for a ground-based experiment at colatitude $\ch$
the frequency is proportional to $\be_L \sin\ch$,
which is the tangential speed of the laboratory
relative to the Earth's axis of rotation. 
This speed is an order of magnitude smaller
than the speed $\be_s$ of a satellite such as the ISS relative to the Earth.
This shows that an experiment realized on a space platform
is more sensitive to this type of variation
as well as offering access to more coefficient components
that a ground-based counterpart.

\subsection{$nS_{1/2}$-$n'S_{1/2}$ and $nS_{1/2}$-$n'P_{1/2}$ transitions}
\label{The nSnP transitions}

We next turn attention 
to the effects of Lorentz and CPT violation
on high-precision studies 
of the hydrogen transitions with $J=1/2$ and $\De J=0$,
and in particular the transitions
$nS_{1/2}$-$n'S_{1/2}$ and $nS_{1/2}$-$n'P_{1/2}$.
The most prominent of these is perhaps the $1S$-$2S$ transition,
which has recently been measured to a relative uncertainty 
of $4.2 \times10^{-15}$ 
\cite{1s2s}.  
Other transitions of this type 
that are measured to high precision include
\cite{codata} 
the classical $2S_{1/2}$-$2P_{1/2}$ Lamb shift 
\cite{lamb}, 
the $1S_{1/2}$-$3S_{1/2}$ transition 
\cite{1s3s}, 
several $2S_{1/2}$-$n S_{1/2}$ transitions 
\cite{we95,de97,bo96},
and the $2S_{1/2}$-$4 P_{1/2}$ transition 
\cite{be95}. 
In this subsection,
we first present a general expression for the frequency shifts
due to Lorentz and CPT violation.
We then outlining the extraction of constraints 
by matching theoretical expectations to experimental results
and by studying sidereal and annual variations involving boosts.

\subsubsection{Frequency shift}
\label{Frequency shift two}

In searching for most effects of Lorentz and CPT violation,
the absolute sensitivity of an experiment 
is of more significance than its relative precision
because all nonrelativistic coefficients 
for Lorentz and CPT violation carry mass dimensions.
It is therefore reasonable to neglect 
the contribution from the spin-dependent coefficients to Eq.\ \rf{1Sf} 
for any of the hydrogen transitions of interest here,
as the attainable absolute sensitivity is significantly below 
that accessible to the hyperfine Zeeman transitions.
For example,
the long lifetime of the $2S_{1/2}$ state and the 
impressive relative precision achieved on the $1S$-$2S$ transition
\cite{1s2s}
yields the lowest absolute uncertainty of about 10 Hz
among the optical transitions in hydrogen,
but this remains four or more orders of magnitude
below the absolute sensitivity reached in hyperfine measurements. 
Studies of optical transitions 
involving variations with sidereal time and colatitude
at zeroth order in the boost
are therefore of lesser interest.

In contrast,
the $nS_{1/2}$-$n'S_{1/2}$ and $nS_{1/2}$-$n'P_{1/2}$ transitions
offer sensitivity to isotropic coefficients for Lorentz and CPT violation
that cannot be acessed via Zeeman hyperfine transitions.
For example,
Table \ref{rest} reveals that only coefficients with $j=0$ and $j=1$ 
contribute to the $1S$-$2S$ transition.
Effects involving the coefficients with $j=1$ can be neglected as above,
but those involving the isotropic components with $j=0$
are of definite interest.
We therefore proceed in this subsection
under the assumption that any transition with $\De J=0$ and $J=1/2$ 
is sensitive only to isotropic coefficients in the laboratory frame.

Within this scenario,
we find that in the laboratory frame the frequency shift 
of any hydrogen transition $n,L$-$n',L'$ with $J=1/2$, $\De J=0$
due to Lorentz and CPT violation can be written as
\bea
2\pi \de \nu &=&
2\mr (\ve_{n}-\ve_{n'}) 
\sum_\f 
(\cnrfc{\f,2}-\anrfc{\f,2})
\nonumber \\
&&
\hskip -30pt
- 4 \mr^2
\Bigg[
\ve_{n}^2
\left(\fr{8n}{2L+1} - 3 \right)
-\ve_{n'}^2 
\left(\fr{8n'}{2L'+1} - 3 \right)
\Bigg] 
\nonumber \\
&&
\hskip 70pt
\times 
\sum_\f 
(\cnrfc{\f,4}-\anrfc{\f,4}),
\label{Jhalf}
\eea
where $\ve_{n}\equiv -\al^2 \mr/2n^2$. 
Note that the quantities $\Vnrf{\f}{k00}$ contain only isotropic coefficients,
all of which are absent in the analogous expression \rf{1Sf}
for the frequency shift of the hyperfine Zeeman levels. 
Also,
only contributions from coefficients with $k\geq 2$ occur,
a result consistent with previous conclusions that minimal coefficients 
have no effect on the $1S$-$2S$ transition at leading order 
in Lorentz and CPT violation
\cite{bkr,yo12}.
We remark in passing that contributions to the $1S$-$2S$ transition 
are known to appear when higher-order corrections 
in minimal coefficients are included
\cite{ba10,yo12}.

\subsubsection{Self-consistent analysis}
\label{Self-consistent analysis}

At leading-order in $\be_\oplus$,
the transformation of isotropic coefficients 
from the laboratory frame to the Sun-centered frame
is the identity map,
$\K_{k00}^{\rm lab}\rightarrow \K_{k00}^{\rm Sun}$.
The expression \rf{Jhalf} for the frequency shift 
of any hydrogen transition $n,L\rightarrow n',L'$ with $J=1/2$, $\De J=0$
therefore also holds in the Sun-centered frame.
At zeroth order in the boost,
the result represents a constant shift in the transition frequency.  
However,
a tiny constant frequency shift 
is challenging to measure experimentally.

One approach to studying the shift \rf{Jhalf}
is to compare the experimental data
to the theoretical prediction for conventional Lorentz-invariant physics. 
To date,
the available experimental data all appear consistent
with theoretical expectations within the 10 Hz absolute uncertainty. 
However,
making a definitive theoretical prediction
requires knowledge of constants
such as the Rydberg constant and the proton radius,
which at present are also determined via hydrogen spectroscopy.
For example,
the contribution due to the coefficients 
$\Vnrf{\f}{200}$ acts to produce a shift $\de R_\infty$
in the Rydberg constant,
given by
\beq
\de R_\infty = 
\fr{4\pi\mr^2}{m_e}
R_\infty \sum_\f
(\cnrfc{\f,2}-\anrfc{\f,2}).
\label{rydshi}
\eeq
Analogously,
the contribution due to the coefficients $\Vnrf{\f}{400}$ 
produces a change $\de\nu_{\rm Lamb}$ in the classical Lamb shift,
given by
\beq
2\pi \de\nu_{\rm Lamb} = 
-\frac 23 (\al\mr)^4 
\sum_\f 
(\cnrfc{\f,4}-\anrfc{\f,4}),
\label{lamshi}
\eeq
which could change the proton radius 
determined by hydrogen spectroscopy.   
This presents a self-consistency issue
for direct comparison of experiment with theory,
as the theoretical prediction based on Lorentz-invariant physics
cannot be immediately disentangled 
from Lorentz-violating effects on the hydrogen spectrum. 

Techniques to avoid this issue are possible,
at least in principle.
One option could be to measure 
the coefficients for Lorentz and CPT violation 
by comparing several transitions.
For example,
a best fit to the shift \rf{Jhalf} could be performed.
A related option is to perform a careful self-consistent comparison.  
In practice,
the present limiting absolute uncertainty of order 10 kHz
on the various transitions
is likely to lead to maximal attainable sensitivities
of about $10^{-7}$ GeV$^{-1}$ on coefficients with $k=2$ 
and of about $10^{5}$ GeV$^{-3}$ on ones with $k=4$.
Performing an analysis of this type 
remains an interesting open possibility
to set first or improved constraints on several coefficients.
Moreover,
the efforts underway to improve 
the data from hydrogen spectroscopy 
with an eye to a more precise determination 
of the Rydberg constant and the proton radius 
\cite{be13,be13a,pe13,ha94,sc99}
offer the potential for substantially improved 
future sensitivities on Lorentz and CPT violation.

\subsubsection{Sidereal and annual variations due to boost corrections}
\label{Sidereal and annual variations due to boost corrections}

As described in Sec.\ \ref{Annual variations and parity-odd operators},
isotropic terms in the laboratory frame 
can be used to study anisotropies in the Sun-centered frame
by incorporating boost corrections in the analysis.
Consider,
for example,
the $1S$-$2S$ transition.  
Using the expression \rf{Suniso},
the frequency shift $\de \nu$ due to Lorentz and CPT violation
in the laboratory frame
can be converted to the Sun-centered frame,
giving
\beq
2\pi\de \nu =
2\pi \de\nu^{\rm Sun} 
+ \sum_{\f d} V^{(d)J}_{\f} (\bevec_\oplus + \bevec_L)^J.
\label{1s2svar}
\eeq
The first term on the right-hand side 
is the constant shift discussed 
in Sec.\ \ref{Self-consistent analysis}.
The second term is suppressed by boost factors
but offers interesting prospects for measuring
anisotropic coefficients in the Sun-centered frame.
Analogous results for other transitions
can also be obtained.

The boost factors in Eq.\ \rf{1s2svar} 
generate time variations in the $1S$-$2S$ frequency. 
The dependence of $\de\nu$ on the Earth's velocity $\bevec_\oplus$
introduces annual variations given by 
\bea
V^{(d)J}_\f \bevec_\oplus^J &=& 
\be_\oplus \big[
\sin\Om_\oplus T~V^{(d)X}_{\f}
\nonumber\\
&&
- \cos\Om_\oplus T 
(\cos\et ~V^{(d)Y}_{\f} + \sin\et ~V^{(d)Z}_{\f} )
\big],
\qquad
\label{1s2sann}
\eea 
where $\Om_\oplus$ is the Earth orbital frequency.  
The dependence on the laboratory velocity $\bevec_L$
produces sidereal variations,
given by
\beq
\V^{(d)J}_{\f} \bevec_L^J =
\be_L 
( \cos\om_\oplus T_\oplus ~V^{(d)Y}_\f
- \sin\om_\oplus T _\oplus ~V^{(d)X}_\f), 
\label{1s2ssid}
\eeq
where $\om_\oplus$ is the Earth sidereal frequency.  

\renewcommand\arraystretch{1.7}
\begin{table}
\caption{Values of the vectors $V^{(d)J}_\f$ 
for the electron and proton in atomic hydrogen
with $5\leq d \leq 8$.}
\setlength{\tabcolsep}{7 pt}
\begin{tabular}{cc}
\hline
\hline
$d$ & $V^{(d)J}_\f$ 
\\
\hline
5 & $\frac 34 (\al \mr)^2 
(2{a_\f}^{TTJ}_{\rm eff}+{a_\f}^{KKJ}_{\rm eff})$ \\
6 & $-3(\al \mr)^2 m_\f 
({c_\f}^{TTTJ}_{\rm eff}+{c_\f}^{TKKJ}_{\rm eff})$ \\
7 & $\frac 52 (\al \mr)^2 m_\f^2 
(2{a_\f}^{TTTTJ}_{\rm eff}+3{a_\f}^{TTKKJ}_{\rm eff})$ \\
& $+\frac{67}{16} (\al \mr)^4
{a_\f}^{KKLLJ}_{\rm eff}$ \\
8 & $-\frac {15}2 (\al \mr)^2 m_\f^3 
({c_\f}^{TTTTTJ}_{\rm eff}+2{c_\f}^{TTTKKJ}_{\rm eff})$ \\
& $-\frac{201}{8} (\al \mr)^4 m_\f
{c_\f}^{TKKLLJ}_{\rm eff}$ \\
\hline\hline
\end{tabular}
\label{VJ}
\end{table}

\renewcommand\arraystretch{1.5}
\begin{table}
\setlength{\tabcolsep}{7pt}
\caption{
Sensitivities to the absolute value of 
nonminimal cartesian coefficients for $5\leq d \leq 8$.}
\begin{tabular}{lcll}
\hline
\hline
Coefficient& $J$ & Electron & Proton \\
$\mathcal{K}^{(d)\nu\mu_1...\mu_{d-3}}_{\rm eff}$ &  & $\left(\text{GeV}^{4-d}\right)$ &  $\left(\text{GeV}^{4-d}\right)$
\\\hline
$a^{(5)TTJ}_{\rm eff}$ & $X$& $< 3.4\times 10^{-8}$ & $< 3.4\times 10^{-8}$ \\
                                            & $Y$& $< 5.6\times 10^{-8}$ & $< 5.6\times 10^{-8}$\\
                                            & $Z$& $< 1.3\times 10^{-7}$ & $< 1.3\times 10^{-7}$\\
$a^{(5)KKJ}_{\rm eff}$ & $X$ & $< 6.7\times 10^{-8}$ & $< 6.7\times 10^{-8}$ \\
                                            & $Y$& $< 1.1\times 10^{-7}$ & $< 1.1\times 10^{-7}$\\
                                            & $Z$& $< 2.5\times 10^{-7}$ & $< 2.5\times 10^{-7}$\\
$c^{(6)TTTJ}_{\rm eff}$ & $X$ & $< 3.3\times 10^{-5}$ & $< 1.8\times 10^{-8}$ \\
                                            & $Y$& $< 5.5\times 10^{-5}$ & $< 3.0\times 10^{-8}$\\
                                            & $Z$& $< 1.3\times 10^{-4}$ & $< 6.9\times 10^{-8}$\\
$c^{(6)TKKJ}_{\rm eff}$ & $X$ & $< 3.3\times 10^{-5}$ & $< 1.8\times 10^{-8}$ \\
                                            & $Y$& $< 5.5\times 10^{-5}$ & $< 3.0\times 10^{-8}$\\
                                            & $Z$& $< 1.3\times 10^{-4}$ & $< 6.9\times 10^{-8}$\\
$a^{(7)TTTTJ}_{\rm eff}$ & $X$ & $<  3.9\times 10^{-2}$ & $< 1.1\times 10^{-8}$ \\
                                            & $Y$& $< 6.5\times 10^{-2}$ & $< 1.9\times 10^{-8}$\\
                                            & $Z$& $< 0.15$ & $< 4.4\times 10^{-8}$\\
$a^{(7)TTKKJ}_{\rm eff}$ & $X$ & $<  2.6\times 10^{-2}$ & $< 7.6\times 10^{-9}$ \\
                                            & $Y$& $< 4.3\times 10^{-2}$ & $< 1.3\times 10^{-8}$\\
                                            & $Z$& $< 0.99$ & $< 2.9\times 10^{-8}$\\
$a^{(7)KKLLJ}_{\rm eff}$ & $X$ & $< 9.2\times 10^{2}$ & $< 9.2\times 10^{2}$ \\
                                            & $Y$& $< 1.5\times 10^{3}$ & $< 1.5\times 10^{3}$\\
                                            & $Z$& $< 3.5\times 10^{3}$ & $< 3.5\times 10^{3}$\\
$c^{(8)TTTTTJ}_{\rm eff}$ & $X$ & $< 84$ & $< 1.4\times 10^{-8}$ \\
                                            & $Y$& $< 140$ & $< 2.3\times 10^{-8}$\\
                                            & $Z$& $<  320$ & $< 5.2\times 10^{-8}$\\
$c^{(8)TTTKKJ}_{\rm eff}$ & $X$ & $< 42$ & $< 6.8\times 10^{-9}$ \\
                                            & $Y$& $< 70$ & $< 1.1\times 10^{-8}$\\
                                            & $Z$& $< 160$ & $< 2.6\times 10^{-8}$\\
$c^{(8)TKKLLJ}_{\rm eff}$ & $X$ & $< 1.7\times10^{5}$ & $< 92$ \\
                                            & $Y$& $< 2.8\times 10^{5}$ & $< 150$\\
                                            & $Z$& $< 6.5\times 10^{5}$ & $< 360$\\
\hline\hline
\end{tabular}
\label{cartcoeff}
\end{table}

Table \ref{VJ} provides explicit expressions 
for $V^{(d)J}_\f$ with $5\leq d\leq 8$ 
in terms of 
the rest masses $m_\f$ of the particles of flavor $\f=e$ and $\f=p$,
the fine-structure constant $\al$,
the reduced mass $\mr$ of the system,
and the effective cartesian coefficients for Lorentz violation
defined in Eq.\ (27) of Ref.\ \cite{km13}.
Only leading-order contributions from each coefficient are included.
Note that
all the spin-independent minimal coefficients 
leave unaffected the $1S$-$2S$ frequency
at leading order in the nonrelativistic limit
\cite{bkr,yo12},
so both $V_\f^{(3)J}$ and $V_\f^{(4)J}$ vanish at this order.

Studies at subleading nonrelativistic order are also of interest.
For example,
an experimental search for annual variations 
of the $1S$-$2S$ transition frequency
has been used to measure the coefficients $c_e^{(TJ)}$ in the minimal SME 
to parts in $10^{11}$
\cite{ma13}.
We can reinterpret the results in terms of the  
nonrelativistic coefficients and thereby extract
first measurements of a variety of nonminimal coefficients.
At this order,
the restriction of $V_\f^{(4)J}$ to 
the coefficients $c_e^{(TJ)}$ in the electron sector
of the minimal SME gives
\beq
\sum_{\f d}V_\f^{(d)J} = \frac 54 \al^2 m_e c_e^{(TJ)}.
\label{ma13match}
\eeq
Adopting this relation,
the results in Eq.\ (4) of Ref.\ \cite{ma13} generalize to
\beq
\sum_{\f d}
V_\f^{(d)X} = 
-(5.3\pm 3.2) \times 10^{-19} {\rm ~GeV}
\label{Xcon}
\eeq
and
\beq
\sum_{\f d} \left(2.3 V_\f^{(d)Y} + V_\f^{(d)Z}\right) =
-(1.1\pm 2.3)\times 10^{-18} {\rm ~GeV}.
\label{Ycon}
\eeq
We can now use the results in Table \ref{VJ} 
to extract attained sensitivities to nonminimal cartesian coefficients
in the electron and proton sectors.
Table \ref{cartcoeff}
displays the resulting sensitivities
to the absolute values of cartesian $a$- and $c$-type coefficients
for $5 \leq d \leq 8$. 
As before,
we adopt the standard assumption 
that only one coefficient is nonzero at a time.
The first column of this table lists the cartesian coefficient
and the second column its component.
The third and fourth columns contain the resulting constraints
in the electron and proton sectors,
respectively.

In contrast to tests using annual variations,
sidereal-variation studies of the $1S$-$2S$ transition 
remain unexplored to date.
While this type of experiment is expected 
to be about two order of magnitude less sensitive 
to the vectors $V^{(d)J}_\f$,
different combinations of coefficients for Lorentz and CPT violation
are involved.
Pursuing this possibility remains an interesting open avenue
for future research.

\subsection{$nS_{1/2}$-$n'P_{3/2}$ and  $nS_{1/2}$-$n'D$ transitions}
\label{The nSnD transitions}

The interest in improving the experimental values
of the Rydberg constant and the proton radius
has spurred the development of high-precision spectroscopy
with atomic hydrogen.
Experiments have measured or plan to study the transitions 
$2S_{1/2}$-$nP_{3/2}$ 
\cite{ha94,be95,be13,be13a},
$1S_{1/2}$-$3D$ 
\cite{pe13},
and 
$2S_{1/2}$-$nD$ 
\cite{we95,bo96,de97,sc99}.
The absolute uncertainties achieved for the
corresponding frequencies are typically in the 10 kHz range,
reaching values as low as about 1 kHz in some cases
\cite{codata}.

In the context of searching for Lorentz and CPT violation,
the sensitivities of these measurements 
to the nonrelativistic spherical coefficients with $j=0$ and $j=1$
are weaker than those 
from hyperfine Zeeman and $1S$-$2S$ transitions. 
However,
a glance at Table \ref{rest} reveals that
the involvement in a transition 
of a level with $J\geq 3/2$ or $F\geq 3/2$ 
means that nonrelativistic spherical coefficients with $j\ge2$
can be measured.
For example,
a transition to a state $nD$ with $F=3$
could be sensitive to all the coefficients with $k\leq 4$ 
contributing to the matrix element \rf{mel}.
The $nS_{1/2}$-$n'P_{3/2}$ and  $nS_{1/2}$-$n'D$ transitions
therefore offer excellent prospects
for studying certain effects from Lorentz and CPT violation
that otherwise are difficult to observe. 

The Lorentz-violating perturbative corrections 
to a specific frequency of interest
depend on the particular details of the experiment.  
For example,
the magnitudes of applied fields and the nature of the measurement
need to be considered to obtain expressions for the corrections.
For an experiment sensitive to the hyperfine structure
with the hyperfine energy dominating all perturbations,
the Lorentz-violating corrections may be obtained
from the matrix elements \rf{mel}.
Comparatively simple expressions can be obtained in some cases,
as illustrated for a weak applied magnetic field
in Sec.\ \ref{Weak magnetic field}.
In this scenario,
the signals for Lorentz and CPT violation are similar to those discussed
in previous sections of this work,
including sidereal and annual variations of the measured frequency.

For definiteness and simplicity,
we limit attention here to the scenario with a weak applied magnetic field.
The analysis of experiments with more involved configurations,
which can often include large applied Zeeman or Stark fields,
is of substantial interest but lies outside our present scope.
Nonetheless,
the discussion here demonstrates the potential
for discovery in these types of experiments
and serves to motivate future investigations 
of Lorentz- and CPT-violating signals
using other experimental configurations.

High-precision spectroscopy of atomic hydrogen
typically concerns transitions involving the $1S$ ground state
or the metastable $2S$ state.
The Lorentz- and CPT-violating corrections to these levels 
have been discussed in previous subsections,
so the discussion here focuses on the energy corrections
to the  $nP_{3/2}$ and $nD$ states.
The nonrelativistic spherical coefficients that can contribute
to the corrections are displayed in Table \ref{rest}.
In the Sun-centered frame and at zeroth order in the boost,
the explicit form of the energy corrections 
in the presence of a weak magnetic field
is given by Eq.\ \rf{rotgen}.
The weights $\AM_{jm}^{\rm Sun}(nFJL)$
are specified by the result \rf{Ajm}
with the factors $\Agen$ given for $J\leq 5/2$ in Table \ref{LaCoeff},
while the expectation values $\vev{\pmag^k}_{nL}$ 
are provided in Eq.\ \rf{radialExp}.  

The expression \rf{rotgen} displays the sidereal variations
in the energy shifts.
Any nonrelativistic spherical coefficient $\K_{kjm}^{\nr}$
contributing to the shift
introduces oscillations at the $m$th harmonic
of the sidereal frequency $\om_\oplus$.
The allowed harmonics are determined by $J$ and $F$,
as described in Sec.\ \ref{Sidereal variations}.
As an explicit example,
consider a transition to a state $nD_{5/2}^{F=2}$.  
Table \ref{rest} shows that contributions arise 
from spin-independent terms with $j=0,2,4$
and spin-dependent ones with $j=1,3$.
We therefore can expect variations
up to the fourth harmonic of the sidereal frequency.  
The first harmonic receives contributions 
from coefficients with $1\leq j\leq 4$, 
the second from $2\leq j\leq 4$, 
the third from $3\leq j\leq 4$,
and the fourth only from $j=4$.  
Table \ref{rest} also shows the relation 
between the $k$ and $j$ indices.
For example,
only coefficients with $k\geq 4$ can contribute to the fourth harmonic.  
 
\renewcommand\arraystretch{1.5}
\begin{table}[t]
\caption{Potential sensitivities to the moduli 
of the real and imaginary parts 
of electron and proton nonrelativistic coefficients 
from sidereal variations.}
\setlength{\tabcolsep}{6 pt}
\begin{tabular}{llll}
\hline
\hline
$\K_{kjm}^{\nr}$ & $K$ values & $nL_{J}^{F}$            & Sensitivity\\\hline
$\anr{22m}$, $\cnr{22m}$        &  $J\geq3/2$         & $2P_{3/2}^{2}$      &  $6\times 10^{-8}$ GeV$^{-1}$\\ 
$\HzBnr{23m}$, $\gzBnr{23m}$    &  $F\geq2$         & $2P_{3/2}^{2}$         & $2\times 10^{-7}$ GeV$^{-1}$\\  
$\HoBnr{23m}$, $\goBnr{23m}$    &  $F\geq2$         & $2P_{3/2}^{2}$         & $1\times 10^{-7}$ GeV$^{-1}$\\  
$\anr{42m}$, $\cnr{42m}$         &  $J\geq3/2$         & $2P_{3/2}^{2}$      &  $7\times 10^{3}$ GeV$^{-3}$\\ 
$\HzBnr{43m}$, $\gzBnr{43m}$    &  $F\geq2$         & $2P_{3/2}^{2}$         & $2\times 10^{4}$ GeV$^{-3}$\\  
$\HoBnr{43m}$, $\goBnr{43m}$    &  $F\geq2$         & $2P_{3/2}^{2}$         & $1\times 10^{4}$ GeV$^{-3}$\\ 
$\anr{44m}$, $\cnr{44m}$        &  $J\geq5/2$         & $3D_{5/2}^{3}$      &  $7\times 10^{4}$ GeV$^{-3}$\\ 
$\HzBnr{45m}$, $\gzBnr{45m}$   &  $F\geq3$         & $3D_{5/2}^{3}$         & $7\times 10^{4}$ GeV$^{-3}$\\  
$\HoBnr{45m}$, $\goBnr{45m}$   &  $F\geq3$         & $3D_{5/2}^{3}$         & $3\times 10^{4}$ GeV$^{-3}$\\ 
\hline\hline
\end{tabular}
\label{ExpSid}
\end{table}

Using this information,
we can form estimates of the potential sensitivities
to nonrelativistic coefficients
from searches for sidereal variations
in $nS_{1/2}$-$n'P_{3/2}$ and  $nS_{1/2}$-$n'D$ transitions.
Table \ref{ExpSid} provides the results obtained
under the assumption that the absolute experimental uncertainty 
for these variations is 10 kHz.
The first column of the table displays 
the relevant nonrelativistics spherical coefficients,
generically denoted as $\K_{kjm}^{\nr}$.
The second column shows the range of $K\equiv J,F$
for the relevant transitions.  
The third column presents the values of $nL_{J}^{F}$      
for the excited energy levels used in obtaining  the specific estimates.
The smallest value of $n$ producing contributions is chosen for these levels,
as an experiment with fixed absolute uncertainty
is less sensitive to coefficients with $k\ne 0$ and larger $n$.
The final column lists the potential sensitivities
to the moduli of the real and imaginary parts of the coefficients
taken one at a time,
derived with values of $m_F$ and $\ch$ 
chosen to maximize the sensitivity.
Note that the coefficients shown in Table \ref{ExpSid}
remain unmeasured in any experiments to date. 
Note also that sensitivity to both electron and proton coefficients
is achieved despite the mass difference between the particles,
which can be traced to their equal but opposite angular momenta.

\section{Antihydrogen}
\label{Antihydrogen}

The techniques developed in the previous sections
to search for Lorentz and CPT violation
using spectroscopy of atomic hydrogen
can also be applied to other hydrogenic systems.
In this section,
we turn attention to the emerging field of antihydrogen spectroscopy.
A number of collaborations have as goal the precision spectroscopy 
of antihydrogen,
including 
the Antihydrogen Laser Physics Apparatus 
(ALPHA)
collaboration
\cite{alpha11},
the Atomic Spectroscopy and Collisions Using Slow Antiprotons 
(ASACUSA) 
collaboration
\cite{asacusa15},
and the Antihydrogen Trap 
(ATRAP)
collaboration
\cite{ga02}.
Several studies of the gravitational response of antihydrogen
are under development,
including ones by 
the Antihydrogen Experiment: Gravity, Interferometry, Spectroscopy
(AEGIS)
collaboration
\cite{aegis},
the ALPHA collaboration
\cite{alphagrav},
and the Gravitational Behavior of Antihydrogen at Rest
(GBAR)
collaboration
\cite{gbar}.
A proposal for an Antimatter Gravity Experiment (AGE)
also exists 
\cite{age}.

Since CPT violation in realistic effective field theory
necessarily comes with Lorentz violation
\cite{ck,owg},
which implies the breaking of rotation and boost symmetry, 
a natural question to ask is whether
experiments with antihydrogen spectroscopy
can attain sensitivities to new physics
that is inaccessible or impractical to access with experiments 
using rotated or boosted ordinary matter.
The answer is affirmative,
as might intuitively be expected.
Indeed,
the form of Eq.\ \rf{cpt}
already reveals that coefficients for Lorentz violation
for a given species always appear in summed pairs,
one controlling CPT-odd and one CPT-even operators,
and this feature holds in the full relativistic theory as well
\cite{km13}.
As a result,
experiments with strictly nonrelativistic electrons or protons 
in any combination cannot explore the full parameter space 
for the coefficients
and hence cannot study the full range of possible physical effects.
For example,
although the individual nonrelativistic spherical coefficients
modifying the antihydrogen spectrum
are the same as those for hydrogen listed in Table \ref{rest},
the nonrelativistic spectral modifications 
involve disparate coefficient combinations
and so experiments on both are necessary to discern 
the relevant CPT-violating physics.
Situations can even be envisaged in which
no effect exists in nonrelativistic hydrogen
but a large signal occurs in antihydrogen,
such as the isotropic invisible model 
discussed in Sec.\ IX B of Ref.\ \cite{akjt},
which allows comparatively large effects 
in the antihydrogen hyperfine structure
while damping those in hydrogen.

In principle,
high-precision experiments with heavily boosted electrons and protons
offer additional options for a complete coverage of possible effects
because the combinations $\Vnrf{\f}{kjm}$ and ${\T_\f}^{\nr(qP)}_{kjm}$ 
in Eq.\ \rf{cpt}
involve coefficients of different dimensions
accompanied by distinct momentum dependences.
However, 
precision measurements involving significant boosts 
come with additional experimental challenges.
Typical analyses take advantage 
of the comparatively small boost $\sim 10^{-4}$ 
due to the Earth's orbital motion
\cite{sunframe, boost,ma13}.
In this scenario,
for example,
the dominant sensitivities to nonminimal coefficients
available to antihydrogen spectroscopy 
are enhanced by about eight orders of magnitude  
relative to those of hydrogen spectroscopy using annual variations
due to the parity selection rules 
described in Sec.\ \ref{Annual variations and parity-odd operators}.
In practice,
a comprehensive search for Lorentz and CPT violation 
therefore requires performing experiments with positrons and antiprotons 
in various combinations as well.

Among studies of Lorentz and CPT symmetry using positrons and antiprotons,
antihydrogen has distinctive sensitivity 
due to its intrinsic spherical symmetry and flavor content.
The symmetries of other experiments,
such as the cylindrical symmetry of ones 
trapping and studying individual positrons or antiprotons 
in Penning traps
\cite{de99,ga99,pttheory,base,chip},
make them sensitive to different sets of coefficients
and thus different physical effects
\cite{dk15}.
Positronium and protonium do have spherical symmetry but 
involve C-invariant particle-antiparticle combinations of only one flavor
and hence also have distinct physical sensitivities.
Moreover,
other intrinsic factors can enhance the difference between 
various types of experiments.
For example,
certain coefficients for Lorentz and CPT violation
are accompanied by factors of the particle momentum,
which is about $m_e \al \simeq 3.7$ keV for antihydrogen
but differs in other types of experiments. 
In short,
spectroscopy of antihydrogen represents a unique tool
to probe Lorentz and CPT violation,
and one that is essential 
for the definitive and unambiguous detection of CPT violation
involving the nonrelativistic spherical coefficients 
considered in this work. 

In this section,
we begin with a description of the implementation
of the CPT transformation on the hydrogen spectrum.
We then address the effects of nonminimal coefficients
on hyperfine and $1S$-$2S$ transitions.
Finally,
we offer some comments on experiments testing
the gravitational response of antihydrogen.

\subsection{Basics}
\label{hbar basics}

The form of the leading-order Lorentz- and CPT-violating perturbation
$\de h_\atmb^\nr$
to the nonrelativistic hamiltonian for free antihydrogen
is similar to that for hydrogen,
\beq 
\de h_\atmb^\nr = \de h^\nr_\eb + \de h^\nr_\pb ,
\label{pertb}
\eeq 
involving the sum of perturbative contributions 
from the positron $\eb\equiv e^+$ and the antiproton $\pb$. 
The individual perturbations
are given by expressions similar to Eqs.\ \rf{nr}-\rf{FEHSD}
for hydrogen,
\beq
\de h_\fb^\nr
= h_{\fb 0}+ h_{\fb r} \sivec\cdot\ephat_r
+ h_{\fb +} \sivec\cdot\ephat_-
+ h_{\fb -} \sivec\cdot\ephat_+,
\label{nrb}
\eeq
where $\fb$ represents either $\eb$ or $\pb$.
The spin-independent term is
\beq
h_{\fb 0} =
-\sum_{kjm} \pmag^k 
~\syjm{0}{jm}(\phat) 
\Vnrf{\fb}{\k jm},
\label{FEHSIb}
\eeq
while the spin-dependent ones are
\bea
h_{\fb r} &=&
-\sum_{kjm} \pmag^k 
~\syjm{0}{jm}(\phat) 
\TzBnrf{\fb}{kjm},
\nonumber \\
h_{\fb\pm} &=&
\sum_{kjm} \pmag^\k 
~\syjm{\pm 1}{jm}(\phat) 
\left(i\ToEnrf{\fb}{kjm} \pm \ToBnrf{\fb}{kjm}\right).
\nonumber\\
\label{FEHSDb}
\eea
In these equations,
the quantities $\Vnrf{\fb}{kjm}$ and ${\T_\fb}^{\nr(qP)}_{kjm}$
are CPT-transformed versions of those given for hydrogen in Eq.\ \rf{cpt},
\bea
\Vnrf{\fb}{kjm} &=&
\cnrf{\f}{kjm} + \anrf{\f}{kjm},
\nonumber \\
{\T_\fb}^{\nr(qP)}_{kjm} &=&
-{g_\f}^{\nr(qP)}_{kjm} - {H_\f}^{\nr(qP)}_{kjm}.
\label{cptb}
\eea 
These expressions include operators of arbitrary mass dimension $d$.
When restricted to the minimal-SME coefficients,
the above equations reduce to those used 
in the previous literature on CPT violation in antihydrogen
\cite{bkr}. 

The physical effects of Lorentz and CPT violation in antihydrogen 
are determined by the matrix elements of $\de h_\atmb^\nr$
in the unperturbed states.
The coefficient selection rules for hydrogen
presented in Sec.\ \ref{Coefficient selection rules}
are valid for antihydrogen,
and in particular the nonrelativistic spherical coefficients
contributing to modify the antihydrogen spectrum
are those listed in Table \ref{rest}.
The methods used in Sec.\ \ref{Matrix elements} 
to derive the matrix elements for hydrogen
can also be applied,
but the corrections to the antihydrogen spectrum
must be obtained by performing a CPT transformation
on the hydrogen matrix elements.
This involves both using 
the antihydrogen perturbative hamiltonian $\de h_\atmb^\nr$
and the antihydrogen states,
which are CPT transformations of the hydrogen ones.
Specifically,
the CPT counterpart of an energy state $\ket{nFJLm_F}$ in hydrogen 
is the state $\ket{nFJL(-m_F)}$ in antihydrogen,
as the net result of the CPT transformation is to replace 
the atom with the antiatom 
and to invert the direction of the total angular momentum $F$.  

To illustrate the idea,
consider the antihydrogen energy shift $\de\epb(nFJLm_F)$ 
in the presence of a weak uniform magnetic field.
For hydrogen,
the energy shift $\de \ep(n F J L m_F)$ of the Zeeman levels 
is provided by Eq.\ \rf{defe}.
For antihydrogen,
noting that the uniform magnetic field 
and the magnetic dipole moment are both invariant under CPT,
we find instead 
\bea
\de \epb(n F J L m_F) &=& 
\sum_{j} \AMb_{j0}(n F J L) \vev{F (-m_F) j 0|F (-m_F)}
\nonumber\\
&=& 
\sum_{j} (-1)^j \AMb_{j0}(n F J L) \vev{F m_F j 0|F m_F}, 
\nonumber\\
\label{defeb}
\eea
where the weights ${\AMb}_{j0}$
are given by Eq.\ \rf{Ajm} 
with the replacement $\f\to \fb$ throughout.
In the second line,
we have used the Wigner-Eckart theorem 
and the properties of the Clebsch-Gordan coefficients. 
As shown in Sec.\ \ref{Coefficient selection rules},
the weights $\AM_{jm}$ can acquire contributions
for even $j$
only from coefficients associated with spin-independent operators
and for odd $j$ 
only from coefficients associated with spin-dependent operators.
This reveals a simple relationship between
the shifts of the hydrogen and antihydrogen spectra:
given the expression for the shift in a hydrogen energy level,
the shift of the corresponding antihydrogen level
is obtained by implementing the replacements
\beq
\anrf{\f}{kjm} \to -\anrf{\f}{kjm},
\qquad
{H_\f}^{\nr(qP)}_{kjm} \to - {H_\f}^{\nr(qP)}_{kjm}.
\label{spectralmap}
\eeq 
Comparing this rule to the operator transformations 
listed in Table \ref{rest},
we infer that the antihydrogen spectral shifts
can be obtained by charge conjugation
of the hydrogen ones.
This result extends the minimal-SME result obtained in Ref.\ \cite{bkr}.

The reader is cautioned that the spectral map \rf{spectralmap}
is a formal statement of correspondence between energy levels,
which depends on the labeling of the states.
In the above example,
the spectra are described 
using the orientation of the total angular momentum
relative to the applied magnetic field,
which is a C-invariant notion. 
If instead the spectra are described 
using the orientation of the magnetic moment relative to the magnetic field, 
which is a CPT-invariant notion,
then the two spectra would be related by a CPT transformation.
Moreover,
the spectral map is distinct from observable quantities
such as frequency differences,
which in practical scenarios may depend on other factors.
For instance,
magnetically trapped states in hydrogen
have opposite values of $m_F$ 
from those in antihydrogen,
so frequency comparisons of trapped atoms and antiatoms
amount to measuring the effect of the CPT replacements
\beq
\anrf{\f}{kjm} \to -\anrf{\f}{kjm},
\qquad
{g_\f}^{\nr(qP)}_{kjm} \to - {g_\f}^{\nr(qP)}_{kjm}
\label{freqmap}
\eeq 
instead of the C replacements \rf{spectralmap}.  
This is intuitively reasonable for tests of the CPT theorem,
which specifically concerns invariance under CPT transformations
but makes no statement about invariance under C transformations.

\subsection{Hyperfine transitions}
\label{hbar Hyperfine Zeeman transitions}

The application of a comparatively weak external magnetic field 
to antihydrogen 
splits the two $1S_{1/2}$ levels 
into four distinct hyperfine Zeeman sublevels,
one with $F=0$ and three with $F=1$.
These splittings can be in principle be studied experimentally.
For example,
the ASACUSA collaboration 
plans to measure the corresponding hyperfine transitions
using an ultracold beam of antihydrogen atoms
\cite{ku14}.  
For simplicitly,
we neglect any boost effects in what follows,
and we work in the strict Zeeman or Paschen-Back regimes
so that the magnetic mixing of states in intermediate regimes
can be neglected.
Extensions of the results below to include these more general cases
are possible and may be of interest for some future applications
but lie beyond our present scope.

The Lorentz- and CPT-violating shifts in the antihydrogen energies
can be found from the expression \rf{1S} for hydrogen
by implementing the coefficient map \rf{spectralmap}.
The resulting hyperfine frequency shifts $2\pi\de\nub$ 
for transitions with a given $\De m_F$ take the form 
\bea
2\pi\de\nub &=&
-\fr{\De m_F}{2\sqrt{3\pi}}
\sum_{q=0}^2
(\al \mr)^{2q}(1+4\de_{q2})
\nonumber\\ 
&&
\hskip 30pt
\times
\sum_\f 
\big[ \gzBnrf{\f}{(2q)10}+\HzBnrf{\f}{(2q)10}
\nonumber\\ 
&&
\hskip 60pt
+2\goBnrf{\f}{(2q)10}+2\HoBnrf{\f}{(2q)10}
\big]
\qquad
\label{1Sfb}
\eea
in the laboratory frame.
In the minimal-SME limit,
the combination of nonrelativistic spherical coefficients
appearing in this expression
reduces to minimal cartesian coefficients according to
\bea
&&
\hskip -10pt
\gzBnrf{\f}{010} + 2\goBnrf{\f}{010} + \HzBnrf{\f}{010} + 2\HoBnrf{\f}{010} 
\nonumber\\
&& 
\to
2\sqrt{3\pi} [b^\f_3 + m_\f d^\f_{30} + H^\f_{12}
- m_\f g^{\f(A)}_3 + m_\f g^{\f(M)}_{120}],
\nonumber\\
\eea
where the superscripts (A) and (M)
denote the irreducible axial and irreducible mixed-symmetry
combinations of the coefficients $g^{\f}_{\ka\la\nu}$,
respectively
\cite{krt08,fittante}. 
The result \rf{1Sfb} therefore reproduces and extends
the minimal-SME expression obtained in Ref.\ \cite{bkr}
under the assumption that only the cartesian coefficients
$b^\f_\mu$, $d^\f_{\mu\nu}$, and $H^\f_{\mu\nu}$
are nonzero,
with the $g$-type coefficients set to zero 
in accordance with their expected additional suppression
due to the breaking of the electroweak SU(2)$\times$U(1) symmetry
\cite{ck}.

The laboratory-frame coefficients appearing in Eq.\ \rf{1Sfb} 
are time dependent by virtue of the rotation of the Earth
and its revolution about the Sun.
As a result,
all the signals for Lorentz and CPT violation discussed for hydrogen
in Sec.\ \ref{$1S$ hyperfine-Zeeman}
have counterparts in antihydrogen experiments.  
The measured hyperfine Zeeman frequencies in antihydrogen
can exhibit sidereal and annual time variations
and can be sensitive to the orientation of the magnetic field
and the colatitude of the laboratory. 
For example,
at zeroth boost order,
the relation between the coefficients 
in the laboratory and Sun-centered frames
is given by Eq.\ \rf{rot},
revealing that the frequency shifts $\de\nub$ 
undergo sidereal variations at the Earth's rotation frequency $\om_\oplus$.

In addition to searching for the above hyperfine Zeeman signals 
of Lorentz and CPT violation in antihydrogen alone,
interesting prospects for focusing specifically on CPT violation
are offered by direct comparisons of measurements
with hydrogen and antihydrogen.
Note that some caution is required in performing these comparisons,
as differences between hydrogen and antihydrogen
involving only CPT-even effects can appear 
unless the assorted time variations 
and orientation and colatitude dependences
are carefully incorporated in the analysis.

As an illustration of a direct comparison,
consider the hyperfine Zeeman frequency difference  
between hydrogen and antihydrogen
for transitions with $\De m_F = 1$,
which in the presence of CPT violation is given by 
\bea
2\pi\De\nu &\equiv & 2\pi\de\nu - 2\pi\de\nub
\nonumber\\
&=& -\fr{1}{\sqrt{3\pi}}
\sum_{q=0}^2
(\al \mr)^{2q}(1+4\de_{q2})
\nonumber\\ 
&&
\hskip 40pt
\times
\sum_\f 
(\gzBnrf{\f}{(2q)10}
+2\goBnrf{\f}{(2q)10}).
\qquad
\label{DeNuhf}
\eea
This expression depends only on coefficients
controlling CPT-odd operators in the perturbation hamiltonians
for hydrogen and antihydrogen.
In the minimal-SME limit,
this result reduces to
\bea
\De\nu &\to&  
-\fr 1 \pi 
\big(
b^e_3 - m_e g^{e(A)}_3 + m_e g^{e(M)}_{120}
\nonumber\\
&&
\hskip 20pt
+ b^p_3 - m_p g^{p(A)}_3 + m_p g^{p(M)}_{120}
\big),
\eea
which extends the result presented in Ref.\ \cite{bkr}
to include $g$-type coefficients.
Note that the result \rf{DeNuhf} 
is expressed in the laboratory frame
and therefore still generically depends on time.
For example,
time variations at the Earth's sidereal frequency
are given by converting the coefficients to the Sun-centered frame
using Eq.\ \rf{rot}.
This reveals that only the $g$-type components 
involving Sun-frame coefficients with $kjm = 010$, 210, and 410 
are associated with signals independent of sidereal time.

Studies of the antihydrogen spectrum in the presence
of a strong external magnetic field 
are also of experimental interest.
For example,
the ALPHA and ATRAP collaborations
plan to perform spectroscopy on antihydrogen
trapped in the Paschen-Back limit of strong fields
\cite{alpha,atrap}.  

For the Paschen-Back splitting of the $1S_{1/2}$ levels,
the total angular momentum $F$ 
is no longer a good quantum number.
Instead,
the states can be labeled by the spins 
$S_{\eb}=\pm 1/2$ and $S_{\pb}=\pm 1/2$ 
of the positron and antiproton, 
respectively.  
Incorporating perturbative Lorentz and CPT violation as before,
the hyperfine Paschen-Back frequency shifts $\de\nub$
for given $\De S_\f$ in antihydrogen are found to be
\bea
2\pi\de\nub &=&
-\fr 1{\sqrt{3\pi}}
\sum_{q=0}^2
(\al \mr)^{2q}(1+4\de_{q2})
\nonumber\\ 
&&
\hskip 20pt
\times
\sum_\f {\De S_\f} 
\big[ \gzBnrf{\f}{(2q)10}+\HzBnrf{\f}{(2q)10}
\nonumber\\ 
&&
\hskip 60pt
+2\goBnrf{\f}{(2q)10}+2\HoBnrf{\f}{(2q)10}
\big]
\qquad
\label{pasbacb}
\eea
in the laboratory frame.
This agrees with the minimal-SME result in Ref.\ \cite{bkr}
in the appropriate limit.

Converting the coefficients to the Sun-centered frame
leads to antihydrogen frequency signals 
similar to those in the Zeeman limit,
including sidereal and annual time variations
and dependences on the magnetic-field orientation
and laboratory colatitude.
A significant difference between the Zeeman shift \rf{1Sfb}
and the Paschen-Back shift \rf{pasbacb}
is the lack of sensitivity of the latter 
to coefficients of one flavor in certain transitions,
depending on the specific values of $S_{\eb}$ and $S_{\pb}$ .
For example,
the frequency difference
$\De\nu_{c\to d} \equiv \de\nu_{c\to d} - \de\nub_{c\to d}$
for the transition $\ket{c}\to\ket{d}$
essentially involves a proton spin flip
because $\ket{c}$ contains highly polarized electron and proton spins
with $m_{S_e} = 1/2$ and $m_{S_p} = -1/2$.
We find
\bea
\De\nu_{c\to d} &=&
-\fr{1}{\sqrt{3\pi}}
\sum_{q=0}^2
(\al \mr)^{2q}(1+4\de_{q2})
\nonumber\\ 
&&
\hskip 60pt
\times
(\gzBnrf{p}{(2q)10}
+2\goBnrf{p}{(2q)10}).
\qquad
\label{DeNu}
\eea
This reduces in the minimal-SME limit to
\bea
\De\nu_{c\to d} &\to&
-\fr 2 \pi 
\big(
b^p_3 - m_p {g^p}{}^{(A)}_3 + m_p {g^p}{}^{(M)}_{120} \big),
\eea
in agreement with and extending the result found in Ref.\ \cite{bkr}.

\subsection{$1S$-$2S$ and $nL_{1/2}$-$n'L'_{1/2}$ transitions}
\label{hbar 1S2S}

In searching for CPT violation in the minimal SME,
hyperfine spectroscopy of antihydrogen 
has a theoretical advantage over optical spectroscopy
because the $1S$-$2S$ transition
is insensitive to minimal-SME coefficients in free antihydrogen
and exhibits only suppressed sensitivity in magnetically trapped antihydrogen
\cite{bkr}.
Here,
we show this situation changes for nonminimal operators:
optical spectroscopy offers access to nonminimal SME coefficients
with unsuppressed sensitivity.
Moreover,
some of these coefficients are inaccessible to hyperfine spectroscopy.

The derivation of the relevant spectral shifts for free antihydrogen 
parallels the one for free hydrogen
outlined in Sec.\ \ref{The nSnP transitions}.
Restricting attention to isotropic effects as in the hydrogen case,
we find that any antihydrogen transition
$nL$-$n'L'$ with  $J=1/2$, $\De J =0$
experiences a frequency shift $\de \nub$ 
due to Lorentz and CPT violation given in the laboratory frame by
\bea
2\pi \de \nub &=&
2\mr (\ve_{n}-\ve_{n'}) 
\sum_\f 
(\cnrfc{\f,2}+\anrfc{\f,2})
\nonumber \\
&&
\hskip -30pt
- 4 \mr^2
\Bigg[
\ve_{n}^2
\left(\fr{8n}{2L+1} - 3 \right)
-\ve_{n'}^2 
\left(\fr{8n'}{2L'+1} - 3 \right)
\Bigg] 
\nonumber \\
&&
\hskip 70pt
\times 
\sum_\f 
(\cnrfc{\f,4}+\anrfc{\f,4}),
\label{Jhalfb}
\eea
where $\ve_{n}\equiv -\al^2 \mr/2n^2$. 
Note that this result contains only contributions with $k\geq 2$,
confirming the absence of unsuppressed effects 
from the minimal SME
\cite{bkr}.
In contrast,
the nonminimal coefficients carry negative mass dimensions
and so appear accompanied by powers of the relativistic energy of the states.
For frequencies,
this involves the relativistic energy difference between two levels 
and hence at leading order a nonzero contribution proportional
to powers of the particle masses.

Converting this expression to the Sun-centered frame
introduces sidereal and annual variations
along with dependences on the orientation of the magnetic field
and the laboratory colatitude.
In principle,
all the types of searches for Lorentz and CPT violation
discussed in the hydrogen context
in Secs.\ \ref{The nSnP transitions} and \ref{The nSnD transitions}
are of interest in the antihydrogen context as well.
In addition,
direct comparisons of results for hydrogen and antihydrogen
could permit the extraction of clean constraints
on coefficients for CPT violation.
This includes not only coefficients entering the $1S$-$2S$ transition
but also coefficients involved with high-$J$ levels,
thereby providing access to coefficients with large values of $j$
that are inaccessible to hyperfine spectroscopy.

As an example of a direct comparison between hydrogen and antihydrogen,
consider the isotropic coefficients 
generating frequency shifts for the $1S$-$2S$ transition. 
The shift for antihydrogen is given by
\bea
2\pi \de \nub_{1S2S} &=&
\frac 3 4 {(\al\mr)^2} 
\sum_\f 
(\cnrfc{\f,2}+\anrfc{\f,2})
\nonumber \\
&&
+ \fr{67}{16} (\al\mr)^4
\sum_\f 
(\cnrfc{\f,4}+\anrfc{\f,4}).
\label{1S2Sb}
\eea
The frequency difference
$\De \nub_{1S2S} \equiv \de \nu_{1S2S} - \de \nub_{1S2S}$
between the $1S$-$2S$ transitions in hydrogen and antihydrogen
is therefore
\beq
\De \nub_{1S2S} =
-\fr{1}{8\pi}
\sum_\f 
\big[
12 (\al\mr)^2 \anrfc{\f,2}
+ 67 (\al\mr)^4 \anrfc{\f,4}
\big].
\label{De1S2Sb}
\eeq
This permits a clean measurement of isotropic CPT-violating effects
with unsuppressed signals 
in the comparison between trapped atoms and antiatoms.
If attained,
an absolute uncertainty of 1 Hz in the $1S$-$2S$ transition
for both hydrogen and antihydrogen
would yield constraints of order 
$10^{-12}$ GeV$^{-1}$ on the coefficients $\anrfc{\f,2}$ 
and of order
$10^{-2}$ GeV$^{-3}$ on the coefficients $\anrfc{\f,4}$ 
in the electron and proton sectors.  

Note that in principle
the anisotropic coefficients in the general result \rf{Jhalfb}
also contribute to $\de \nub_{1S2S}$.
In the weak-field regime,
the spin-dependent contributions for $\De m_F = 0$ are 
\bea
2\pi \de \nub_{1S2S} &\supset&
\fr{m_F}{8}\sqrt{\fr{3}{\pi}} 
\sum_{q=1}^2 (\al \mr)^{2q}
(1+\frac {67}{12}\de_{2q})
\nonumber\\
&&
+\sum_{\w}
\Big(\gzBnrf{\w}{(2q)10}+\HzBnrf{\w}{(2q)10}
\nonumber \\
&& \hskip 30 pt
+2\goBnrf{\w}{(2q)10}+2\HoBnrf{\w}{(2q)10}\Big).
\qquad
\eea
Note these are nonvanishing only for $m_F \neq 0$.
In the strong-field regime and for $\De S_\f = 0$,
the spin-dependent contributions are
\bea
2\pi \de \nub_{1S2S} &\supset&
\sqrt{\fr{3}{16\pi}} 
\sum_{q=1}^2 (\al \mr)^{2q}
(1+\frac {67}{12}\de_{2q})
\nonumber\\
&&
+\sum_{\w} S_{\w}
\Big(\gzBnrf{\w}{(2q)10}+\HzBnrf{\w}{(2q)10}
\nonumber \\
&& \hskip 40 pt
+2\goBnrf{\w}{(2q)10}+2\HoBnrf{\w}{(2q)10}\Big).
\qquad
\eea
However,
at present the planned ASACUSA measurement,
which is sensitive to the anisotropic coefficients,
is expected to reach an absolute uncertainty of about 100 Hz or better,
while measurements of the $1S$-$2S$ transition in antihydrogen
appear unlikely to approach this benchmark in the near future.
It is therefore reasonable at present to disregard contributions 
from the anisotropic coefficients to the $1S$-$2S$ transition.
Nonetheless,
as antihydrogen is intrinsically a stable antiatom
and the natural linewidth of the $2S$ state is about 1 Hz,
the $1S$-$2S$ transition may offer 
the most interesting long-term prospects for sub-Hz sensitivities.

In parallel with the hydrogen case
discussed in Sec.\ \ref{Self-consistent analysis},
the presence of Lorentz and CPT violation can also cause apparent shifts 
in various fundamental constants measured in antihydrogen experiments.
For instance,
the apparent shift $\de \Rb_\infty$
of the Rydberg constant in antihydrogen
due to Lorentz and CPT violation is given by
\beq
\de \Rb_\infty =
\fr{4\pi\mr^2}{m_e} R_\infty 
\sum_\f 
(\cnrfc{\f,2}+\anrfc{\f,2}).
\eeq
A direct comparison of a measurement of the Rydberg constant \rf{rydshi}
performed using hydrogen with one using antihydrogen
therefore can be expected to reveal a discrepancy $\De R_\infty$
given by
\beq
\De R_\infty \equiv 
R_{\infty}-\Rb_\infty =
-\fr{8\pi\mr^2}{m_e} R_\infty \sum_\f\anrfc{\f,2}.
\eeq
This difference depends purely on CPT-odd effects.
Other fundamental constants may similarly be affected.
For example,
if future experiments can perform high-precision spectroscopy 
of the $2S$-$2P$ transition to determine the classical Lamb shift
in antihydrogen,
\beq
2\pi \de\nub_{\rm Lamb} = 
-\frac 23 (\al\mr)^4
\sum_\f 
(\cnrfc{\f,4}+\anrfc{\f,4}),
\label{lamshib}
\eeq
or of the two-photon transitions $2S$-$nD$ in antihydrogen,
then the radius of the antiproton could be determined.
Since Lorentz and CPT violation produces an apparent shift
in these transitions,
a discrepancy between the proton and antiproton radii could emerge.

\subsection{Antihydrogen and gravity}
\label{hbar gravity}

A long-standing question is
whether antiparticles and particles interact identically with gravity
\cite{ng91}.
Several experiments have been proposed to test this idea directly
using antihydrogen,
including AEGIS
\cite{aegis},
GBAR
\cite{gbar},
ALPHA
\cite{alphagrav},
and AGE
\cite{age}.
While the present work is focused on the spectroscopic effects 
of nonminimal operators for Lorentz and CPT violation in flat spacetime,
we offer in this subsection
a few comments about the role of nonminimal operators
in the gravitational couplings of antihydrogen.

A theoretical model in which the gravitational response of antihydrogen
differs from that of hydrogen is presented 
in Sec.\ IX B of Ref.\ \cite{akjt}.
The model,
called the isotropic parachute model (IPM),
is an effective quantum field theory,
constructed as a subset of the gravitationally coupled minimal SME
\cite{akgrav}.
The IPM overcomes various objections 
to theories with different antimatter and matter couplings to gravity,
demonstrating explicitly that energy can be conserved,
that the binding-energy content can be largely irrelevant 
to the gravitational response,
and that restrictions from other systems such as neutral kaons can be evaded.

In the IPM,
the anomalous gravitational response of antimatter compared to matter
is a consequence of CPT violation
and hence of Lorentz violation in the underlying effective field theory.
The IPM is an isotropic theory,
formulated in an asymptotically Minkowski spacetime
with a weak gravitational field 
and designed to produce a predominantly null effect in matter
by cancellation of CPT-even and CPT-odd effects in the minimal SME.
The physical Lorentz and CPT violation in hydrogen
is therefore countershaded from detection
\cite{countershading}.
The anomalous response in antihydrogen arises
because for antimatter the signs of the CPT-odd contributions change,
disrupting the cancellation.
Explicitly,
in the IPM
the uniform constant background pieces $\ab^\f_T$ and $\cb^\f_{TT}$ 
of the isotropic minimal-SME cartesian coefficients $a^\f_T$ and $c^\f_{TT}$ 
are related by
\beq
\al \ab^\f_T = \frac 13 m \cb^\f_{TT},
\label{lfccondition}
\eeq
for particles of flavor $\f$,
where $\al$ is a model-dependent quantity
determined by the gravitational coupling for Lorentz-violating effects. 

In the context of hydrogen and antihydrogen,
the conditions \rf{lfccondition} for $\f=e$ and $\f=p$
represent two constraints on four independent coefficients,
so the IPM is a two-parameter model.
The resulting inertial and gravitational masses
of hydrogen are equal while those of antihydrogen differ,
\beq
m_i^H = m_g^H,
\quad
m_i^\Hb \neq m_g^\Hb.
\eeq
However,
the strength of the anomalous gravitational response of antimatter
in the IPM
has recently been constrained to parts in $10^7$ 
by an analysis combining data from 
torsion-balance tests, matter-wave interferometry, 
and microwave, optical, and M\"ossbauer clock-comparison experiments,
and by taking advantage of the differing bound kinetic energies of nuclei
\cite{hcpm11}.
Any IPM effects in antihydrogen are therefore beyond
the reach of the currently proposed antihydrogen experiments. 

The IPM uses only isotropic minimal operators in the SME.
However,
the gravitational sector of the SME includes
not only minimal pure-gravity and matter-gravity couplings,
but also nonminimal couplings
\cite{akgrav}
that have definite experimental signatures
\cite{gravnonmin}. 
These nonminimal couplings are also of potential relevance
in the present context,
and in particular we expect them to enhance substantially the prospects 
for a strong anomalous gravitational response of antihydrogen.
A detailed study of nonminimal effects in this context
is challenging and lies well outside our present scope,
although it is likely to offer interesting insights.
Nonetheless,
we can provide some intuition 
by following the conceptual path presented in Ref.\ \cite{akjt}
in the special limit where only isotropic nonminimal coefficients 
in the matter-gravity sector contribute,
keeping only zeroth-order nonrelativistic effects
and first-order gravitational couplings.

In this comparatively simple limit,
starting with the generalized Dirac equation 
incorporating both operators for Lorentz and CPT violation 
of arbitrary mass dimension and gravitational couplings,
the corresponding perturbative hamiltonian
contains no momentum-dependent Lorentz violation
and the Lorentz-violating energy dependence
involves only the particle mass.
The calculation therefore proceeds with 
the minimal-SME background coefficients 
$\ab^\f_T$ and $\cb^\f_{TT}$ 
now accompanied by a series of terms
involving nonminimal background coefficients and the particle mass.
Since these quantities are all constants,
the derivation has the same algebraic structure 
as that presented in Ref.\ \cite{akjt},
up to possible numerical factors 
due to the increased multiplicity of indices on nonminimal coefficients.

Noting that only the nonminimal isotropic coefficients with $k=0$
contain the minimal-SME isotropic coefficients
and using Eqs.\ (93), (111), (129), and (130) of Ref.\ \cite{km13},
we can deduce that the net result of the calculation 
involves the replacements
\bea
\ab^\f_T &\to &
\ring{\ab}{}^{\rm NR}_{\f,0} =
\ab^\f_T 
+ \sum_{{\rm odd~} d\geq 5} N_{a,\f}^d m_\f^{d-3} \afcb{d}{0},
\nonumber\\
\cb^\f_{TT} &\to &
\ring{\cb}{}^{\rm NR}_{\f,0} =
\cb^\f_{TT}
+ \sum_{{\rm even~} d\geq 6} N_{c,\f}^dm_\f^{d-4} \cfcb{d}{0},
\qquad
\eea
where $N_{a,\f}^d$ and $N_{c,\f}^d$ are numerical factors.
We can then make these replacements 
in the discussion in Sec.\ IX B of Ref.\ \cite{akjt}
and conclude that the vertical acceleration $a$ 
of an antihydrogen atom 
of inertial mass $m_i^\Hb$ and gravitational mass $m_g^\Hb$ obeys
\beq
a = \fr{m_g^\Hb}{m_i^\Hb} g 
\equiv \left(1 + \fr {\de g} g \bigg\vert_\Hb\right) g ,
\label{antihg}
\eeq
with
\beq
\fr{\de g} g \bigg\vert_\Hb = 
\fr 2{m_\Hb}
\sum_\f
(\al \ring{\ab}{}^{\rm NR}_{\f,0} 
+ \frac 13 m_\f \ring{\cb}{}^{\rm NR}_{\f,0}),
\qquad
\label{dgghbar}
\eeq
where $\f$ takes the values $e$ and $p$
and $m_\Hb$ is a constant
equal to the inertial mass of an antihydrogen atom
in the absence of Lorentz and CPT violation.

The above derivation suggests introducing a generalized IPM
via the definition
\beq
\al \ring{\ab}{}^{\rm NR}_{\f,0} 
- \frac 13 m_\f \ring{\cb}{}^{\rm NR}_{\f,0} = 0.
\eeq
The corresponding vertical acceleration for hydrogen is then unaffected,
\beq
\fr{\de g} g \bigg\vert_H = 
\fr 2{m_H}
\sum_\f
(\al \ring{\ab}{}^{\rm NR}_{\f,0} 
- \frac 13 m_\f \ring{\cb}{}^{\rm NR}_{\f,0}) = 0,
\qquad
\eeq
while the gravitational response \rf{antihg} of antihydrogen is anomalous.
Note that the presence of the nonminimal coefficients
implies two new degrees of freedom at each dimension $d$.
This provides intuition about the connection between nonminimal coefficients 
and renewed prospects for a comparatively large 
anomalous gravitational response in antihydrogen.
Note also that a complete derivation
can be expected to generate a tensor relation between the acceleration
of a test body and the acceleration due to gravity,
with horizontal components of the acceleration affected.
The relation involves spatial components of nonminimal coefficients,
along with momentum factors as well.
In general,
the motion of a freely falling antihydrogen atom
in the presence of Lorentz and CPT violation
is expected to follow a geodesic in a pseudo-Finsler geometry
determined by the Riemann metric and the SME coefficients,
while its motion in the IPM follows a geodesic 
in pseudo-Randers spacetime
\cite{finsler1}.

On the experimental side,
the above discussion reveals 
that studies of the gravitational couplings of antihydrogen
probe distinct effects from the spectroscopic tests discussed in this work,
as none of the latter can detect isotropic spherical coefficients with $k=0$
for reasons discussed in Sec.\ \ref{Coefficient selection rules}.
Also,
we can use simple dimensional analysis
to provide an estimate of the sensitivity 
of gravitational experiments with antihydrogen
to nonminimal coefficients for CPT violation.
A generic nonrelativistic spherical coefficient ${\K_\f}_{kjm}^\nr$
has mass dimension $1-k$,
so taking one coefficient nonzero at a time as before
and neglecting momentum effects
yields expected constraints of order 
\beq
|{\K_\f}_{kjm}^\nr| \lsim
m_\f^{1-k} ~\fr{\de g} g \bigg\vert_\Hb .
\eeq
For an experiment with 10\% uncertainty,
this gives constraints 
of order $10^{-4-3k}$ GeV$^{1-k}$ on nonrelativistic coefficients
in the electron sector
and of order $10^{-1}$ GeV$^{1-k}$ on ones in the proton sector.

\section{Deuterium}
\label{Deuterium}

The differing nuclear and spin structures 
of the various hydrogen isotopes 
imply these systems have distinct sensitivities 
to Lorentz and CPT violation.
We focus here on deuterium,
a stable fermionic system that has been widely studied
since its discovery in the early 1930s
\cite{ubm}.
Tritium and the higher hydrogen isotopes
are unstable and challenging to handle experimentally,
although an investigation of the spectroscopic properties
of these systems could be worthwhile as well.
Note that tritium decays are of interest 
in the context of precision measurements of the neutrino mass
\cite{katrin}
and the associated searches for Lorentz and CPT violation 
in the neutrino sector
\cite{neutrinolv}.

The isotope shift for the $1S$-$2S$ transition
between deuterium and hydrogen has been measured
with a potentially competitive absolute uncertainty of about 15 Hz
\cite{pa10a},
while the presence of the neutron in the deuteron core
changes the angular-momentum couplings and
opens opportunities for additional sensitivities
to coefficients in the neutron sector of the SME.
Moreover,
a deuterium maser 
\cite{wr72}
could in principle be used to study the deuterium hyperfine structure
at mHz sensitivity or better.
In this section, 
we consider these possibilities in turn.
We outline an approach to the perturbative hamiltonian,
obtain relevant frequency shifts,
and summarize some implications for experimental studies. 

Note that our analysis here disregards 
the gravitational couplings of deuterium.
Although antideuterons were first created in the laboratory
about 50 years ago
\cite{antid},
antideuterium remains only a theoretical possibility at present.
It is expected to be stable, 
and comparisons of its gravitational response
with that of deuterium could conceivably be of interest 
for at least two theoretical reasons. 
Both deuterium and antideuterium incorporate neutron coefficients 
for Lorentz and CPT violation
and therefore a comparison of their gravitational properties
would further extend tests of models
such as the generalized IPM discussed in Sec.\ \ref{hbar gravity}.
Also,
deuterium and antideuterium are fermions,
and as such their behavior in weak gravitational fields
involves a different set of spin-dependent 
coefficients for Lorentz and CPT violation
\cite{akgrav,akjt,yb13}.
However,
the production, trapping, and experimental manipulation of antideuterium 
remains at present a futuristic challenge,
so detailed theoretical considerations
of the free fall of antideuterium lie outside our present scope.

\subsection{Isotropic Lorentz-violating perturbations}
\label{Isotropic Lorentz-violating perturbations}

Since the deuteron is a bound state of two hadrons,
for which exact expressions for the energy levels are lacking,
the perturbative methods developed above for atomic hydrogen
cannot be applied directly.
Nonetheless,
the dominant contributions from isotropic Lorentz and CPT violation
can be obtained within plausible assumptions.
These are of interest in the context of $1S$-$2S$ and similar transitions
in deuterium.

As a reasonable first approximation to the hamiltonian $H_{\rm D}$
governing the dominant deuterium physics of relevance here, 
we can write the three-body expression
\beq
H_{\rm D} \approx
\fr{\mbf{p}_e^2}{2m_e}+\fr{\mbf{p}_p^2}{2m_p}+\fr{\mbf{p}_n^2}{2m_n}
+ V(\mbf{r}_{ep}) + U(\mbf{r}_{\pd}),
\eeq
where $\mbf p_\f$ is the three-momentum of the particle
of flavor $\f=e$, $p$, $n$,
$m_\f$ is the corresponding mass,
$\mbf r_{ep}$ is the relative position of the electron and proton,
and 
$\mbf r_{\pd}$ is the relative position of the proton and neutron.
The potential $V$ accounts for the electromagnetic interaction
between the proton and electron,
while $U$ describes the nuclear interactions between the proton and neutron.
For simplicity,
we work in the zero-momentum inertial frame of the deuterium atom.

To separate the hamiltonian while keeping 
the dominant Lorentz-invariant physics,
we can reinterpret the dynamics of the proton and neutron
in terms of the motion of the deuteron
and the motion of the proton and neutron
relative to the deuteron center of mass.
It is therefore convenient to define
$\mbf p \equiv \mbf p_p + \mbf p_n = - \mbf p_e$
and 
$\mbf p_\pd \equiv (\mbf p_p - \mbf p_n)/2$,
with the latter being the momentum of the proton 
relative to the center of mass of the deuteron.
It is also a sufficient approximation for present purposes
to take $m_n\approx m_p$ 
and $\mbf{r}_{ep}\approx \mbf{r}_d \equiv \mbf{r}_{ep}+\mbf{r}_{\pd}/2$.
The vector $\mbf{r}_d$ can be viewed as 
the approximate position of the deuteron center of mass 
with respect to the electron.  
It follows that $V(\mbf r_{ep})\approx V(\mbf r_d)$.
With these definitions and approximations,
the hamiltonian $H_{\rm D}$ takes the form
\beq
H_{\rm D} \approx
\fr{\mbf{p}^2}{2\mr} + V(\mbf{r}_{d})
+\fr{\mbf{p}_{\pd}^2}{m_p} + U(\mbf{r}_{\pd}),
\eeq
where $\mr\approx 2m_p m_e/(2m_p+m_e)$
is the reduced mass of deuterium.  
This expression is separable,
so its solution is the tensor product of the solutions
of the two individual systems. 

The next step is to express the Lorentz-violating perturbation 
in terms of these variables.  
Following the scenario introduced in
Sec.\ \ref{Basics},
we can suppose that each of the three particles $e$, $p$, $n$
experiences a perturbation $\de h^\nr_\f$ 
of the form \rf{nr}.
As discussed in Sec.\ \ref{Frequency shift two},
the $1S$-$2S$ and similar transitions are of interest primarily
in the context of measuring isotropic coefficients for Lorentz violation,
so we restrict attention here to the quantities
$\Vnrf{\f}{kjm} \equiv \cnrf{\f}{kjm} - \anrf{\f}{kjm}$
defined in Eq.\ \rf{cpt},
but now with $\f=e,p,n$.

Under these assumptions,
we find that the isotropic part 
of the perturbation hamiltonian $\de h^\nr_{\rm D}$ 
can be written in the form
\bea
\de h^\nr_{\rm D} = 
-\fr{1}{\sqrt{4\pi}}
\sum_{k= 2,4}
\big(
\Vnrf{e}{k00} ~\pmag^k
+ \Vnrf{p}{k00} ~|\half \mbf{p}+\mbf{p}_{\pd}|^k
\nonumber\\
+ \Vnrf{n}{k00} ~|\half \mbf{p}-\mbf{p}_{\pd}|^k
\big).
\qquad
\label{deutpert}
\eea
This operator expression describes the leading-order perturbative effects 
arising from isotropic Lorentz and CPT violation.

\subsection{$1S$-$2S$ transition}
\label{Deuterium transitions}

The deuterium energy-level shifts are given by
expectation values of the perturbation hamiltonian $\de h^\nr_{\rm D}$ 
in the Lorentz-invariant states.
In performing the calculations,
only the cross terms 
$\mbf{p}\cdot\mbf{p}_{\pd}$ and $(\mbf{p}\cdot\mbf{p}_{\pd})^2$ 
that couple both systems
could in principle be challenging to handle.
However,
the former is odd under a parity transformation
of either momentum and hence yields zero contribution
at leading order.
To treat the quadratic term,
we can plausibly assume the two systems are sufficiently decoupled 
so that
$\vev{(\mbf{p}\cdot\mbf{p}_{\pd})^2}
\approx \vev{\mbf{p}^2}\vev{\mbf{p}_{\pd}^2}/2$.
Also,
the contribution to the $1S$-$2S$ and other $nL$-$n'L'$ transitions 
can be expected to depend on a nonzero power of $\mbf p$.
Moreover,
the magnitude of $\mbf p_{\pd}$ is roughly $100$ MeV 
while that of $\mbf p$ is about 1 keV,
so $\vev{{\mbf p}^2} \vev{{\mbf p}_{\pd}^2} \gg \vev{{\mbf p}^4}$.  

Combining these considerations
and using the expectation values \rf{radialExp},
we find that the frequency shift $\de\nu_{\rm D}$ 
of the $nL$-$n'L'$ transition in deuterium 
due to isotropic Lorentz and CPT violation 
is given by
\bea
2\pi \de \nu_{\rm D} &=&
\fr \mr {\sqrt{\pi}}
(\ve_{n'}-\ve_{n}) 
\bigg[
\Vnrf{e}{200} 
+ \frac 14 
\left(
\Vnrf{p}{200} + \Vnrf{n}{200}
\right)
\nonumber\\
&&
\hskip 60pt
+ \vev{\mbf p_{\pd}^2}
\left(
\Vnrf{p}{400} + \Vnrf{n}{400}
\right)
\bigg]
\nonumber \\
&&
\hskip -40pt
- \fr {2 \mr^2} {\sqrt{\pi}} 
\Bigg[
\ve_{n'}^2
\left(\fr{8n'}{2L'+1} - 3 \right)
-\ve_{n}^2 
\left(\fr{8n}{2L+1} - 3 \right)
\Bigg] 
\nonumber\\
&&
\hskip 20pt
\times
\left(\Vnrf{e}{400}
+\frac 1 {16} (\Vnrf{p}{400}
+\Vnrf{n}{400})
\right) ,
\label{degenf}
\eea
where $\ve_{n}\equiv -\al^2 \mr/2n^2$ 
and $\vev{\mbf p_{\pd}^2}\simeq 10^4$ MeV$^2$.
This expression generalizes the result \rf{Jhalf} for hydrogen
and reduces to it in the limit where the proton and neutron
are taken to have identical momenta,
each of magnitude half that of the electron momentum. 

The $1S$-$2S$ transition in deuterium
provides interesting sensitivity to anisotropic coefficients
in the Sun-centered frame
via boost corrections
that produce sidereal and annual variations.
At leading order in the boost parameter,
the deuterium $1S$-$2S$ transition frequency
takes the same form \rf{1s2svar}
as its hydrogen counterpart,
except that the sum over flavors
now includes also the neutron.
The leading-order contributions for the electron vectors $V^{(d)J}_e$ 
have the same form as for hydrogen
and so can be found in Table \ref{VJ}.
The leading-order contributions for the proton and neutron vectors 
$V^{(d)J}_p$ and $V^{(d)J}_n$ 
with $5\leq d\leq 8$ 
can be obtained from Table \ref{VJD}.
In this table,
$m_\f$ represents the rest masses of the proton $\f=p$ and neutron $\f=n$.
As before,
$\al$ is the fine-structure constant,
$\mr$ is the reduced mass of the system,
and the effective cartesian coefficients for Lorentz violation
are defined in Eq.\ (27) of Ref.\ \cite{km13}.
The minimal-SME spin-independent coefficients 
$V^{(3)J}_\f$ and $V^{(4)J}_\f$ vanish 
at leading order in the nonrelativistic limit
and so have no effect on the $1S$-$2S$ frequency,
in parallel with the hydrogen case
\cite{bkr,yo12}.

\renewcommand\arraystretch{1.7}
\begin{table}
\caption{Values of the vectors $V^{(d)J}_\f$ 
for the proton and neutron in deuterium 
with $5\leq d \leq 8$.}
\setlength{\tabcolsep}{7 pt}
\begin{tabular}{cc}
\hline
\hline
$d$ & $V^{(d)J}_\f$ 
\\
\hline
5 & $\frac{3}{16} (\al \mr)^2 (2{a_\f}^{TTJ}_{\rm eff} + {a_\f}^{KKJ}_{\rm eff})$\\
6 & $-\frac{3}{4} (\al \mr)^2 m_\f ({c_\f}^{TTTJ}_{\rm eff} + {c_\f}^{TKKJ}_{\rm eff})$\\
7 & $\frac 58 (\al \mr)^2 m_\f^2 (2{a_\f}^{TTTTJ}_{\rm eff} + 3{a_\f}^{TTKKJ}_{\rm eff})$\\
   &  $+\frac{3}{4} (\al \mr)^2 \vev{\pvec^2_\pd} {a_\f}^{KKLLJ}_{\rm eff}$\\
8 & $-\frac {15}8 (\al \mr)^2 m_\f^3 ({c_\f}^{TTTTTJ}_{\rm eff} + 2{c_\f}^{TTTKKJ}_{\rm eff})$\\
    &$-\frac{9}{2} (\al \mr)^2 m_\f \vev{\pvec_\pd^2} {c_\f}^{TKKLLJ}_{\rm eff}$\\
\hline\hline
\end{tabular}
\label{VJD}
\end{table}

The time variations in the deuterium $1S$-$2S$ transition 
are determined by the same expressions as for hydrogen,
namely,
Eq.\ \rf{1s2sann} for the annual frequency $\Om_\oplus$
and Eq.\ \rf{1s2ssid} for the sidereal frequency $\om_\oplus$.
However,
the deuterium transition offers some advantages 
over its hydrogen counterpart.
One is the sensitivity of deuterium to neutron coefficients.
A more subtle advantage  
is that the motion of the proton in the nucleus 
makes the deuterium experiment substantially more sensitive 
to some of the proton coefficients.  
The point is that in hydrogen the proton
is a comparatively placid object with momentum
opposite that of the electron,
with magnitude $\sim\al m_e$ of a few keV. 
In contrast,
the proton and neutron in deuterium together have total momentum
opposite that of the electron,
but each nucleon has momentum of over 100 MeV,
producing an expectation value 
$\vev{\pvec^2_\pd}\sim 10^4$~MeV$^2$.
As a result,
measurements of the deuterium $1S$-$2S$ transition 
offer about a billionfold
greater sensitivity than hydrogen to the proton coefficients
${a_p}^{KKLLJ}_{\rm eff}$ and ${c_p}^{TKKLLJ}_{\rm eff}$,
as can be deduced from the entries for $d=7,8$
in Table \ref{VJD}.

We remark in passing that 
a study of subleading effects in the deuterium $1S$-$2S$ frequency 
along the lines of the experiment with hydrogen 
performed in Ref.\ \cite{ma13}
could also be used to constrain minimal SME coefficients 
in the neutron sector.
The analogue of the minimal-SME match \rf{ma13match} for deuterium
involves both proton and neutron coefficients,
\beq
\sum_{\f d}V_\f^{(d)J} = \frac 54 \al^2 \mr
[c^{(TJ)}_e + \frac 14 (c^{(TJ)}_p + c^{(TJ)}_n )],
\eeq
and this expression could be used to determine sensitivities 
to nonminimal coefficients in all three sectors $\f = e$, $p$, $n$
as for the hydrogen case.

\subsection{Comparative analyses}
\label{Deuterium self consistent}

In atomic hydrogen,
the isotropic coefficients produce 
an effective shift of the Rydberg constant
given by Eq.\ \rf{rydshi}.
An analogous effect occurs in deuterium,
but the shift is instead given by 
\bea
\de R_{\infty,{\rm D}} &=& 
\fr {2\sqrt{\pi}\mr^2}{m_e} R_\infty 
\bigg[
\Vnrf{e}{200} 
+ \frac 14 
\left(
\Vnrf{p}{200} + \Vnrf{n}{200}
\right)
\nonumber\\
&&
\hskip 50pt
+ \vev{\mbf p_{\pd}^2}
\left(
\Vnrf{p}{400} + \Vnrf{n}{400}
\right)
\bigg].
\qquad
\eea
Since the shift $\de R_{\infty}$ in hydrogen
and the shift $\de R_{\infty,{\rm D}}$ in deuterium are distinct,
any difference between
the values obtained for the Rydberg constant
in experiments with hydrogen and with deuterium
could be a signal for Lorentz and CPT violation.
Similarly,
the change $\de \nu_{\rm Lamb,D}$ in the classical Lamb shift
($2S_{J=1}$-$2P_{J=1}$)
in deuterium,
\beq
2 \pi\de \nu_{\rm Lamb,D} =
-\fr {(\al \mr)^4} {3\sqrt{\pi}}
\big[
\Vnrf{e}{400} 
+ \frac 1{16} 
\left(
\Vnrf{p}{400} + \Vnrf{n}{400}
\right)
\big],
\eeq
differs from the change $\de \nu_{\rm Lamb}$ in the hydrogen Lamb shift
given by Eq.\ \rf{lamshi}.
A signal for Lorentz and CPT violation would therefore be
an observed discrepancy between experimental values
obtained in the two systems.

The above comparative analyses
can also be extended to physical effects in other transitions.
These could include,
for example,
the transitions $2S$-$4D$, $2S$-$8D$, and $2S$-$12D$,
all of which have been measured in deuterium 
\cite{we95,de97,sc99}.
To illustrate the extraction of constraints
on coefficients for Lorentz violation via this method,
we consider here a comparative analysis 
of the two experimental values
of the difference between the proton and deuteron radii
obtained in Refs.\ \cite{we95} and \cite{je11}.
As before,
only the contributions from isotropic coefficients 
for Lorentz and CPT violation are included,
as other types of searches are more sensitive to anisotropic coefficients.

Consider first the weighted difference 
\beq
\De = \nu_{2S4S}- \frac 14 \nu_{1S2S}
\eeq
between the $2S$-$4S$ frequency $\nu_{2S4S}$
and the $1S$-$2S$ frequency $\nu_{1S2S}$,
measured for hydrogen and deuterium in Ref.\ \cite{we95}.
The change in the isotope shift $\de\nu_{\rm shift}$ 
between the two measurements 
is governed by the difference $\de(r_d^2-r_p^2)$
between the square of the charge radii
of the proton and deuteron,
\beq
2\pi\de\nu_{\rm shift} = 
-\fr {7\pi \al^4 R_\infty}{24 r_e^2} \de(r^2_d-r^2_p),
\eeq
where $r_e$ is the classical electron radius. 
The frequency shift $\de\nu_{\rm LV}$ of $\De$ 
due to Lorentz and CPT violation is given by
\beq
2\pi\de\nu_{\rm LV} =
-\fr {89(\al\mr)^4} {4096\sqrt{4\pi}}
(15 \Vnrf{p}{400}-\Vnrf{n}{400}).
\eeq
Assuming the observed value of $\de\nu_{\rm shift}$ 
arises entirely from $\de\nu_{\rm LV}$,
we find that the difference between the square of the charge radii 
determined in Ref.\ \cite{we95}
is approximately given by
\beq
\de(r_d^2-r_p^2)\approx
(2\times 10^{-6})
(15 \Vnrf{p}{400}-\Vnrf{n}{400})
{\rm ~GeV}^{3} {\rm ~fm}^2,
\label{diff1}
\eeq
where the coefficients for Lorentz and CPT violation
have units of GeV$^{-3}$.

Next,
consider the difference between the square of the charge radii 
obtained in Ref.\ \cite{je11}
from measurements of the $1S$-$2S$ frequency $\nu_{1S2S}$
in hydrogen and deuterium.
In this case,
the change in the isotope shift $\de\nu_{\rm shift}$ is given by 
\beq
2\pi\de\nu_{\rm shift} =
\fr{7\pi \al^4 R_\infty}{3 r_e^2}\de(r^2_d-r^2_p),
\eeq
while the Lorentz-violating shift $\de\nu_{\rm LV}$ is
\bea
2\pi\de\nu_{\rm LV} &=&
\fr{3(\al\mr)^2}{16\sqrt{4\pi}}
\big[
3\Vnrf{p}{200}-\Vnrf{n}{200}
\nonumber\\
&&
\hskip 40pt
-4\vev{\mbf{p}_{pd}^2}(\Vnrf{p}{400}+\Vnrf{n}{400})
\big].
\qquad
\eea
The assumption that the observed frequency shift
originates from Lorentz violation
now gives the difference between the square of the charge radii 
as approximately 
\bea
\de(r_d^2-r_p^2) &\approx &
(2\times 10^{5})
\big[3 \Vnrf{p}{200}-\Vnrf{n}{200}
\nonumber\\
&&
\hskip -5pt
- 4 \vev{\mbf{p}_\pd^2}
(\Vnrf{p}{400}+\Vnrf{n}{400})\big]
{\rm ~GeV}^{3} {\rm ~fm}^2.
\qquad
\label{diff2}
\eea
We see that this expression and the result \rf{diff1}
provide two distinct measures of $\de(r_d^2-r_p^2)$
in terms of coefficients for Lorentz and CPT violation. 

The two reported experimental values are essentially in agreement.
Note that the above results make use of values of the Rydberg constant
and the mass ratios of the electron to the proton and to the deuteron.
In principle, 
these quantities could be shifted by Lorentz and CPT violation,
so we take a conservative value of less than 0.02 fm$^2$
for the uncertainty in $|\de(r_d^2-r_p^2)|$ .
Disregarding coefficients in the electron sector,
which give contributions proportional to
the difference of powers of the reduced masses of deuterium and hydrogen
and so are suppressed by four or more orders of magnitude
relative to coefficients in the proton and neutron sectors,
we can finally extract the constraint
\bea
\big|
3 \anrfc{p,2}-\anrfc{n,2}
-4 \vev{\mbf{p}_\pd^2} (\anrfc{p,4}+\anrfc{n,4})
\hskip 20pt
&&
\nonumber\\
- 3 \cnrfc{p,2}+\cnrfc{n,2}
-4 \vev{\mbf{p}_\pd^2} (\cnrfc{p,4}+\cnrfc{n,4})
\big|
&&
\nonumber\\
&&
\hskip -60pt
< 2\times10^{-7} {\rm ~GeV}^{-1}
\qquad
\label{deutconst}
\eea
on coefficients for Lorentz and CPT violation.

Similar comparative analyses can be performed
in other systems.
An example discussed in Ref.\ \cite{gkv14}
is a comparison of radii using the isotope shift
between muonic hydrogen and muonic deuterium.
A complete analysis of this system
would place constraints on muon coefficients for Lorentz and CPT violation
as well as proton and neutron coefficients.

\subsection{Deuterium maser}
\label{Deuterium maser}

The successful construction and operation of a deuterium maser
with absolute frequency uncertainty around 1 mHz 
\cite{wr72}
implies that high-precision spectroscopy 
of the hyperfine structure of deuterium
is a realistic possibility.
The high momenta of the proton and neutron
in the deuteron core,
which as described in Sec.\ \ref{Deuterium transitions}
leads to a billionfold gain in sensitivity 
to certain coefficients for Lorentz and CPT violation
in $1S$-$2S$ spectroscopy,
can similarly be expected to enhance the sensitivity
to coefficients affecting the hyperfine transitions 
of the deuterium maser relative to the hydrogen one.
Moreover,
the deuteron core exists in an admixture
of orbital angular momentum 0 and 2,
so the deuterium maser also provides access 
to coefficients with larger values of $j$. 

To study the corrections to the hyperfine structure in deuterium,
a reasonable description of the unperturbed ground state is required.
The angular part of this state can be obtained
by coupling the spin $S_d$ of the deuteron to the spin $S_e$ of the electron.  
The deuteron component involves the coupling
of the triplet state of the proton and the neutron
with a superposition of $L=0$ and $L=2$ orbital states
of the nuclear motion
\cite{ma01}.
As a result,
the unperturbed wavefunction for the ground state can be expressed as
\beq
\vev{\mbf{p},\mbf{p}_\pd|F m_F} =
\ps_{10}(p) 
\sum_{S_e S_d} 
\vev{\frac 12 S_e 1S_d|Fm_F}
\ket{S_e}\vev{\mbf{p_\pd}|S_d},
\eeq
where $\ps_{10}$ is the spin-independent piece,
$S_e$ is the electron spin,
and $S_d$ is the deuteron spin.
The deuteron spin wavefunction takes the form
\beq
\vev{\mbf{p}_\pd|S_d} = 
\sum_{l=0}^1 \Ps_{2l}(p_\pd)
\sum_{qm}
\vev{1q(2l)m|1S_d} Y_{(2l)m}(\phat_\pd) \ch_q,
\eeq
where $\Ps_{2l}$ contains the radial piece
and $\ch_m$ is the spin-triplet wavefunction 
constructed from the proton and neutron spin states.

Using this wavefunction,
we can determine the perturbative shifts
by calculating the expectation values
of the full perturbative hamiltonian 
$\de h^\nr_{\rm D}$
obtained following the discussion in Sec.\ \ref{Basics},
assuming each particle $e$, $p$, $n$
experiences a pertubation of the form \rf{nr}.
In the hyperfine Zeeman regime,
the effects from isotropic coefficients cancel as usual.
Neglecting these coefficients,
we find the energy-level shifts are
\bea
\de \ve(F,m_F) &=&
-\fr{m_F}{3\sqrt{3\pi}}
\sum_{q=0}^2 \fr{(\al \mr)^{2q}} {2(F-1)}(1+4\de_{q2}) 
\nonumber\\
&&
\hskip 40pt
\times
( \TzBnrf{e}{(2q)10}+2\ToBnrf{e}{(2q)10} ).
\qquad
\eea
Comparison of this expression with
the energy-level shifts \rf{1S} and frequency shifts \rf{1Sf} 
for the hydrogen maser shows that both types of maser are sensitive
to the same combination of electron coefficients.
The deuterium maser therefore has no particular advantage
over the hydrogen maser in this regard.
Note also that the two deuterium-maser transitions
$F=3/2\to F=1/2$ with $m_F=\mp 1/2\to m_F=\pm1/2$,
which are the most independent of fluctuations in the applied magnetic field,
are insensitive to Lorentz and CPT violation at leading order.
This parallels the result for $F=1\to F=0$ with $m_F=0$
for the hydrogen maser.
 
The calculation of the perturbative shifts
due to the proton and neutron coefficients is more involved.
Since the deuteron spin is a good quantum number,
the coefficient selection rules
discussed in Sec.\ \ref{Coefficient selection rules}
imply that only proton and neutron coefficients with $j\leq2$ can contribute,
so only the cases $j=1$ and $j=2$ need be considered.
Some identities are useful to evaluate the factors involving the momentum
and spin-weighted spherical harmonics.
Writing $\mbf{p}\equiv \mbf{p}_a+\mbf{p}_b$,
we find for $j=1$ the identities
\bea
\pmag ~\syjm{s}{1m}(\phat) &=&
|\mbf{p}_a| ~\syjm{s}{1m}(\phat_a) 
+|\mbf{p}_b| ~\syjm{s}{1m}(\phat_b), 
\nonumber \\
\pmag \si^s(\phat) &=&
|\mbf{p}_a| \si^s(\phat_a) + |\mbf{p}_b| \si^s(\phat_b).
\label{spinmod}
\eea
For $j=2$,
we obtain 
\bea
\pmag^2 Y_{20}(\phat) &=& 
|\mbf{p}_a|^2 Y_{20}(\phat_a)
+ |\mbf{p}_b|^2 Y_{20}(\phat_b)
\nonumber \\
&&
\hskip -30pt
+\half |\mbf{p}_a| |\mbf{p}_b|
\sqrt{\dfrac{5}{\pi}}
(3\cos{\th_a}\cos{\th_b}-\cos{\ga}),
\qquad
\eea
where $\cos\ga\equiv\phat_a\cdot\phat_b$.  

Armed with these identities,
we can determine the perturbative level shifts.
Consider first the expectation value 
of the spin-dependent terms with $j=1$.  
Choosing as before the applied magnetic field
to be aligned with the laboratory $z$ axis,
we require the expectation value of Eq.\ \rf{nr} for $jm=10$.
We find
\bea
\de\ve(F,m_F) &=& 
-\fr{1}{3\sqrt{6\pi}}
\fr{m_F}{2^{F-2}}
\sum_k \vev{\mbf p^{2k}_\pd} 
\nonumber\\
&&
\hskip 15pt
\times
\sum_{w}
( \TzBnrf{\w}{(2q)10} + 2\ToBnrf{\w}{(2q)10}),
\quad
\qquad
\eea
where $\w$ takes the values $p$ and $n$.

For states with $F=1/2$,
only proton and neutron coefficients with $j=1$ contribute
to the hyperfine Zeeman frequencies. 
However,
when $F=3/2$,
contributions from coefficients with $j=2$ can appear. 
In this case,
we find the contribution to the energy-level shift is
approximately
\beq
\de\ep(F,m_F) = 
\fr{1}{\sqrt{5\pi}}
\fr{2F-1}{(8m_F^2-10)}
\sum_{q=0}^2
\vev{\mbf p_{pd}^{2q}}'
\sum_\w
\Vnrf{\w}{(2q)20},
\eeq
where
\beq
\vev{\mbf  p^{2q}_\pd}' \equiv
\vev{\Ps_2|\mbf p^{2q}_\pd|\Ps_2}
- \sqrt{8}~\Re {\vev{\Ps_2|\mbf p^{2q}_\pd|\Ps_0}} .
\eeq
This depends on the specific model used
for the radial deuteron wavefunctions $\Ps_0$, $\Ps_2$.

Combining the above results for electron, proton, and neutron coefficients,
we find that the anisotropic shifts 
in the deuterium hyperfine Zeeman levels are given by
\bea
\de\ep(F,m_F) &=& 
\fr{1}{\sqrt{5\pi}}
\fr{2F-1}{(8m_F^2-10)}
\sum_{q=0}^2
\vev{\mbf p_\pd^{2q}}'
\sum_\w
\Vnrf{\w}{(2q)20}
\nonumber\\
&&
\hskip -60pt
-\fr{1}{3\sqrt{6\pi}}
\fr{m_F}{2^{F-2}}
\sum_{q=0}^2
\vev{\mbf p^{2q}_\pd} 
\sum_\w
( \TzBnrf{\w}{(2q)10}+2\ToBnrf{\w}{(2q)10} )
\nonumber\\
&&
\hskip -60pt
-\fr{m_F}{3\sqrt{3\pi}}
\sum_{q=0}^2
\fr{(\al m_r)^{2q}}{2(F-1)} (1+4\de_{q2}) 
( \TzBnrf{e}{(2q)10}+2\ToBnrf{e}{(2q)10} ),
\nonumber \\
\eea
where the sum over $\w$ spans the values $p$ and $n$.
In this expression,
the quantities 
$\Vnrf{\w}{(2q)20}$,
$\TzBnrf{\f}{(2q)10}$, 
and $2\ToBnrf{\f}{(2q)10}$
are given in terms of the coefficients for Lorentz and CPT violation
by Eq.\ \rf{cpt}.
Note that the proton and neutron coefficients
contribute to all possible hyperfine Zeeman transitions,
including the ones that are the most independent 
of fluctuations in the external magnetic field. 
Another key feature of this result
is the appearance of coefficients with $j=2$,
which implies experimental signals at the second harmonic $2\om_\oplus$
of the sidereal frequency
that can be measured in hyperfine Zeeman transitions
with $\De F \neq 0$.
Moreover,
the dependence on the expectation values ${\vev{\mbf p^{2q}_\pd}}$
acts to enhance the sensitivity 
to the coefficients for Lorentz and CPT violation
by factors of a billionfold for coefficients with $k=2$
and by $10^{18}$-fold for coefficients with $k=4$.
Generically,
this suggests attainable sensitivities
to nonrelativistic coefficients $\K_{kjm}$ with even $k=2,4$ 
at the level of $|\K|\lsim 10^{-27+k}$ GeV$^{1-k}$,
representing an impressive potential improvement
over the results in Table \ref{tanr1S}
as well as sensitivity to numerous coefficients
unmeasurable using a hydrogen maser. 

\renewcommand\arraystretch{1.7}
\begin{table}
\caption{Values of the pseudotensor $T_{\w}^{(d)JKL}$ for $5\leq d\leq 8$.}
\setlength{\tabcolsep}{7 pt}
\begin{tabular}{cc}
\hline
\hline
$d$ & $T_\w^{(d)JKL}$ 
\\
\hline
5 & $-\frac 1{5} \vev{\mbf{p}^2_\pd} 
(2{a_\w}^{TTJ}_{\rm eff} \de^{KL} + {a_\w}^{JKL}_{\rm eff})$
\\
6 & $\frac 45 \vev{\mbf{p}^2_\pd} 
m_\w ({c_\w}^{TTTJ}_{\rm eff} \de^{KL} + {c_\w}^{TJKL}_{\rm eff})$
\\
7 & $-\frac 23 \vev{\mbf{p}^2_\pd}
m_\w^2 (2{a_\w}^{TTTTJ}_{\rm eff} \de^{KL} + 3{a_\w}^{TTJKL}_{\rm eff})$
\\
&  $- \frac 2{7} \vev{\mbf{p}^4_\pd}
{a_\w}^{MMJKL}_{\rm eff}$
\\
8 & $ 2\vev{\mbf{p}^2_\pd}
m_\w^3 ({c_w}^{TTTTTJ}_{\rm eff} \de^{KL} + 2{c_\w}^{TTTJKL}_{\rm eff})$
\\
& +$\frac 6{7} \vev{\mbf{p}^4_\pd}
m_\w{c_w}^{TMMJKL}_{\rm eff}$
\\
\hline\hline
\end{tabular}
\label{TJd}
\end{table}

This remarkable potential reach naturally suggests 
investigating the possibility of additional constraints 
from an analysis of the deuterium hyperfine Zeeman transitions
that incorporates the boost relative to the Sun-centered frame,
in analogy to the discussion for hydrogen 
in Sec.\ \ref{Boost corrections and parity-odd operators}. 
Adopting a similar notation,
we find that the first-order shifts $\de \ep^{(1)}(F,m_F)$ 
of the energies due to the boost correction
take the form
\bea
\de \ep^{(1)}(F,m_F) &=& 
\fr{m_F}{3(F-1)}
\sum_{d}
T^{(d)JK}_e R^{zJ} 
(\be^{K}_\oplus+\be^{K}_L)
\nonumber\\
&&
\hskip -5pt
+ \fr{(2F-1)}{5-4m_F^2} 
\sum_{\w d}
T^{(d)JKL}_\w M^{JK}
(\be^{L}_\oplus+\be^{L}_L)
\nonumber\\
&&
\hskip -5pt
+ \fr{\sqrt{2}}{3}
\fr{m_F}{2^{F-2}}
\sum_{\w d}
{\Tb}^{(d)JK}_\w R^{zJ}
(\be^{K}_\oplus+\be^{K}_L),
\nonumber\\
\eea
where the sums over $\w$ are over the flavors $p$ and $n$.
In this expression,
the quantities $M^{IJ}$ represent combinations of rotations
given by
\beq
M^{JK}=2R^{zJ}R^{zK}-R^{xJ}R^{xK}-R^{yJ}R^{yK}.
\eeq
The pseudotensors $T_{e}^{(d)JK}$ in the electron sector
are given by Table \ref{Amm} with $\f=e$,
while the pseudotensors $\Tb_{w}^{(d)JK}$ in the proton and neutron sectors
are also obtained from this table with the substitutions
$(\al \mr)^2 \to \vev{\mbf{p}^2_\pd}$ 
and 
$(\al \mr)^4\to \vev{\mbf{p}^4_\pd}/5$.  
The explicit forms of the pseudotensors $T_{\w}^{(d)JKL}$ 
are listed in Table \ref{TJd}
in terms of the expectation values ${\vev{\mbf p^{2q}_\pd}}$,
the rest masses $m_\w$,
the fine-structure constant $\al$,
the deuterium reduced mass $\mr$,
and the effective cartesian coefficients for Lorentz violation
defined in Eqs.\ (27) and (28) of Ref.\ \cite{km13}.
For each coefficient, 
only the leading-order nonrelativistic contributions are provided.

In parallel with the results for the boost-independent terms,
the expectation values ${\vev{\mbf p^k_\pd}}$
enhance the attainable sensitivity to spin-dependent coefficients
relative to the hydrogen maser by factors of about a billion for $k=2$ 
and about $10^{18}$ for $k=4$.
Moreover,
the reach for the $a$- and $c$-type coefficients in Table \ref{TJd}
is also substantially enhanced relative to 
related measurements of the $1S$-$2S$ transitions 
in hydrogen and deuterium.
The prospects for these boosted measurements with a deuterium maser
therefore appear excellent as well.

\section{Positronium}
\label{Positronium}

Positronium is another hydrogenic system
with potential for measurable signals 
from Lorentz and CPT violation.
Studies of CPT violation in positronium decay
have been published for both experiment
\cite{ve04}
and theory
\cite{ad10}.
Here,
we offer some remarks on the potential role of positronium spectroscopy
the search for Lorentz and CPT violation.

The perturbative corrections to the hydrogen spectrum
obtained in Secs.\ \ref{Theory} and \ref{Applications}
cannot generically be applied to determine
shifts in the positronium spectrum
because the large magnetic moment of the positron
implies the fine and hyperfine structures in positronium are comparable 
and so the hierarchy of angular-momentum couplings 
of hydrogen and positronium are different.  
However,
the $nS_{1/2}$ levels present an exception to this,
as the two schemes for angular-momentum couplings 
coincide when $L=0$.
We therefore focus here on experimental scenarios
involving transitions among the $nS_{1/2}$ levels.
In particular,
we consider potential signals for Lorentz and CPT violation
in the hyperfine splitting of the $1S$ ground-state levels
and in the $1S$-$2S$ transition. 

The quantum states of free parapositronium ($S=0$)
or orthopositronium ($S=1$)
are eigenstates of the charge-conjugation operator C,
so only C-even Lorentz-violating operators can contribute 
to the energy shifts. 
Examining Table \ref{rest} reveals that 
only the electron coefficients 
$\cnrf{e}{kjm}$	
can contribute to spin-independent shifts 
of the positronium ground-state splitting,
while only 
$\gzBnrf{e}{kjm}$ and $\goBnrf{e}{kjm}$	
can contribute to spin-dependent ones.
This basic feature means that positronium
naturally disentangles CPT-even and CPT-odd operators
in the electron sector in the nonrelativistic limit.

In the limit of a weak applied magnetic field,
the frequency shift $\de\nu_{\rm Z}$ 
of the hyperfine Zeeman transitions in positronium is given by
\bea
2\pi\de\nu_{\rm Z} &=&
-\fr{\De m_F}{\sqrt{3\pi}}
\sum_{q=0}^2
(\al \mr)^{2q}(1+4\de_{q2})
\nonumber\\ 
&&
\hskip 50pt
\times
\left( \gzBnrf{e}{(2q)10} + 2\goBnrf{e}{(2q)10}\right).
\qquad
\label{1Sfpos}
\eea
As expected,
only contributions from coefficients controlling CPT violation appear.
In contrast,
in a strong applied magnetic field
the quantum states of the system 
are no longer eigenstates of the charge-conjugation operator
due to mixing of the entangled spin-triplet and spin-singlet levels.
In this Paschen-Back limit, 
both CPT-odd and CPT-even Lorentz-violating operators
contribute to the frequency shift,
\bea
2\pi\de\nu_{\rm PB} &=&
-\fr 1{\sqrt{3\pi}}
\sum_{q=0}^2
(\al \mr)^{2q}
(1+4\de_{q2})
\nonumber\\ 
&&
\times
\Big[ 
(\De S_e + \De S_\eb) 
\left(
\gzBnrf{e}{(2q)10} +2\goBnrf{e}{(2q)10}
\right)
\nonumber\\ 
&&
-
(\De S_e - \De S_\eb) 
\left(
\HzBnrf{e}{(2q)10} +2\HoBnrf{e}{(2q)10}
\right)
\Big],
\nonumber\\ 
\label{pasbacbpos}
\eea
where $\De S_e$ and $\De S_\eb$
denote the electron and positron spin changes,
respectively.

Current precision measurements of the positronium hyperfine structure
lie in the ppm range,
with an absolute uncertainty of order 1 MHz 
\cite{positronium}.
For example,
this implies a potential reach of about $10^{-17}$ GeV
for the coefficients $\gzBnrf{e}{010}$ and $2\goBnrf{e}{010}$,
about $10^{-6}$ GeV$^{-1}$
for $\gzBnrf{e}{210}$ and $2\goBnrf{e}{210}$,
and about $10^{5}$ GeV$^{-3}$
for $\gzBnrf{e}{410}$ and $2\goBnrf{e}{410}$.
While these sensitivities are about nine orders of magnitude
below those presented in Table \ref{tanr1S}
obtained via spectroscopy with a hydrogen maser
taking only one coefficient nonzero at a time,
the actual combinations of coefficients
in the positronium and hydrogen observables are distinct.
This confirms that positronium hyperfine measurements can be used
to separate CPT-even and CPT-odd spin-dependent effects
in the electron sector.

Positronium also has the advantage of being a purely leptonic atom,
allowing precision tests of quantum electrodynamics or new physics
via the direct comparison between experiment and theory.
Paralleling the discussion for hydrogen,
isotropic coefficients for Lorentz violation 
can be expected to shift the value of experimental measurements
of the $1S$-$2S$ transition in positronium
relative to the Lorentz-invariant theory.  
Comparing experiment to theory therefore provides 
a constraint on Lorentz violation in the electron sector.

The Lorentz-violating frequency shift 
for the $1S$-$2S$ transition in positronium
is given by
\beq
2\pi \de \nu =
\frac 32 (\al\mr)^2
\left(\cnrfc{e,2}
+ \frac{67}{12} (\al\mr)^2 \cnrfc{e,4}\right).
\label{1S2posfreq}
\eeq
As expected,
it contains only CPT-even effects.
The observed difference between the theoretical and experimental values
of the positronium $1S$-$2S$ frequency is
$5.8 \pm 3.3$ MHz 
\cite{pa99}.
Identifying this difference with the frequency shift \rf{1S2posfreq}
yields the result
\beq
\cnrfc{e,2} + \frac{67}{12} (\al \mr)^{2} \cnrfc{e,4}
\simeq (4.5\pm 2.5)\times 10^{-6} {\rm ~GeV}^{-1},
\label{1S2Spos}
\eeq
representing a 1.8 sigma effect.
It can conservatively be taken as indicating
an experimental reach of about $10^{-5}$ GeV$^{-1}$
to the coefficients $\cnrfc{e,2}$ 
and about $10^{5}$ GeV$^{-3}$ 
to $\cnrfc{e,4}$.
Improvements of about a factor of five in the experimental sensitivity
are within reach of future experiments
\cite{cr11}.

Measuring the free-fall acceleration 
of positronium has been proposed as a test 
of the gravitational couplings of matter and antimatter
\cite{cr14},
in a spirit similar to the proposals for antihydrogen
discussed in Sec.\ \ref{hbar gravity}.
Following the line of reasoning leading to Eq.\ \rf{dgghbar},
we find that nonzero isotropic coefficients for Lorentz violation
lead to a fractional change in the gravitational acceleration
of positronium given by
\beq
\fr{\de g} g \bigg\vert_{Ps} \approx 
\frac 83 \ring{\cb}{}^{\rm NR}_{e,0}.
\qquad
\label{dggpos}
\eeq
Note this depends only on C-even Lorentz violation.
The prospective measurements of the positronium gravitational acceleration 
at the 10\% level could therefore either provide direct sensitivity 
to CPT-even Lorentz violation in the electron sector
or help disentangle CPT-even and CPT-odd effects obtained
in other experiments.

\section{Hydrogen molecules}
\label{Hydrogen molecules}

High-precision molecular spectroscopy 
presents an interesting alternative potential arena
for tests of Lorentz and CPT symmetry,
albeit one that remains largely unexplored to date. 
Although the primary focus of the present work is hydrogenic systems,
some of the tools developed here can be applied
in the context of comparatively simple molecules and molecular ions
such as $H_2$, $H_2^{+}$, $HD$, and $HD^+$.
In this section,
we offer a few comments about the prospects
for measuring nonrelativistic spherical coefficients
for Lorentz and CPT violation in these systems.

Corrections to the energy levels and internuclear distances
of $H_2$, $H_2^{+}$, $HD$, and $HD^+$
arising from Lorentz and CPT violation in the electron sector
of the minimal SME 
have previously been studied by M\"uller \etal\ 
\cite{mu04}.
In this work,
the unperturbed molecular states are approximated
by a wavefunction ansatz of the form
$\ph_\ga(\mbf{r}_{a1})\ph_\ga(\mbf{r}_{b2})
+ \ph_\ga(\mbf{r}_{b1})\ph_\ga(\mbf{r}_{a2})$ 
for $H_2$ or $HD$
and 
$\ph_\ga(\mbf{r}_{a1})+\ph_\ga(\mbf{r}_{b1})$  
for $H_2^+$ or $HD^+$,
where $\ph_\ga (r)\equiv\exp(-\ga r)$
and the displacements between the pointlike nuclei $f=a,b$
and the electrons $j=1,2$ 
are denoted by $r_{fj}$. 
The electron wavefunctions depend on two parameters, 
the bond length $R$ of the molecule 
and the fall-off parameter $\ga$,
both of which are fixed by minimizing
the expectation value of the electron hamiltonian
in the Born-Oppenheimer approximation.  

Here,
we extend this methodology
to nonrelativistic spherical coefficients with $k=2$ in the electron sector, 
which includes operators of arbitrary nonminimal dimension $d$.
Incorporating coefficients with $k >2$ also of interest in principle,
but it turns out that the higher powers of the momentum operator
accompanying the larger values of $k$ 
become unbounded with the simple wavefunction ansatz adopted here.
A more sophisticated ansatz is likely to overcome this issue
and would be of interest to investigate
but lies beyond our present scope.

The relevant perturbation hamiltonian $\de h^\nr_e$
for Lorentz and CPT violation in the electron sector 
is given by Eqs.\ \rf{nr}-\rf{cpt} with $\f=e$ and $k=2$.
Using the approproate ansatz for the wavefunction 
and working in a frame in which the $z$ axis is aligned 
along the displacement $\mbf{R}$ 
between the positions of the two nuclei,
we find the energy shift of the ground state of $H_2$ 
due to Lorentz and CPT violation 
is given by
\beq
\vev{\de h^\nr_e}_{H_2} =
\fr{-1}{\sqrt{\pi}}
\Big\langle
p^2 \Vnrf{e}{200} + \sqrt{5}( p_z^2-p_x^2) \Vnrf{e}{220}
\Big\rangle,
\label{mol1}
\eeq
while the shift for $H_2^+$ takes the form 
\bea
\vev{\de h^\nr_e}_{H^+_2} &=&
\fr{-1}{\sqrt{4\pi}}
\Big\langle
p^2 \Vnrf{e}{200} + \sqrt{5} (p_z^2-p_x^2) \Vnrf{e}{220}
\Big\rangle
\nonumber \\
&&
\hskip -40pt
- \vev{\si^3} \sqrt{\fr{3}{4\pi}}
\Big\langle 
\TzBnrf{e}{010} 
+ p_z^2 \TzBnrf{e}{210}
+ 2p_x^2 \ToBnrf{e}{210} 
\nonumber\\
&&
\hskip 20pt
+\sqrt{\frac 73}(p_z^2-p_x^2) \TzBnrf{e}{230}
\Big\rangle.
\label{mol2}
\eea
In the above equations,
$\Vnrf{e}{kj0}$,
$\TzBnrf{e}{kj0}$, 
and $\ToBnrf{e}{kj0}$
are related to the electron coefficients for Lorentz and CPT violation
via Eq.\ \rf{cpt}.
The derivation of these results takes advantage 
of axial symmetry to replace factors of $p_y^2$ with $p_x^2$
for convenience.

Numerical values for the expectation values of the momenta 
in the above expressions
are compiled in Table I of Ref.\ \cite{mu04}.
These tabulated values must be divided by a factor of two 
before substitution in Eq.\ \rf{mol1}
to match our use of $\mbf p$ for the electron momentum,
but they can be used directly in Eq.\ \rf{mol2}.
For $H_2^+$,
the expectation values of the electron spin operator $\mbf\si$
appearing in Eq.\ \rf{mol2} 
can be taken in the electron spin state to an excellent approximation
because the two protons are in a symmetric singlet.
Note that in principle operators with $E$-type parity might contribute
to the shift \rf{mol2}.
However,
these are associated with the difference
between $\vev{\si^1}$ and $\vev{\si^2}$.
The symmetry of the system suggests equality of these expectation values
and hence a zero contribution.
This symmetry might fail in a more realistic model,
but any corresponding effects are likely to be suppressed.

The analogue of the shift \rf{mol2} for $HD^+$
is also of potential interest.
However,
the spin state of the electron in $HD^+$ is nontrivial,
so expectation values involving all the components of $\mbf\si$ play a role.
The contributions from $\si^m$ with $m=\pm 1$ for this case are given by 
\bea
\vev{\de h^\nr_e}_{HD^+} &\supset&
-\sqrt{\fr 3{4\pi}}
\sum_{m}
\vev{\si^{m}}
\Big\langle
\sum_{k} p_x^{k} \TzBnrf{e}{k1m}
\nonumber \\
&&
\hskip 80pt
+(p_z^2-p_x^2)\ToBnrf{e}{21m}
\Big\rangle
\nonumber \\
&&
-\sqrt{\fr 7 {6\pi}}
\sum_{m}
\vev{\si^{m}}
\big\langle 
(p_z^2-p_x^2) \TzBnrf{e}{23m}
\big\rangle,
\nonumber\\
\eea
where the sum over $m$ spans the values $m=\pm 1$
and the sum over $k$ the values $k=0,2$.
The expectation values are understood 
to be taken in the electron part of the full spin wavefunction.

In addition to shifting the ground-state energy,
the presence of Lorentz and CPT violation
also modifies other physical quantities
\cite{mu04}.
The bond length $R$ is changed by an amount $\de R$,
which can be expressed as
\beq
\de R = -\fr1 {\prt_{R}^2 \ep_0}
\prt_{R} \vev{\de h^\nr_e},
\eeq
where the unperturbed ground-state energy $\ep_0$
can be taken as the miminum of the expectation value
of the hamiltonian in the absence of Lorentz violation.
The vibrational spectrum of the molecule within the electronic ground state
is also shifted.
An expression for this shift can be obtained 
by approximating the vibrating molecule as a harmonic oscillator
and calculating the effective change $\de \om_v$
in the resonance frequency $\om_v$ 
due to Lorentz and CPT violation.
This yields
\beq
\de \om_v = \fr{\om_v}{2\prt_{R}^2 \ep_0}
( \prt_{R}^3\ep_0 ~\de R
+ \prt^2_R \vev{\de h^\nr_e}).
\eeq
The rotational spectrum within the electronic ground state
is shifted by the Lorentz and CPT violation as well.
In the rigid rotor approximation,
this shift can be understood as an effective change 
$\de \om_r$
in the rotation frequency $\om_r$ given by
\beq
\de \om_r = -\fr{2\om_r}R ~\de R.
\eeq

In the limit of zero nonminimal coefficients,
all the above results reduce to the minimal-SME expressions
presented in Ref.\ \cite{mu04}.
The nonminimal terms introduce several qualitatively novel features.
One is the dependence of the bond length of $H_2^+$
on the electron spin state in the presence of Lorentz and CPT violation. 
Another noteworthy effect is the occurrence of contributions
from coefficients with $j=2$ and $j=3$
to the ground-state energies of all the molecular species.
This implies,
for example,
a signal involving sidereal variations 
at the third harmonic of the sidereal frequency,
which could be detected in measurements of suitable rovibrational transitions. 
Including terms with values $k>2$ would result 
in effects from coefficients with $j\geq 4$ as well,
together with the concomitant sidereal signals at higher harmonics.
In contrast,
as discussed in Sec.\ \ref{Matrix elements},
the ground state in atomic hydrogen only receives contributions
from coefficients with $j\leq 1$.

Rovibrational transitions with unchanged electronic state 
in $HD^+$ are dipole allowed.  
Current experiments with $HD^+$ have reached 
an impressive relative uncertainty of about $10^{-9}$ 
\cite{koe07}. 
The long lifetime of rovibrational excited states suggests 
considerable room for improvement remains,
and indeed it is believed possible in principle 
for future experiments to achieve relative uncertainties of 
about $5 \times 10^{-17}$ for $H_2^+$
and of order $10^{-18}$ for $HD^+$
\cite{sbk14}.
For illustrative purposes here,
consider future prospective relative uncertainties
of about $10^{-14}$ in frequency measurements. 
Using the expressions \rf{mol1} and \rf{mol2},
we find this corresponds to experimental sensitivities
of about $10^{-9}$ GeV$^{-1}$ 
to the nonrelativistic spherical coefficient with $k=2$.
For the coefficients with $j=2$ and $j=3$,
a glance at Table \ref{ExpSid} reveals that 
this represents a sharper measurement
by at least one order of magnitude
than would be available using atomic hydrogen. 
 
The discussion in this subsection suffices to confirm
that spectroscopy of hydrogen molecules 
has the potential to provide competitive searches 
for Lorentz and CPT violation in coming years.
Several improvements on the theoretical treatment can be envisaged.
A comparatively straightforward one
would be to adopt improved unperturbed electron ground states.
For example,
a more detailed form of the $H_2^+$ wavefunction 
has been available for many decades
\cite{jo41},
and high-accuracy computations of various systematic effects now exist 
\cite{hi01}.
This or related improvements in the ground-state wavefunctions
could also make feasible the calculations
for coefficients with $k>2$.
Another improvement would be to incorporate contributions
from the nucleons.
The perturbative rovibrational shifts established above
can be traced to the effective shift of the nucleon separation $\mbf R$.
However,
in reality these transitions drastically change the states of the nucleons,
which could plausibly lead to signals allowing sensitive measurements 
of coefficients in the proton and neutron sectors of the SME.

\section{Summary}
\label{Summary}

In this paper,
we studied spectroscopic searches
for Lorentz and CPT violation
using hydrogen, antihydrogen,
deuterium, positronium, and hydrogen molecules
and molecular ions.
Our considerations begin in Sec.\ \ref{Theory}
with a treatment of theoretical aspects for hydrogen spectroscopy.
The leading-order perturbative hamiltonian \rf{pert}
is constructed in the nonrelativistic limit,
incorporating coefficients for Lorentz and CPT violation
of arbitrary mass dimension $d$.
These coefficients can be expressed in a spherical basis
and separated into ones controlling CPT-even and CPT-odd effects,
as presented in Eq.\ \rf{cpt}.
The symmetries of the hydrogen atom restrict
the possible perturbative contributions to certain coefficients,
listed in Table \ref{rest} 
along with some of their key features.
The formalism permits determining the general matrix elements \rf{mel}
of the full perturbation hamiltonian $\de h_\atm^\nr$.
The result is used in Sec.\ \ref{Energy shifts for $F=0$ and $F=1$}
to establish analytical expressions for the perturbative energy shifts.
We follow this with a discussion of the general features
of effects on the hyperfine Zeeman transitions,
including notably the sidereal variations \rf{rotgen}
and the annual variations discussed in
Sec.\ \ref{Annual variations and parity-odd operators}.

The potential applications of our methodology
to measurements using hydrogen spectroscopy
are discussed in Sec.\ \ref{Applications}.
We first address the issue of possible signals for free hydrogen
in Sec.\ \ref{no field},
and then consider spectroscopy in an external magnetic field.
We derive the explicit formula \rf{1Sf}
for the Lorentz- and CPT-violating shift of the hyperfine Zeeman frequency,
and we combine it with published results
from experiments searching for sidereal variations using a hydrogen maser  
\cite{hu00,maser,hu03}
to place the constraint \rf{1Sb}
on a combination of nonrelativistic coefficients
controlling spin-dependent effects.
Taken one coefficient at a time,
this constraint yields the results displayed 
in Table \ref{tanr1S}.
The prospects for hyperfine Zeeman measurements using annual variations
and studies on a space-based platform are discussed in 
Secs.\ \ref{Boost corrections and parity-odd operators}
and \ref{Space-based experiments}.
We then turn to precision spectroscopy
with $nL$-$n'L'$ transitions in hydrogen.
The frequency shift 
for any hydrogen transition of this type with $J=1/2$, $\De J=0$ 
arising from isotropic Lorentz and CPT violation 
is given by Eq.\ \rf{Jhalf}.
In Sec.\ \ref{Sidereal and annual variations due to boost corrections},
we consider various options for using annual variations 
of the $1S$-$2S$ transition frequency
to measure coefficients for Lorentz and CPT violation.
Results from an existing experiment of this type 
\cite{ma13} 
are used to place first constraints 
on various nonminimal cartesian coefficients
with mass dimensions $5\leq d\leq 8$,
as reported in Table \ref{cartcoeff}.
In Sec.\ \ref{The nSnD transitions}
the possibility of measuring nonrelativistic coefficients with $j\geq 2$ 
using these types of transitions 
is discussed,
and estimates for the reach of future analyses 
are presented in Table \ref{ExpSid}.

Antihydrogen spectroscopy 
in the context of the search for Lorentz and CPT violation 
is the topic of Sec.\ \ref{Antihydrogen}.
The implementation of the CPT transformation
on the hydrogen spectrum is provided in Sec.\ \ref{hbar basics}.
For hyperfine Zeeman transitions,
the induced frequency shift is presented in Eq.\ \rf{1Sfb},
while for Paschen-Back transitions 
it is given in Eq.\ \rf{pasbacb}.
The $1S$-$2S$ transition in free antihydrogen 
is shown to depend only on nonminimal coefficients
for Lorentz and CPT violation,
with the corresponding frequency shift specified in Eq.\ \rf{Jhalfb}.
For all these cases,
we provide estimates of attainable sensitivities
both from direct measurements
and from comparisons with hydrogen spectroscopy.
In Sec.\ \ref{hbar gravity},
the prospects for detecting an anomalous gravitational response 
of antihydrogen is considered. 
Insight into the role of nonminimal operators
is obtained by constructing a generalization
of the isotropic parachute model
\cite{akjt}.
Sensitivity estimates for future experiments are obtained.

Deuterium spectroscopy is the focus of Sec.\ \ref{Deuterium}.
We obtain the corrections to the $1S$-$2S$ transition frequency
arising from isotropic Lorentz and CPT violation
in Eq.\ \rf{degenf}.
Associated signals are discussed in Sec.\ \ref{Deuterium self consistent}.
The observed difference between the square of the charge radii
of the proton and deuteron
\cite{we95,je11}
is used to derive the constraint \rf{deutconst}
on nonrelativistic coefficients.
Another interesting approach to using deuterium spectroscopy
to search for Lorentz and CPT violation
is performing hyperfine Zeeman measurements with a deuterium maser.
The implications of this are discussed in Sec.\ \ref{Deuterium maser},
where we show that the Lorentz and CPT reach of a deuterium maser
represents in principle an improvement
of many orders of magnitude over that of a hydrogen maser.

Some aspects of positronium spectroscopy 
in the context of the search for Lorentz and CPT violation
are considered in Sec.\ \ref{Positronium}.
We obtain expressions for the measurable frequency shifts
and use the observed difference between 
theoretical and experimental values of the $1S$-$2S$ transition frequency
to deduce the measurement \rf{1S2Spos}
of isotropic coefficients in the electron sector.
Finally,
spectroscopy with hydrogen molecules and molecular ions 
is the subject of Sec.\ \ref{Hydrogen molecules}.
We determine the energy shifts \rf{mol1} and \rf{mol2}
for the ground states of these systems,
and we discuss the sensitivities
and potential advantages of the corresponding frequency measurements. 

The discussions in this paper 
provide a working summary of the prospects
for observing Lorentz and CPT violation
using precision spectroscopy of a number of comparatively simple systems.
Although our analysis has led to a variety of
new or improved constraints,
many coefficients for Lorentz and CPT violation
remain unmeasured at present.
The numerous potential signals identified 
and impressive attainable sensitivities in future experiments
offer strong motivation for further analyses,
along with encouragement for a potential breakthrough discovery
in this foundational subject.

\section*{Acknowledgments}
\bigskip

This work was supported in part
by the Department of Energy under grant number {DE}-SC0010120
and by the Indiana University Center for Spacetime Symmetries.

\end{document}